\documentclass[
    aps, pra, reprint,
    twocolumn,
    superscriptaddress,
    showkeys,
    xcolor, groupedaddress,
    nofootinbib
    ]{revtex4-2}

\usepackage[utf8]{inputenc}
\usepackage{bbm}
\usepackage{appendix}

\usepackage{lipsum}

\usepackage{tabstackengine}
\usepackage{cancel}
\usepackage{enumitem}
\setlist{nosep}

\makeatletter
\renewcommand\@seccntformat[1]{%
  \ifcsname the#1\endcsname
    \csname the#1\endcsname\quad
  \fi}
\makeatother
\makeatletter
\renewcommand{\p@subsubsection}{}
\makeatother

\usepackage{amsmath}
\usepackage{amssymb}
\usepackage{braket,nicefrac}
\usepackage{siunitx,upgreek}
\usepackage{mathtools}
\usepackage{bm}

\usepackage[table]{xcolor} 
\usepackage{hhline}
\usepackage{tabularx}
\usepackage{diagbox}
\usepackage{dcolumn}
\usepackage{multirow}
\usepackage{makecell}
\usepackage{bigdelim}
\usepackage{blkarray}
\usepackage{array}

\usepackage{tabularray}
\UseTblrLibrary{diagbox}

\usepackage{qcircuit}
\usepackage{graphicx}
\usepackage{float}
\usepackage{rotating}
\usepackage{tikz}
\usepackage{svg}

\usepackage{hyperref}
\hypersetup{
final=true,
plainpages=false,
pdfstartview=FitV,
pdftoolbar=true,
pdfmenubar=true,
bookmarksopen=true,
bookmarksnumbered=true,
breaklinks=true,
linktocpage,
colorlinks=true,
linkcolor=black,
filecolor=black,      
urlcolor=blue,
citecolor=black,
anchorcolor=black
}
\usepackage[nameinlink]{cleveref}
\crefname{equation}{Eq.}{Eqs.}
\crefname{align}{Eq.}{Eqs.}
\crefname{figure}{Fig.}{Figs.}
\crefname{table}{Table}{Tables}
\crefname{tabular}{Table}{Tables}
\crefname{section}{Sec.}{Secs.}
\crefname{appendix}{App.}{Apps.}

\pretocmd{\appendix}{\crefalias{section}{appsec}}{}{}

\crefname{appsec}{App.}{Apps.}
\crefname{appchapter}{App.}{Apps.}

\crefname{algorithm}{Alg.}{Algs.}
\creflabelformat{equation}{#2#1#3}

\usepackage{xurl}

\usepackage{color}
\usepackage{colortbl}

\definecolor{qctrl_primary}{HTML}{680Ce9}
\definecolor{qctrl_secondary}{HTML}{BF04DC}
\definecolor{qctrl_noise}{HTML}{7B7479}
\definecolor{qctrl_axis_labels}{HTML}{514B4F}
\definecolor{qctrl_borders}{HTML}{CFCBCE}
\definecolor{qctrl_blue}{HTML}{4177D8}
\definecolor{qctrl_aqua}{HTML}{32A4A8}
\definecolor{qctrl_green}{HTML}{32A857}
\definecolor{qctrl_lime_green}{HTML}{A2A933}
\definecolor{qctrl_orange}{HTML}{D6742F}
\definecolor{qctrl_red}{HTML}{D84144}
\definecolor{qctrl_fuchsia}{HTML}{D84190}

\makeatletter
\newcommand{\shorteq}{%
  \settowidth{\@tempdima}{\,--}
  \resizebox{\@tempdima}{\height}{\,=}%
}
\let\oldsqrt\sqrt
\def\sqrt{\mathpalette\DHLhksqrt}
\def\DHLhksqrt#1#2{%
\setbox0=\hbox{$#1\oldsqrt{#2\,}$}\dimen0=\ht0
\advance\dimen0-0.2\ht0
\setbox2=\hbox{\vrule height\ht0 depth -\dimen0}%
{\box0\lower0.4pt\box2}}

\definecolor{tNR}{HTML}{FADBD8} 
\definecolor{tYG}{HTML}{D5F5E3} 
\definecolor{tPO}{HTML}{FAE5D3} 
\NewTblrTableCommand{\tN}{\SetCell{bg=tNR}}
\NewTblrTableCommand{\tY}{\SetCell{bg=tYG}}
\NewTblrTableCommand{\tP}{\SetCell{bg=tPO}}

\newcommand{\csubsubsec}[1]{\vspace{0.5em} \noindent \textbf{#1}}

\usepackage{enumitem}

\newlist{cseclist}{description}{1}
\setlist[cseclist]{
  topsep=0.2\baselineskip,
  font=\normalfont\itshape,
  partopsep=0pt
}

\begin{document}

\title{Heterogeneous architectures enable a 138x reduction in physical qubit requirements for fault-tolerant quantum computing under detailed accounting}

\author{Pranav S. Mundada}
\altaffiliation{Contributed equally}
\email{pranav.mundada@q-ctrl.com}
\author{Aleksei Khindanov}
\altaffiliation{Contributed equally}
\author{Yulun Wang}
\altaffiliation{Contributed equally}
\author{Claire L. Edmunds}
\author{Paul Coote}
\author{Michael J. Biercuk}
\author{Yuval Baum}
\author{Michael Hush}

\affiliation{Q-CTRL, Los Angeles, CA USA and Sydney, NSW Australia}
\date{\today}

\begin{abstract}
Quantum computer hardware performance and scale have accelerated rapidly over the past few years, enabling real workloads to be run using hundreds of physical qubits, and industry roadmaps predicting systems with hundreds of thousands of qubits coming online in the next decade. Moving to this scale requires major advances in quantum error correction (QEC) and its useful integration into real hardware systems. Despite significant theoretical and experimental QEC progress, quantum computer architecture has suffered a significant gap, with bottom-up physical-device-driven challenges largely disconnected from top-down QEC-code-driven considerations. In this work, we unify these two views, presenting a complete heterogeneous quantum computing architecture incorporating task-specific hardware selection and QEC encoding, and agnostic to code selection or physical qubit parameters. The architecture is based on the observation that due to the lengthy idling periods encountered in common algorithms, it can be advantageous to separate quantum processing operations from quantum memory, enabling expanded flexibility in qubit modality and code choice between subsystems. Our approach further enables special-purpose processing modules, and includes a full microarchitecture for fault-tolerant implementation of interfaces between quantum processing units and quantum memories.
Using this architecture and a new fully featured compiler functioning across subsystems at the scale of $1,000$ logical qubits, we schedule and orchestrate a variety of algorithms down to hardware-specific instructions; 
a detailed accounting of all operations reveals up to $551\times$ reduction in algorithmic logical error and up to $138\times$ reduction in physical-qubit overhead compared to a monolithic baseline architecture. We then consider the factorization of 2048-bit RSA-integers; using 
an experimentally demonstrated grid-coupling topology, factoring RSA-2048 requires 381k physical qubits and 9.2 days, which can be reduced to 4.9 days via addition of an algorithm-specific accelerator for the Adder subroutine (requiring 439k qubits). Finally, assuming hypothetical long-range coupling, implementing quantum memory using qLDPC codes reduces the resources required for factoring to just 190k qubits and under 10 days. These results and the tooling we have built indicate that heterogeneous quantum-computer architectures can deliver significant, verifiable benefits on realistic hardware.

\end{abstract}

\maketitle

\section{Introduction}
\label{sec:intro}

The realization of large-scale quantum computing, implying the use of fault-tolerant error-corrected protocols, is widely recognized as a strategic capability with transformative potential across materials science, defense, finance, and energy \cite{WhiteHouse2022}. Over the past few years the community has seen significant technical progress towards the implementation of quantum error correction using superconducting, trapped-ion, and neutral atom systems \cite{AtomFaultTolerant2025, QueraFaultTolerant2025, GoogleFaultTolerant2025, Yamamoto2025}. These demonstrations have validated many of the scientific concepts underpinning fault tolerance, and spurred heightened interest in the pathway to building large-scale machines at cryptographically relevant scales. 

Scaling from today's machines with $\sim100-200$ qubits \cite{AtomFaultTolerant2025, QueraFaultTolerant2025, GoogleFaultTolerant2025} to the hundreds of thousands or more expected to be required for the fully fault-tolerant implementation of high-value algorithms presents foundational challenges in device architecture. The quantum industry is now confronting its own ``tyranny of numbers'' \cite{Franke2018}; in the 1950s this term described the linear growth of wiring and interconnect complexity with the number of discrete components, a regime that ultimately forced the transition to integrated circuits and heterogeneous architectures\textsuperscript{\ref{fn:rent}}.
A comparable structural-scaling challenge is now evident in quantum hardware; in most platforms, increasing qubit count entails a near-proportional expansion of control wiring, readout channels, and qubit-shuttling overhead. 
\footnotetext{\label{fn:rent}In well engineered classical ICs, empirical observations later formalized as Rent’s rule show that the number of external terminals scales sub-linearly with system size \cite{Franke2018}. Today, many current quantum-hardware platforms rely on near-linear scaling of control and readout connections with qubit number.}

The impacts of such pressures on contemporary device design have been explicitly identified in silicon spin~\cite{Reilly2015}, superconducting~\cite{Krinner2019,Kawabata2026}, and trapped-ion \cite{Malinowski2023} systems. At the thousand-qubit scale, these constraints are no longer theoretical: IBM’s 1,121-qubit Condor processor \cite{Castelvecchi2023} required approximately a mile of cryogenic flex I/O wiring within a single dilution refrigerator \cite{IBMOldRoadmap}. The industry’s subsequent shift toward multi-chip modular systems, including IBM’s 
Nighthawk series~\cite{IBMCurrentRoadmap}, and systems from Rigetti~\cite{RigettiMultiChip2024}, Pasqal~\cite{Pasqal2025}, IonQ~\cite{IonQ2026}, and Xanadu~\cite{Xanadu2026}, reflects a recognition that further system scaling will require architectural reorganization rather than simple scaling of monolithic hardware. These architectural approaches are supported by activity in the development of physical interconnects to transfer quantum states via teleportation protocols between modules~\cite{Caleffi2024, Nickerson2014, Haug2025, Jacinto_2026, Singh2025modular}, partially addressing system I/O bottlenecks. Modular quantum computation connected by interconnects was first demonstrated experimentally by Monroe \emph{et al.}~\cite{Monroe2014}, and interest in the role of interconnects has since accelerated: ETH Zurich~\cite{Magnard2020} and IBM~\cite{Heya2025} have demonstrated meter-scale microwave interconnects; dedicated interconnect companies have formed (Nu Quantum \cite{NuSite}, Aliro~\cite{AliroSite}, Qunnect \cite{QunnectSite}); and hardware vendors have acquired photonic interconnect companies, including Aeponyx by Pasqal \cite{Pasqal2025} and Lightsynq by IonQ \cite{IonQ2026}.

Independent of these bottom-up physical-device design challenges, top-down architectural analyses have been driven largely by theoretical considerations arising from the QEC community. For instance, the large-scale RSA-resource-requirement analyses of Gidney \emph{et al.} \cite{GidneyEkera2021RSA,Gidney2025RSAUpdate} focus on logical metrics and then inform device architecture primarily through code recommendation. This approach reveals many orders-of-magnitude variability in overall hardware and algorithmic-runtime estimates based on small variations in device-performance assumptions and code selection. Even when extended to homogeneous modular architectures as a new ``multi-core'' concept~\cite{Caleffi2024}, the key structural and analytic ``code-first'' approaches have been maintained~\cite{Marqversen2025,Strikis2023,Sutcliffe2025}. 

There is a persistent gap between these two approaches in considering quantum computer architecture. In top-down QEC-driven analyses, hardware complexity is compressed into a small set of scalar parameters -- for instance, physical error rate and logical cycle time -- reporting outcomes in terms of physical qubit numbers, code distances, and asymptotic error scaling. By construction, system-level constraints such as the classical control stack, synchronization overhead, and interconnect scaling are excluded from the cost model. These abstractions can obscure or even conflict with the previously established infrastructure pressures that ultimately constrain physical system architectures. For instance, recent work by Webster \emph{et al.} \cite{Webster2026Iceberg} reports QEC-code-driven reduction in physical overheads bringing the resource estimates for factoring a 2048-bit number to $\sim10^5$ qubits, by assuming qubits can have arbitrary non-local couplings that are not typically available or planned on many dominant platforms. Another recent work by Cain \emph{et al.} \cite{cain2026} reports a further reduction in the qubit count, to $\sim10^4$, at the price of unfeasible runtime of $\sim$120 years.
New insights arising from top-down approaches such as this can certainly motivate future hardware development and, of course, code selection. Nonetheless, optimizing the QEC encoding approach in isolation does not answer the question of how such a device should realistically be built given physical-layer constraints in device fabrication, connectivity, and modular interconnects.

In this work, we attempt to bridge these divergent analytical starting points and provide a novel solution to the tyranny of numbers in quantum computing hardware. 
First, we embrace hardware heterogeneity at the architectural level, defining modules with distinct functional specialization that are matched to quantum information processing tasks. We specifically recognize that there is significant performance benefit to be derived by segregating the storage of idling quantum information from active processing; for instance, even in the efficient implementation of Shor provided in Refs.~\cite{GidneyEkera2021RSA,Gidney2025RSAUpdate}, on average each qubit is inactive for $\sim96-97\%$ of logical clock cycles. Further, we identify opportunities to achieve enhanced performance by incorporating application-specific modules for magic-state distillation, and introducing a fault-tolerant quantum bus for communications and information routing between modules. Within this modular architecture we further embrace heterogeneity of qubit modality and QEC encoding for each module, recognizing that device-performance variations as well as operational differences in code requirements may bias selection for use in different modules (e.g., transversality is not relevant in memory where logical operations between encoded qubits are not anticipated). %
Second, we describe and deliver a complete micro-architecture and compiler that explicitly models timing, connectivity, scheduling, buffering, and transfer latency across subsystems operating at different clock rates in order to allow effective resource estimation directly connected with physical-device parameters and constraints. Rather than treating execution as constituting uniform logical cycles parameterized by a single error rate and clock speed, we model the explicit sequence of machine instructions required for fault-tolerant execution under realistic hardware constraints.

By co-designing the architecture at every level --- logical compilation, physical micro-architecture, and hardware --- we demonstrate substantial reductions in both error and physical-qubit overhead for algorithms at the scale of $1,000$ logical qubits, using physically realistic assumptions about available hardware devices and the quantum bus. For the Quantum Fourier Transform (QFT) --- a subroutine critical to many implementations of Shor’s algorithm \cite{Kutin2006} --- our heterogeneous framework achieves a $42-59\times$ reduction in logical error and a $60-138\times$ reduction in physical-qubit overhead relative to a homogeneous baseline using the same device parameters. For dynamic simulations of the Fermi–Hubbard model and for arithmetic (Adder), we observe reductions in logical errors over $100\times$. 

We then turn our attention to the factorization of a 2048-bit RSA integer using the sequential implementation of Gidney's algorithm as presented in Ref.~\cite{Gidney2025RSAUpdate}\footnote{Parallelizing the outer loop may provide more beneficial space-time tradeoffs; we consider that to be an algorithmic innovation beyond the scope of this paper.}. Our estimation fully compiles and schedules the three underlying subroutines in the $1,399$-logical-qubit program, $33$-bit Adder, $6$-bit Lookup and $6$-bit ``Phaseup'', on a heterogeneous architecture, accurately accounting for idling error, delays in CCZ-state supply, and state-transfer space-time cost. Using an experimentally demonstrated Bell-pair purification scheme and two quantum processor units (QPUs), we find that 2048-bit RSA integers can be factored with only 381k physical qubits and in 9.2 days. Introducing a dedicated $37$-logical-qubit algorithm-specific accelerator to overcome the runtime bottleneck in the Adder subroutine reduces factorization time by $\sim87\%$ to 4.9 days, at a $13\%$ hardware penalty, now requiring 439k physical qubits. These estimates use error rates and cycles times as in Table~\ref{table:compiler input}, fully account for the physical qubits required for fault tolerant state-transfer and high quality T-cultivation, and only assume an experimentally demonstrated grid topology that does not incorporate long-range connectivity. If one then assumes that the nonlocal physical connectivity, required in \cite{Yoder2025} using qLDPC codes, is achievable in memory, the physical resource requirements for factoring a 2048-bit number in our heterogeneous architecture drop to 190k physical qubits and 9.2 days. To our knowledge, these results represent the first demonstrations of circuit-level scheduling and orchestration of a factoring algorithm on a modular architecture, and include the first bottom-up accounting for runtime and physical-device resourcing. 

The remainder of this paper is organized as follows. In \cref{sec:intro-qnexus}, we establish formal design requirements for high-performance heterogeneous architectures used to guide the downstream decisions presented here. \cref{sec:framework} provides detailed operational specifications for each constituent hardware module in what we term the ``Q-NEXUS'' hardware framework, encompassing processing units, state factories, memory units, and quantum buses to facilitate interfaces between the different modules. In \cref{sec:compiler}, we introduce the new compilation pipeline, called ``Q-CHESS'', outlining the synchronization and routing methods used to orchestrate program execution across modules with disparate clock rates. \cref{sec:midnalysis} presents a comprehensive quantitative resource analysis, benchmarking the architecture using the Quantum Fourier Transform, a Quantum Adder, and simulation of Fermi-Hubbard dynamics. In \cref{sec:rsa}, we conduct similar analysis for the task of RSA-2048 integer factorization. \cref{sec:discussion} discusses the broader implications of this stored-program approach for scalable quantum computing, and \cref{sec:conclusion} concludes the work.

\section{Requirements for high-performance heterogeneous quantum architectures}\label{sec:intro-qnexus}

We aim to design a heterogeneous quantum computing architecture that scales by reducing resource overheads relative to monolithic designs. To establish a baseline, we consider the prototypical fault-tolerant monolithic architecture introduced by Litinski \emph{et al.}~\cite{Litinski2019} which has inspired numerous architectural proposals with resource estimates \cite{Kan2025, Guinn2023, Webster2026Iceberg, Robertson2025}. These approaches share three defining characteristics: 
\begin{enumerate}
\item They are \emph{code-first}, deriving architectural structure directly from a single QEC code and associated fault-tolerant implementation. 
\item They assume a monolithic ``sea'' of physical qubits: a contiguous lattice with fixed local connectivity determined by the code, whose total size is not specified a priori but instead grows to meet algorithmic and error-rate requirements. 
\item Functional roles -- such as memory, compute, and routing -- are assigned to lattice ``regions'' during execution, rather than being formally segregated. These regions are typically defined in an algorithm-specific manner, with efficiency arguments tied to particular workloads, implying that their allocation is determined dynamically at runtime \cite{Gidney2025RSAUpdate,Webster2026Iceberg}.
\end{enumerate}

Moving beyond these approaches, incorporating heterogeneity to overcome the limitations of homogeneous designs constitutes a multi-level optimization problem. At the highest level, fault tolerance requires minimizing space-time overhead for high-value algorithms under strict logical error budgets. At the lowest level, physical feasibility is constrained by the tyranny of numbers, where increasing device scale drive growth in wiring density, cryogenic load, and control complexity. These constraints are not captured by qubit counts alone; meaningful resource estimation requires simultaneous consideration of algorithmic demands and hardware-production roadmaps.

To address these competing constraints, we embrace modularity as a core design principle, adopting a fully heterogeneous architecture in which the system is decomposed into distinct modules with well-defined functional roles. Each module is specialized for computation, communication, or storage, rather than attempting to support all functionality within a single homogeneous lattice. By restricting each module to a narrow functional role, implementation complexity is reduced, lowering resource overheads. This decomposition further enables each function to be implemented by the most suitable quantum hardware and the most suitable QEC code.

At a system level, this approach parallels the stored-program model that underpins modern high-performance computing systems, including CPUs and GPUs~\cite{Hennessy2012,Patterson2017}, and traces back to the original proposal by von Neumann \emph{et al.}~\cite{vonNeumann1945}. These systems address the same fundamental challenges of computation, communication, and storage of data. The key distinction in quantum computing lies in the representation of that data: QEC-protected encodings of physical qubits. System-level principles from both modern computing and historical work\textsuperscript{\ref{fn:vNeu}} therefore provide a useful foundation for designing quantum computer architectures at scale.

\footnotetext{\label{fn:vNeu} Von Neumann criticized designs that accelerated computation by ``telescoping'' arithmetic logic -- multiplying arithmetic elements at the cost of increased system complexity \cite{vonNeumann1945}. He instead proposed scaling a central memory storing instructions and data, establishing the stored-program architecture. A similar tradeoff arises in quantum computing: homogeneous modular architectures replicate quantum processing units, whereas our approach scales memory.}

We first define a set of system-level requirements, ensuring that modules are specified by their \emph{function} -- how they operate on quantum data\textsuperscript{\ref{fn:qdata}} -- rather than by any particular implementation. We group these requirements into the following sections:\footnote{\label{fn:qdata}We use the term \emph{quantum data}, as opposed to \emph{quantum information}, to emphasize its role in the architecture. It is distinguished from the \emph{classical instruction} (i.e. the algorithmic quantum circuit) which governs how the quantum data is processed. This terminology mirrors the stored-program paradigm, where data and instructions are distinct.}

\begin{enumerate}[leftmargin=*]
    \item \textbf{Computation:} How modules apply universal quantum logic operations to quantum data.
    \item \textbf{Communication:} How quantum data is transferred between modules.
    \item \textbf{Storage:} How modules store quantum data efficiently at scale.
    \item \textbf{Control:} How classical instructions orchestrate the processing, movement, and timing of quantum data.
    \item \textbf{Representation:} How quantum data is physically embodied and encoded for fault tolerance.
\end{enumerate}

\vspace{0.5em}
In the following we define two classes of requirements. Firm requirements (Requirement [Name]: ``shall'') specify structural properties that must be satisfied by the architecture, while optional requirements (Optional requirement [Name]: ``should'') capture performance-oriented goals. Each requirement is labeled by a three-letter key derived from its description. Firm requirements are verified directly in the architectural specification of Q-NEXUS and its compiler Q-CHESS (Sec.~\ref{sec:compiler}), whereas optional requirements are evaluated through compiled resource estimates in Sec.~\ref{sec:midnalysis}.

\newcommand{\reqfmt}[1]{\textit{#1} \vspace{0.2em}}


\subsection{Computation}
\csubsubsec{Requirement (Fixed-size processor, FXD):}
\reqfmt{Computation of universal fault-tolerant quantum logic on quantum data shall be performed within fixed-size quantum processing unit(s) (QPU).} 

Hardware designed to execute universal logic must remains limited in size, at scales where classical scaling limits and fault-tolerant protocol developments are well in hand~\cite{Zhang2025,Pogorelov2025,Harper2019,Wang2024,Yamamoto2025}. By limiting the physical size of the QPU, we mitigate yield issues and crosstalk that plague large-scale monolithic chips, and reduce the demand for encoded logical qubits with universal functionality. Architecturally, each QPU should be treated as a high-complexity resource with bounded size and \emph{connectivity}. Further limiting physical connectivity to grid topology accounts for otherwise challenging growth in coupler counts, frequency collisions, and calibration burden \cite{GoogleFaultTolerant2025}. Appropriate architectural constraints can account for the opportunities provided by higher-connectivity devices e.g. IBM’s bivariate-bicycle architecture \cite{Yoder2025}.

\csubsubsec{Requirement (Magic state generation, MGC):}
\reqfmt{ Magic states for non-Clifford operations on quantum data shall be generated by a specialized quantum state factory (QSF).} 

Magic state factories~\cite{Rodriguez2025} materially influence system cost and must be architecturally explicit rather than incidental, as Magic state distillation is a dominant overhead in fault-tolerant resource estimates \cite{Gidney2024}. 

\csubsubsec{Optional requirement (Multi-core, MLT):}
\reqfmt{ The architecture should support parallel processing of quantum data across multiple QPU cores when it reduces execution time.}

Modern computing architectures employ multiple processing cores to exploit parallelism and reduce runtime for suitable workloads \cite{Hennessy2012}. A similar approach can benefit quantum programs that expose parallel execution of logical operations. Much like CPU cores, universal fault-tolerant processors are complex resources with substantial overheads so their number should remain bounded. Allowing a small number of processing cores can nevertheless provide significant runtime reductions.

\csubsubsec{Optional requirement (Application-specific accelerator, APP):}
\reqfmt{ The architecture should support APpLication-specific quantum processing units (ASQPU) that apply specialized logic operations to quantum data -- those which occur in a target algorithm with high frequency -- when they reduce execution time or resource overhead.} 

A heterogeneous architecture allows dedicated processing units to be optimized for algorithmic subroutines such as QFT, removing the requirement of full universality. ASQPUs perform a restricted set of logical operations extremely efficiently and can therefore be engineered specifically for their target workload. For algorithms that repeatedly invoke these subroutines -- such as Shor’s algorithm, which relies heavily on QFT -- dedicated accelerators can substantially improve execution efficiency.

\subsection{Communication}
\csubsubsec{Requirement (Micro-architecture, MCR):}
\reqfmt{ The architecture shall include a quantum bus (QB) for communication of quantum information between modules, with a MiCRo-architecture that includes fault-tolerant transfer protocols and resource generation.} 

Interconnects with limited functionality can allow the transfer of a qubit state across dense networks without creating an unrealistic load of local couplers, each requiring individually calibrated local operations~\cite{Nickerson2014, Haug2025}. We define a QB as these interconnects combined with complete \emph{fault-tolerant} transfer protocols across different modules. Resource estimation must explicitly account for the encoding requirements and overheads of each interface. 

\csubsubsec{Optional requirement (Long-range routing, LNG):}
\reqfmt{ LoNG-range routing of quantum data should be handled by the QB when it reduces logical (SWAP) operations.}

Routing is a dominant contributor to fault-tolerant overhead~\cite{Gidney2025RSAUpdate,Litinski2019, Kan2025,Guinn2023,Kobori2025}; long-range communication should be implemented through dedicated interconnects rather than through routing qubits embedded within the compute fabric. Introducing a quantum bus composed of interconnects can shift quantum transport to photonic channels (optical~\cite{AbuGhanem2024} and microwave~\cite{Gu2017, Magnard2020}), where Bell-pair generation enables teleportation-based state transfer. 

\vspace{-0.7em}
\subsection{Storage}
\csubsubsec{Requirement (Idling, IDL):} 
\reqfmt{The architecture shall include a dedicated quantum memory (QM) tier for storing IDLe quantum data.}

Architectures must account for the significant idling penalty implicit in most space-time optimal routines. Many quantum algorithms contain extended idle periods for individual qubits relative to the logical clock cycle. For example, in the implementation of Shor’s algorithm analyzed in Refs.~\cite{GidneyEkera2021RSA,Gidney2025RSAUpdate}, each qubit is inactive for approximately $96$--$97\%$ of logical cycles. It is inefficient to hold idling information in hardware optimized for universal logic or routing. This occurs in monolithic architectures, where memory is implemented, implicitly, as a subset of the ``sea'' of qubits~\cite{Litinski2019}. 
It is inefficient to store idling quantum information in hardware optimized for universal logic or routing. This occurs in monolithic architectures, as memory is not implemented as a separate subsystem~\cite{Litinski2019}. Architectures should therefore engineer an explicit memory tier designed for high-fidelity storage rather than general-purpose computation, building on demonstrations that scalable quantum memories are approaching practical viability~\cite{AtomFaultTolerant2025,QueraFaultTolerant2025,GoogleFaultTolerant2025,Wang2021}.

\csubsubsec{Requirement (Scaling, SCL):}
\reqfmt{System Scaling shall occur predominantly through storage of quantum data in the QM tiers.}

Memory has the simplest functional requirement -- reliable storage -- which reduces constraints and requirements on associated QEC encoding and physical-device connectivity. System growth should therefore occur primarily in memory, while universal logic remains bounded in size. Scaling the simplest subsystem minimizes wiring, cryogenic, and control overhead, recognized previously by Liu \emph{et al.} \cite{Liu2023} and Xu \emph{et al.}\cite{Xu2024}.

\csubsubsec{Optional requirement (Random-access memory, RND):}
\reqfmt{The architecture should include a random-access quantum memory (RAQM) tier capable of storing quantum data with uniform access latency.}

Long-duration storage of logical states requires a memory system that can be accessed efficiently during program execution. A RAQM provides this capability while supporting active QEC for reliable storage. RAQM involves both the use of long-coherence qubit modalities to reduce the required code distance relative to the QPU, and the use of qubits with non-local connectivity to enable higher-density quantum error-correcting codes. A defining property of this tier is that the time required to access stored information is approximately independent of its physical location within the memory. This uniform access latency allows memory capacity to scale without increasing retrieval time and enables predictable interaction between memory and processing units.

\csubsubsec{Optional requirement (Static memory, STC): } 
\reqfmt{The architecture should include a distinct ``Static'' transversal quantum memory (STQM) tier that defers active QEC to the QPU, enabling low-latency temporary storage of quantum data with simplified control.}

There exist qubit modalities with ultra-long coherence times (ULC), such as nuclear spins, where quantum information remains effectively static over timescales relevant to fault-tolerant computation \cite{Wang2025,Hughes2025,Liu2023}. This enables a distinct memory tier in which quantum states are stored without active error correction, provided they are returned to the QPU before accumulated errors exceed correctable thresholds. Here, supporting efforts in quantum-memory operation have already explored practical pathways to minimizing access latency while maximizing coherence preservation via development of appropriate control protocols~\cite{Khodjasteh_LowLatencyMemory}. By eliminating the need for continuous syndrome extraction and feedback, this ``static'' memory tier significantly reduces control complexity, calibration overhead, and interconnect requirements. It is therefore well suited for short-duration storage with minimal latency, acting as a quantum analogue of a classical cache \cite{Patterson2017}. 

\csubsubsec{Optional requirement (Hierarchical memory, HRC):} 
\reqfmt{The architecture should employ a memory Hierarchy when combining memory tiers to maintain low-latency access and high-fidelity long-term storage of quantum data.}

Idle periods in quantum programs span a wide range of durations. Some idle intervals are shorter than the latency required to transfer states to quantum memory technologies~\cite{Liu2023} that offer advantages in density or scalability but have slower write–read times than the QEC cycle time of processing qubits. In such cases transferring states to memory can introduce a net performance penalty due to access overhead. A standard solution in modern computer architecture is the introduction of a hierarchical memory system \cite{Patterson2017}. This approach mitigates latency differences by introducing specialized storage tiers with distinct performance characteristics. Fast tiers located close to the QPU buffer quantum information required for near-term computation, while slower but more scalable tiers store information needed later in program execution.

\subsection{Control}
\csubsubsec{Requirement (Machine-level instruction, MCH):}
\reqfmt{Control of quantum data shall be performed by Q-CHESS: a Quantum Compiler for Heterogeneous Execution Scheduling and Synthesis, which is micro-architecture aware and outputs Machine-level instructions.}

Accurately architecting heterogeneous systems requires the ability to accurately model resource requirements at a fine-grained level, accounting for the additional complexity of scheduling, routing, timing mismatches, buffering, and interface overheads introduced by distinct modules operating at different clock rates. Scaling arguments which abstract scheduling and execution details are not sufficient; without machine-level compilation~\cite{Robertson2025} that incorporates these effects, resource estimates risk misrepresenting true system costs.

\csubsubsec{Requirement (Timing control, TMN):}
\reqfmt{Q-CHESS shall efficiently account for TiMiNg mismatches between modules during processing of quantum data.}

Heterogeneous architectures, especially those using mixed qubit modalities, will operate with different clock cycles, mandating accounting of timing latencies; for example, a superconducting QPU may operate at $T_{\text{QPU}} \approx 1~\upmu$s \cite{GoogleFaultTolerant2025}, while an atomic memory operates at $T_{\text{QM}} \approx 100~\upmu$s--$1$~ms \cite{Wang2025,Hughes2025}. Such mismatches invalidate uniform execution models assuming identical clock cycles $T_{\text{cycle}}$. Furthermore, the use of hierarchical memory, including caches and RAQM, introduces additional compiler decisions on whether a state remains in the QPU, moves to cache, or moves to RAQM based on the expected idle duration and transfer latency. Scheduling decisions at scale therefore also depend on explicit modeling of this timing, buffering, and transfer latency. 

\vspace{-0.5em}
\subsection{Representation}
\csubsubsec{Optional requirement (multiple Qubit modality, QBT):}
\reqfmt{The architecture should support multiple physical QuBiT modalities for representing quantum data, when doing so reduces resource overheads. }

Functional specialization allows each qubit modality to occupy the architectural role best aligned with its intrinsic strengths, mitigating system-level compromises and creating new opportunities for hardware platforms within a heterogeneous computing stack. In heterogeneous architectures, fast, strongly coupled qubits can be applied for universal logic \cite{GoogleFaultTolerant2025}; long-coherence systems can be used for storage \cite{Wang2025,Hughes2025, Liu2023}; and photonic \cite{Monroe2014} or transduction-based \cite{Heya2025} interconnects applied to enable connectivity between modules. This approach breaks the ``one qubit to rule them all'' viewpoint that has shaped much of the discourse in the quantum industry \cite{Ball2024,Frackiewicz2025}, and has recently appeared, for instance, as a feature of DARPA’s Heterogeneous Architectures for Quantum (HARQ) initiative~\cite{DARPA2025HARQ}.

\csubsubsec{Optional requirement (multiple Codes, CDE):} 
\reqfmt{The architecture should support multiple error-correcting CoDEs for encoding quantum data, when it reduces resource overheads.}

Universal logic modules require codes with mature gate constructions (e.g., surface-codes \cite{Litinski2019} and bivariate bicycle codes \cite{Yoder2025}), whereas memory modules benefit from high-rate codes optimized for storage density while sacrificing ease-of-implementation for universal logic (e.g., LDPC-style codes \cite{Bravyi2024}). When modules utilize different QEC codes, the quantum bus should therefore support code conversion~\cite{Pogorelov2025}, enabling reliable transfer of logical information between them.

\subsection{Comparison between this work and prior heterogeneous architecture studies}

\begin{table*}[ht]
\renewcommand{\arraystretch}{1.2}
\begin{tblr}{
width=\textwidth,
colsep=2pt,
colspec={l|X[c]X[c]X[c]X[c]|X[c]X[c]|X[c]X[c]X[c]X[c]X[c]|X[c]X[c]|X[c]X[c]},
}
\hline\hline
\textbf{Q. Data} & %
\SetCell[c=4]{c} \textbf{Computation} & & & &%
\SetCell[c=2]{c} \textbf{Comm.} & &%
\SetCell[c=5]{c} \textbf{Storage} & & & & &%
\SetCell[c=2]{c} \textbf{Control} & &%
\SetCell[c=2]{c} \textbf{Repr.} & \\
\textbf{(Opt.) Req.} &%
\textbf{FXD} &\textbf{MGC} &\textbf{MLT} &\textbf{APL} &%
\textbf{MCR} &\textbf{LNG} &%
\textbf{IDL} &\textbf{SCL} &\textbf{STC} &\textbf{RND} &\textbf{HRC} &%
\textbf{MCH} &\textbf{TMN} &%
\textbf{QBT} &\textbf{CDE} \\%
\hline

Litinski$^*$ \cite{Litinski2019}      & \tP Part$^1$    & \tY Yes & \tN No  & \tN No  & \tP Part$^2$    & \tN No  & \tP Part$^1$    & \tN No  & \tN No & \tN No          & \tN No  & \tY Yes$^3$     & \tN No  & \tN No       & \tN No  \\
Gidney$^*$ \cite{Gidney2025RSAUpdate} & \tP Part$^1$    & \tY Yes & \tN No  & \tN No  & \tN No          & \tN No  & \tP Part$^1$    & \tY Yes & \tN No & \tN No          & \tY Yes & \tY Yes         & \tN No  & \tN No       & \tY Yes \\
Marian. \cite{Mariantoni2011}         & \tY Yes         & \tN No  & \tN No  & \tN No  & \tY Yes         & \tN No  & \tY Yes         & \tN No  & \tN No & \tN No          & \tN No  & \tN No          & \tN No  & \tP Part$^4$ & \tN No  \\
Monroe \cite{Monroe2014}              & \tY Yes         & \tN No  & \tY Yes & \tN No  & \tY Yes         & \tY Yes & \tN No$^5$      & \tN No  & \tN No & \tN No          & \tN No  & \tY Yes         & \tY Yes & \tP Part$^6$ & \tN No  \\
Brandl \cite{Brandl2017}              & \tY Yes         & \tN No  & \tN No  & \tN No  & \tY Yes$^7$     & \tY Yes & \tY Yes         & \tY Yes & \tN No & \tY Yes         & \tN No  & \tN No          & \tN No  & \tP Part$^6$ & \tN No  \\
Gouzien \cite{Gouzien2021}            & \tY Yes         & \tY Yes & \tN No  & \tN No  & \tP Part$^2$    & \tN No  & \tY Yes         & \tY Yes & \tN No & \tN No          & \tN No  & \tN No          & \tY Yes & \tY Yes      & \tN No  \\
Stein \cite{Stein2023}                & \tP Part$^8$    & \tN No  & \tY Yes & \tY Yes & \tY Yes         & \tN No  & \tP Part$^8$    & \tN No  & \tN No & \tN No          & \tY Yes & \tP Part$^9$    & \tY Yes & \tP Part$^4$ & \tY Yes \\
Andres. \cite{Andres2024}             & \tY Yes         & \tN No  & \tY Yes & \tN No  & \tP Part$^{10}$ & \tN No  & \tN No          & \tN No  & \tN No & \tN No          & \tN No  & \tP Part$^{10}$ & \tN No  & \tN No       & \tN No  \\
Xu \cite{Xu2024}                      & \tY Yes         & \tN No  & \tN No  & \tN No  & \tP Part$^2$    & \tN No  & \tY Yes         & \tY Yes & \tN No & \tN No          & \tN No  & \tN No          & \tN No  & \tN No       & \tY Yes \\
Liu \cite{Liu2023}                    & \tY Yes         & \tN No  & \tN No  & \tN No  & \tP Part$^2$    & \tN No  & \tY Yes         & \tY Yes & \tN No & \tY Yes         & \tN No  & \tN No          & \tN No  & \tY Yes      & \tN No  \\
Kobori \cite{Kobori2025}              & \tP Part$^1$    & \tY Yes & \tN No  & \tN No  & \tN No          & \tN No  & \tP Part$^1$    & \tY Yes & \tN No & \tP Part$^{11}$ & \tY Yes & \tP Part$^9$    & \tY Yes & \tN No       & \tN No  \\
Bicycle \cite{Yoder2025}              & \tY Yes         & \tY Yes & \tY Yes & \tN No  & \tP Part$^2$    & \tN No  & \tY Yes         & \tY Yes & \tN No & \tN No          & \tN No  & \tP Part$^9$    & \tN No  & \tN No       & \tN No  \\
Helios \cite{Ransford2025}            & \tY Yes         & \tN No  & \tY Yes & \tN No  & \tY Yes$^7$     & \tY Yes & \tY Yes         & \tY Yes & \tN No & \tY Yes         & \tY Yes & \tP Part$^{10}$ & \tY Yes & \tP Part$^6$ & \tN No  \\
ModEn-Hub \cite{Chen2025}             & \tN No          & \tN No  & \tY Yes & \tN No  & \tY Yes         & \tY Yes & \tN No          & \tN No  & \tN No & \tN No          & \tN No  & \tP Part$^9$    & \tY Yes & \tN No       & \tN No  \\
Fang \cite{Fang2026}                  & \tY Yes         & \tY Yes & \tN No  & \tN No  & \tP Part$^2$    & \tN No  & \tY Yes         & \tY Yes & \tN No & \tN No          & \tN No  & \tY Yes         & \tY Yes & \tY Yes      & \tY Yes \\
Pinnacle \cite{Webster2026Iceberg}    & \tP Part$^{12}$ & \tY Yes & \tY Yes & \tN No  & \tP Part$^2$    & \tN No  & \tP Part$^{12}$ & \tN No  & \tN No & \tN No          & \tN No  & \tY Yes         & \tN No  & \tN No       & \tY Yes \\
\hline
\textbf{Q-NEXUS} %
& \tY \textbf{Yes} & \tY \textbf{Yes} & \tY \textbf{Yes} & \tY \textbf{Yes} & \tY \textbf{Yes} %
& \tY \textbf{Yes} & \tY \textbf{Yes} & \tY \textbf{Yes} & \tY \textbf{Yes} & \tY \textbf{Yes} %
& \tY \textbf{Yes} & \tY \textbf{Yes} & \tY \textbf{Yes} & \tY \textbf{Yes} & \tY \textbf{Yes} \\
\hline
\textbf{Module} &%
\SetCell[c=4]{c} \textbf{QPU/QSF/ASQPU} & & & &%
\SetCell[c=2]{c} \textbf{QB} & &%
\SetCell[c=5]{c} \textbf{QM/STQM/RAQM} & & & & &%
\SetCell[c=2]{c} \textbf{Q-CHESS} & &%
\SetCell[c=2]{c} \textbf{All} & \\
\hline\hline
\end{tblr}
\caption{Comparisons of quantum computing architectures (rows) based on whether they satisfy the (Optional, Opt.) Requirements (Req.) defined for high-performance heterogeneous architectures (columns). Two homogeneous architectures (annotated with $^*$) are included as baselines. We present a comprehensive set of architectures, ordered by publication date, that claim to be modular, heterogeneous, and/or ``von Neumann’’.
The requirements are grouped according to the treatment of quantum data and are labeled by three-letter keys derived from keywords in their definitions.
Computation: 
quantum data is processed by Fixed-size (FXD) QPUs; 
Magic (MGC) states are supplied by QSFs; 
Multi-core (MLT) processing is supported; 
AppLication-specific (APP) logic may be applied via ASQPUs. 
Communication: the architecture uses a quantum bus (QB) with a Micro-architecture (MCR) to communicate between modules using fault-tolerant transfer protocols; 
Long-range (LNG) routing of quantum data is handled by the QB when it minimizes logical SWAP gates. 
Storage: quantum data is stored in QM, which absorbs Idling (IDL) and serves as the primary Scaling (SCL) axis; supports Static (STC) short-term storage, Random-access (RND) long-term storage, and Hierarchical (HRC) organization.
Control: quantum data is orchestrated by Q-CHESS via Machine-level (MCH) instructions, accounting for Timing (TMN) mismatches.
Representation: quantum data may be encoded in multiple Qubit (QBT) modalities and/or QEC Codes (CDE). The Q-NEXUS architecture is presented as the final row, with the modules responsible for meeting the requirements below. 
    $^1$Computation and storage are assigned to lattice regions rather than dedicated QPU/QM modules.
    $^2$Teleportation protocols are discussed, but no resource estimates provided.
    $^3$Litinski architecture compiler provided by Robertson \emph{et al.}~\cite{Robertson2025}.
    $^4$Mixes superconducting device types (qubits for compute and resonators for storage); no distinct qubit modalities.
    $^5$``Memory'' qubits are used for computation; no distinct storage function.
    $^6$Mixes trapped-ion qubit species; no distinct qubit modalities.
    $^7$Uses physical qubit shuttling rather than interconnects.
    $^8$Hardware specialization of compute and storage within modules; no distinct QPU/QM modules.
    $^9$Partial simulation only; no full compilation of algorithms with resource estimation.
    $^{10}$ Not fault-tolerant.
    $^{11}$ Scan-access memory reduces access latency; does not provide random access.
    $^{12}$ Module organization is specific to algorithm, implying runtime allocation.}
\label{table:architecture_comparison}
\end{table*}

The desiderata above provide general architectural guidance for heterogeneous systems and form a natural basis for comparison with prior work in which several consistent patterns emerge, as summarized in Table~\ref{table:architecture_comparison}. Many architectures recognize the need to separate computation and storage, satisfying requirements such as fixed-size processing units (FXD) and dedicated memory (IDL, SCL). Support for fault-tolerant communication (MCR) is also increasingly common, although often without detailed resource modeling. Compilation-aware resource estimation has been explored, but is typically limited to targeted, workload-specific studies -- most notably for RSA factorization -- rather than general-purpose architectural frameworks. Overall, prior work adopts elements of heterogeneity, but typically addresses them in isolation without a unified system-level treatment.

Table~\ref{table:architecture_comparison} also highlights several gaps that are addressed by the development of Q-NEXUS, expressed in detail in the following section. Notably, no prior architecture includes a static memory tier (STC), despite its direct relevance to mitigating idling overheads. Similarly, support for application-specific processing units (APL), mixed qubit modalities (QBT), and dissimilar error-correcting codes (CDE) is occasionally considered, but their system-level impact is not quantitatively evaluated. In contrast, as we will show Q-NEXUS satisfies all listed requirements, combining these features into a coherent architecture centered on the treatment of quantum data. In particular, the Q-CHESS compiler provides a machine-level compilation framework for heterogeneous architectures that operates on general quantum circuits, in contrast to prior approaches that are typically restricted to specific algorithms (e.g., Shor’s algorithm).

\pagebreak
\section{Q-NEXUS Architecture}
\label{sec:framework}

\begin{figure*}[t]
    \centering
    \includegraphics[width=\linewidth]{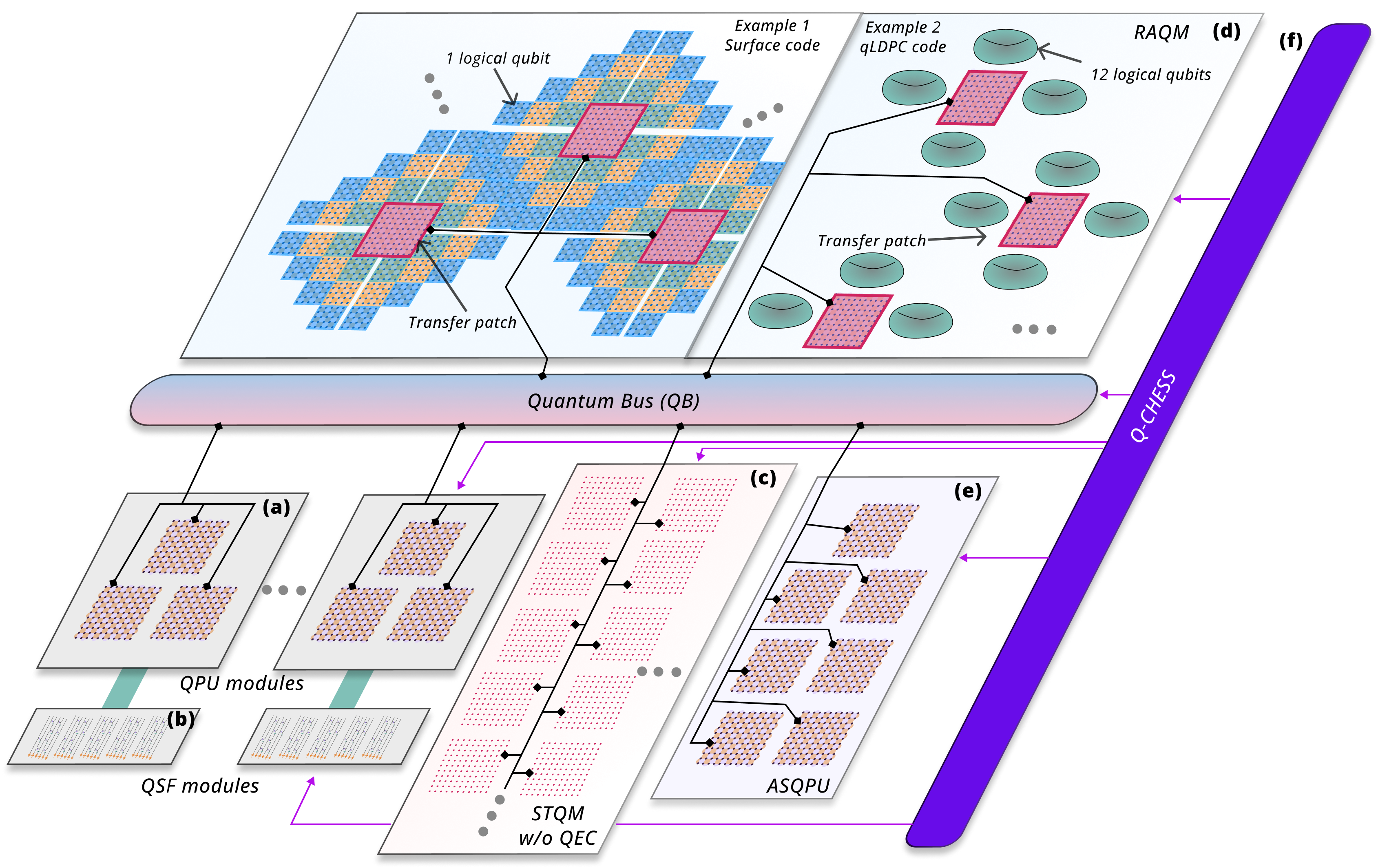}
    \caption{Overview of Q-NEXUS: a heterogeneous architecture made of specialized functional modules connected through an interconnect bus, with compilation and execution orchestration provided by Q-CHESS
    (a) Quantum processing unit (QPU): Modules capable of universal fault-tolerant quantum logic operations with fixed size. Implemented with physical qubit modalities optimized to fast gate speeds (e.g., superconducting/spin qubits). 
    (b) Quantum State Factories (QSF): Specialized units for the rapid generation and distillation of resource states, directly connected to QPU. 
    Quantum Memory (QM): Persistent quantum storage, high-density and scalable. Implemented with ultra-long coherence storage substrates (e.g., rare-earth ions or neutral atoms). Quantum bus protocol between QM and QPU is implemented with two variations based on storage requirements:
    (c) STQM utilized for transient buffering without active QEC. No operations allowed (including syndrome extraction) in this module.
    (d) RAQM for long term storage with active QEC. To facilitate communication with other modules, logical patches must be routed to the nearest transfer patch (large patches) via a sequence of SWAP operations. The concentric color gradient surrounding each transfer patch (teal through blue) quantifies the routing cost, where each color represents the SWAP-distance to a transfer patch. (d-right) An alternative qLDPC RAQM storage option. Toroidal manifolds (donuts) illustrate a module of 12 qubits in the ``Gross" bivariate bicycle code. Four such modules arranged next to a transfer patch offers significant hardware efficiency for large-scale data preservation. 
    (e) An optional accelerator unit. In the case of RSA a 37Q arithmetic Adder accelerator was considered. 
    (f) Control infrastructures which executes and orchestrates the machine code from Q-CHESS (pink arrows). All modules interface with the centralized interconnect bus. The interconnect bus mediates all-to-all logical connectivity via optical connections (black lines), enabling transversal teleportation between the dedicated transfer patches (pink centers) within the RAQM storage array, the RAQM and the QPUs.}
    \label{fig:rsa_architecture}
\end{figure*}

Reflecting the requirements above, we introduce Q-NEXUS (Quantum Networked EXecUtion and Storage), a heterogeneous architecture with explicitly defined functional modules, illustrated in Fig.~\ref{fig:rsa_architecture}. Q-NEXUS separates the computation and storage of quantum data, facilitates communication between modules through a quantum bus, and leverages heterogeneous qubit representations to reduce overheads within a single architectural framework. It further enables system-level evaluation through the Q-CHESS compiler, which produces machine-level instructions, captures interconnect resource requirements, and models timing mismatches between heterogeneous modules.

In the remainder of this section, we describe the core architectural subsystems which we have adopted, based on the formal requirements and recommendations described generically above:
\begin{enumerate}
\item \textbf{QPU}: fixed-size processing units for universal quantum logic.
\item \textbf{QSF}: specialized factories for magic-state production.
\item \textbf{ASQPU}: application-specific accelerators for structured subroutines.
\item \textbf{QB}: a quantum bus providing inter-module communication.
\item \textbf{FTTP}: fault-tolerant transfer protocols for reliable movement of quantum data over the bus.
\item \textbf{QM}: hierarchical memory system comprising static (STQM) and random-access (RAQM) storage.
\end{enumerate}

\subsection{Quantum Processing Units (QPU)}

The QPU is the computational engine, designed for execution speed rather than data retention. It plays an analogous role to a CPU in a classical computer\textsuperscript{\ref{fn:cpuanalogy}}. We define the QPU as a module supporting arbitrary logical operations on a fixed (small) number of logical qubits. Unlike existing modular QPU approaches, this QPU does not store quantum-state information long-term; it receives data, executes a logic block, and exports the result immediately \footnote{\label{fn:cpuanalogy} We use the term QPU to align with common quantum computing terminology \cite{Liu2023}, but its architectural role differs from that of a classical CPU. In modern computers, the CPU manages both computation and program control under the stored-program model, where data \emph{and instruction} reside in memory. In Q-NEXUS, \emph{instruction} execution is handled by the control infrastructure (via the Q-CHESS compiler), while the QPU performs only logical operations. In this sense, the QPU is more closely analogous to the arithmetic logic unit (ALU) than to the full CPU.}.

Q-NEXUS supports a small number of parallel quantum processing units to exploit algorithmic parallelism, addressing the optional multi-core requirement. We treat each core as identical and connected to other modules (including other QPUs) through the quantum bus. The responsibility for managing parallel execution is therefore primarily handled by the Q-CHESS compiler.

We posit that general-purpose quantum computers will employ a small number of universal QPU cores, analogous to modern classical architectures (see~\cref{sec:discussion}).

To make our analysis concrete, in all resource estimates we model the QPU as a superconducting device with microsecond-scale cycle times ($1\,\upmu$s), leveraging their ability to perform fast logical gates \cite{Castelvecchi2023}. We assume that the physical qubits have square-lattice connectivity and implement universal fault-tolerant operations through a surface code \cite{Litinski2019}. These choices are made in order to provide concrete resource estimates, but none of the results we present are specific to, or optimized for, surface codes; they are chosen solely because they provide a mature and efficient set of fault-tolerant logical operations \cite{Litinski2019}.

\subsection{Quantum State Factories (QSF)}

In Q-NEXUS, the generation of non-Clifford resource states is offloaded to dedicated quantum state factory (QSF) modules. These modules provide high-rate production and distillation of magic states (e.g., T-states or CCZ-states), enabling the QPU to consume resource states without allocating logical qubits to in-line distillation. This separation removes a key performance bottleneck in fault-tolerant quantum computation and allows compute resources to remain focused on algorithm execution (see \cref{fig:rsa_architecture}b). 

We envision that photonic circuits can enable multi-qubit magic state generation at rates exceeding $100$ MHz \cite{rudolph2023logical}. However, our conservative quantitative analysis only uses experimentally proven state generation rates. To make our analysis concrete, we consider two representative QSF implementations aligned with the workloads studied here.

\csubsubsec{General-purpose T-state factory:} 
For T-state generation, we adopt catalyzed T-factories~\cite{Gidney2019}, with $N_\textrm{dist}=72$ logical qubits per factory and $N_\textrm{MF}=3$ factories per logical qubit in the QPU, giving a physical-qubit overhead proportional to $N_\textrm{dist}\cdot N_\textrm{MF}$. We consider the injection time for a logical T-gate to be $2d = 30\,\upmu$s, matching the homogeneous baseline used for comparison.

\csubsubsec{Application-specific CCZ factories:} 
In the RSA-2048 analysis, we employ CCZ factories based on T-cultivation. For Shor’s algorithm, CCZ factories are more efficient and leverage recent advances in T-state cultivation, reducing $N_\textrm{dist}$ to $12$ (a $6\times$ reduction in physical qubits) with minimal runtime impact \cite{Gidney2024}.

\subsection{Algorithm-specific accelerators (ASQPU)}

Q-NEXUS supports specialized processing units optimized for frequently occurring subroutines with specific demands on processing performance and architecture. In conventional microprocessor architecture it is typical to segregate a fast arithmetic logic unit from a general purpose floating-point-operation unit, based on the diversity of tasks that must be executed.

Universal QPUs enable arbitrary fault-tolerant logic, but reflective of the demands driving state-of-the-art microprocessor architecture, many algorithms are dominated by a small number of structured subroutines that open opportunities for increased efficiency in a similar way. For example, Gidney's analysis of RSA-2048 factorization \cite{Gidney2025RSAUpdate} shows that $\sim70\%$ of the runtime is spent executing an adder routine.

Several promising hardware approaches exist to implementing such accelerators. As a plausible example, Cuccaro \emph{et al.} proposed a highly efficient ripple-carry adder that requires using the Toffoli as a single non-Clifford primitive~\cite{Cuccaro2004}. This suggests that hardware optimized for efficient implementation of a restricted transversal gate set and the specific adder circuit, could yield significant performance gains. Jiao \emph{et al.} proposed hybrid code constructions enabling transversal $T$ gates with reduced qubit overhead \cite{Jiao2025}, while Haah \emph{et al.} developed efficient $T$-to-Toffoli distillation techniques \cite{Haah2018}. 

Comprehensively designing a physical processing unit optimized for fault-tolerant execution of such routines is beyond the scope of this work. In this work, we abstract these advances into an effective model, assuming a specialized 33-bit adder implemented with 37 logical qubits (matching \cite{Gidney2019}).

\subsection{Quantum Busses (QB)}

A key requirement of any modular framework is the ability to transfer logical information between modules in a fast, efficient, and fault-tolerant manner. In Q-NEXUS this functionality is provided by the quantum bus, which serves as the communication system responsible for transferring quantum information between the processing, memory, and other specialized modules of the architecture.

The quantum bus is composed of a collection of physical interconnects. Interconnects generate Bell pairs between physical qubits in different modules, typically using optical \cite{Monroe2014} or microwave \cite{Magnard2020} photons as mediators. These Bell pairs form a resource that is consumed by teleportation-based protocols to transfer quantum information \cite{Caleffi2024, Nickerson2014, Haug2025} (see Sec.~\ref{sec:busproto}). By mediating interactions through photonic channels, the bus enables long-range connectivity without requiring direct physical coupling between qubits.

The bus includes the physical infrastructure required to generate and distribute Bell pairs, including entangled photon sources, detectors, beam splitters, and phase shifters \cite{Kimble2008,AbuGhanem2024,Gu2017}. These components enable scalable communication through spatial and temporal multiplexing and routing, mitigating control overheads associated with the ``tyranny of numbers''. Mixing modules with different qubit modalities may additionally require transduction between microwave and optical photons \cite{Kaiser2022, Kumar2023}.

\begin{figure}[t]
    \centering
    \includegraphics[width=\linewidth]{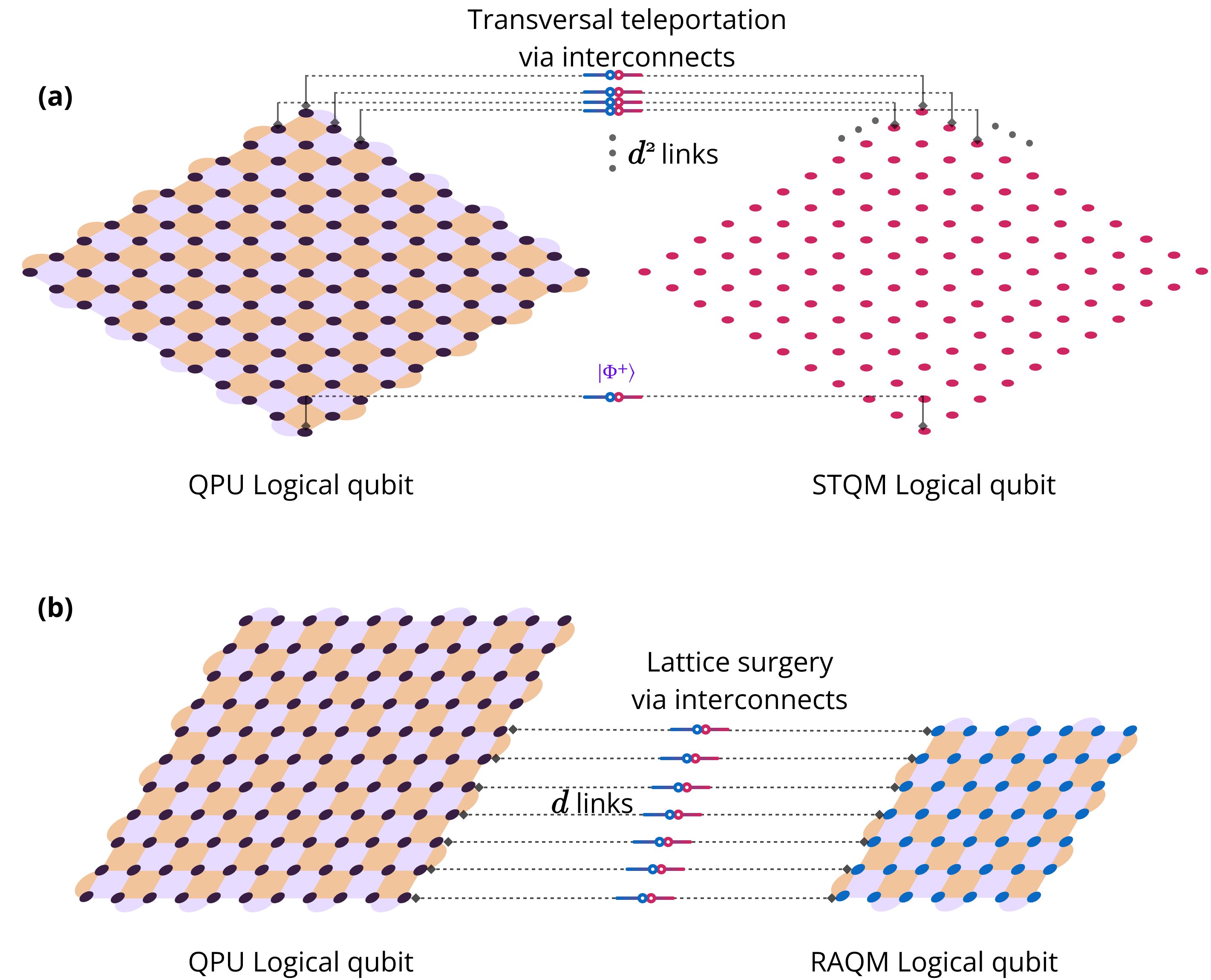}
    \caption{Transferring logical information from the QPU to QMs using two different bus protocols. In both, a distance $d$ rotated surface code is used for the QPU encoding. Dots denote physical qubits: purple in QPU, pink is STQM, blue in RAQM. Dotted lines represent Bell pairs prepared by bus, between corresponding qubits, ready for teleportation. (a) Transfer between the QPU and STQM using the transversal teleportation protocol with $d^2$ Bell pairs. This enables high-bandwidth transfer and simplified control in the STQM (no active QEC). (b) Transfer between the QPU and RAQM using lattice surgery transfer protocol with $d$ Bell pairs. Both the QPU and RAQM require active QEC, and logical state transfer is performed over $d$ QEC cycles.
    }
    \label{fig:protocol}
\end{figure}

\subsection{Fault-tolerant Transfer Protocols (FTTP)}
\label{sec:busproto}

Logical quantum states are transferred across the quantum bus using fault-tolerant bus protocols. These protocols consume Bell pairs generated by the interconnects to move encoded logical information between heterogeneous modules. To ensure a sufficient supply of high-fidelity Bell pairs, the protocols exploit multiple techniques: parallel preparation of Bell pairs (spatial multiplexing), buffering of entanglement resources (temporal multiplexing), and entanglement purification \cite{Heya2025}. In this work, we consider two such protocols: transversal teleportation and lattice surgery transfer, shown in Fig~\ref{fig:protocol}.

\csubsubsec{Transversal teleportation protocol:} This protocol relies on high-fidelity fault-tolerant physical-qubit teleportation \cite{Gottesman1999}. Logical-state transfer is achieved via the parallel application of quantum teleportation across all physical qubits in the encoded state, requiring a strict one-to-one mapping between source and target qubits. Recent experimental demonstrations have realized such fault-tolerant logical teleportation using transversal operations, achieving high process fidelities \cite{RyanAnderson2024}. This protocol enables high-bandwidth state transfer with minimal latency. However, the requirement of one-to-one mapping limits flexibility: it does not support code deformation, and the target encoding inherits the physical-qubit count of the source.

\csubsubsec{Lattice-surgery transfer protocol:} An alternative protocol employs lattice surgery, enabling fault-tolerant transfer between modules with different encoding or code distances \cite{Litinski2019, stein2024architecturesheterogeneousquantumerror}. This protocol supports code deformation and therefore allows communication between heterogeneous modules with heterogeneous encoding schemes. Its total duration is fixed to a multiple of the error correction cycles due to active fault-tolerant operations, which can result in high read-write latency when using slower qubit modalities.

\subsection{Quantum Memories (QM)}

In Q-NEXUS, quantum memory serves as the primary storage layer, deferring all requirements for arbitrary logical operations to the fixed-size QPU. This separation creates a significant opportunity to simplify control hardware and design memory systems that can scale efficiently. 

Further, Q-NEXUS embraces a full hierarchical memory strategy in implementation, based on use of both RAQM and STQM. These two memory types naturally occupy different regions of the latency–duration trade space: RAQM supports long-duration storage with higher access overhead while STQM provides low-latency access for short-duration storage. Implementing this hierarchy in quantum data storage enables low-latency access to actively used quantum information, while accommodating large-scale and long-duration storage. The detailed description of these QM subsystems is provided for the Q-NEXUS implementation in the following.

\subsubsection{Random Access Quantum Memory (RAQM)} 
\label{sec:qram}

The RAQM is used for long-duration storage of quantum information during execution of a program, but not beyond this timing envelope\footnote{Outside the scope of this manuscript is the consideration of a quantum sequential-access memory, analogous to a classical ``hard drive'', which would store quantum information indefinitely.}. To ensure rapid program execution, the defining property of a random-access memory is \emph{uniform access latency}: a logical qubit can be read or written in approximately the same time irrespective of its physical location within the memory \cite{Liu2023,Kobori2025}\textsuperscript{\ref{fn:qramdef}}. This mirrors the defining property of classical random-access memory and is essential for scalable architectures \cite{Patterson2017}.\footnote{\label{fn:qramdef} Our definition of RAQM refers to the physical access of \emph{logical} quantum information using classical address information. This aligns with prior work on heterogeneous quantum architectures \cite{Liu2023,Kobori2025} and with the classical definition of random-access memory \cite{Patterson2017}. It is distinct from the quantum information concept of ``qRAM'', which is an abstract algorithmic primitive: a unitary operation that returns a ``stored'' superposition state given a superposed ``address'' input \cite{Giovannetti2008}.}

We present two strategies for achieving scale in a RAQM while retaining this random-access property. These strategies are closely tied to the implementation of QEC and are therefore classified based on their connectivity relative to the QPU.

\csubsubsec{Matched connectivity:} 
The first strategy assumes that the quantum memory has the same connectivity geometry as the QPU and is shown in Fig.~\ref{fig:rsa_architecture}~(d). 

To achieve scale, we exploit heterogeneity in qubit modalities. Specifically, we consider long-coherence (LC) qubits -- long relative to the modality used in the QPU -- with fully developed control infrastructure for QEC. Prominent candidate hardware modalities include neutral atoms~\cite{Bluvstein2022, Manetsch2025, Barnes2022}, and trapped atomic ions~\cite{Wang2021}, both of whose coherence times exceed those of superconducting QPU cycle times by factors of $\sim 10^7$. These extended coherence times can enable encoding with reduced code distance while maintaining equivalent logical error rates, thus increasing the storage density (see \cref{app: coherence}). Furthermore, LC modalities enable reduced QEC cycle frequency (see \cref{app: coherence}), allowing optimization of storage fidelity through dynamic adjustment of cycle times (Sec.~\ref{sec:clock_Synchronization}).

To make this explicit, we assume both the QPU and RAQM have square-lattice connectivity and employ surface codes. Logical qubits are stored using surface-code patches with distance $d_\text{QM}$ in the memory and $d_\text{QPU}$ in the QPU, where $d_\text{QM} < d_\text{QPU}$ achieves comparable idling error rates due to the longer coherence times.

Quantum information enters the RAQM from the QPU via the quantum bus to a transfer patch using the transversal teleportation protocol (Sec.~\ref{sec:busproto}). The one-to-one physical qubit requirement of this protocol necessitates that the transfer patch uses distance $d_\text{QPU}$. The transfer patch then moves quantum information into surrounding higher-density storage patches of distance $d_\text{QM}$ using lattice-surgery-based local swap operations.

To ensure uniform access latency in a RAQM storing a total of $N$ logical qubits, we distribute $N/(2k^2 + 6k + 1)$ transfer patches throughout the memory. This guarantees that any logical qubit is at most $k$ swap operations away from a transfer patch. The construction is based on Manhattan-distance geometry on a square lattice, where each transfer patch services a bounded neighborhood of nearby qubits \cite{KletteRosenfeld2004}. The absolute latency of the RAQM can be tuned by adjusting the value of $k$.

\csubsubsec{Hybrid connectivity:} 
The second strategy allows the RAQM to employ a different connectivity and encoding from the QPU, enabling the use of heterogeneous error-correcting codes, as shown in Fig.~\ref{fig:rsa_architecture}~(d). This is motivated by innovations in the family of qLDPC codes \cite{Bravyi2024} which have shown strong potential for achieving high-density storage of quantum information, but require non-local coupling in the underlying connectivity graph and exhibit relative immaturity of fault-tolerant gate constructions compared to surface codes \cite{Xu2024,Zhu2025}. Nonetheless their strengths and the simplified requirements of memory-only implementations have motivated significant interest in platforms where such connectivity arises naturally: trapped-ion systems with all-to-all connectivity within modules \cite{Ye2025}, and neutral-atom platforms with controllable long-range coupling through Rydberg interactions and atom rearrangement \cite{Xu2024}. In parallel, this has driven the development of long-range couplers in superconducting systems \cite{Yoder2025}.

To achieve scale with hybrid connectivity, we assume the RAQM supports non-local coupling sufficient to implement qLDPC codes. More explicitly, we consider a QPU based on a square lattice with surface-code encoding, while the memory is composed of toroidal connectivity modules (``donuts'') encoding $N_{\text{GBB}}$ logical qubits using the Gross bivariate bicycle (GBB) code (a specific qLDPC code \cite{Yoder2025}). 

In this architecture, logical state transfer between the QPU and RAQM is performed using the lattice-surgery transfer protocol via a transfer bus. This protocol supports code deformation, enabling logical information to be transferred directly into surrounding toroidal modules. As a result, no routing of the stored $N_{\text{GBB}}$ logical qubits is required within the memory, preserving uniform access latency. This flexibility comes with a trade-off; compared to transversal teleportation, transfer latency increases, as operations must be performed over multiple QEC cycles and are limited by the slower module.

We have presented two RAQM strategies that achieve both increased information density, addressing the scaling requirement, and uniform access latency; Surface codes and qLDPC codes serve as illustrative examples, but these constructions generalize naturally to other codes. For instance, color codes on a hexagonal lattice \cite{Bombin2006} could be used within the matched-connectivity scheme (with the corresponding update to the underlying distance metric). More broadly, the lattice-surgery transfer protocol is compatible with a wide range of code combinations, enabling flexible integration of heterogeneous QEC schemes \cite{stein2024architecturesheterogeneousquantumerror}.

\subsubsection{Static Transversal Quantum Memory (STQM)} \label{sec:qstm}
The STQM addresses the scaling requirement through a drastic simplification of control requirements via both the elimination of operations beyond memory and the elimination of active QEC. Instead of logically encoded qubits the STQM leverages ultra-long-coherence (ULC) qubits whose coherence times exceed the maximum contiguous idling time of any logical qubit. 

Even for long-running programs such as RSA-2048 factorization \cite{Gidney2025RSAUpdate}, qubits idle continuously only for short intervals due to continuous scheduling. Estimates suggest idle times of seconds in baseline implementations, and milliseconds when STQM is used as a cache. Consequently, coherence-time requirements are relatively modest: for a worst-case idle time of $100~\mathrm{ms}$ and physical error rate $p=5\times10^{-4}$, a coherence time of only minutes is sufficient. Longer storage can be supported by periodically refreshing the state via a QEC cycle on the QPU.

The realization of the STQM within Q-NEXUS is shown in Fig.~\ref{fig:rsa_architecture}~(c). Each logical qubit in the QPU is mapped to a corresponding set of physical qubits in the STQM, and the transversal teleportation protocol is used to transfer the encoded state into memory. During storage, no active QEC is applied; instead, errors accumulate either passively or with active unencoded control strategies~\cite{Khodjasteh_LowLatencyMemory} and undergo error correction only upon retrieval when the state is transferred back to the QPU via the same protocol. 

Prominent example technologies of relevance to the STQM are nuclear-spin-based systems, such as rare-earth-ion (REI) quantum memories. REI memories have demonstrated coherence times exceeding $13$ hours \cite{Zhong2015, Wang2025,Ahlefeldt2020}, with predicted hyperfine-state lifetimes of up to \emph{23 days} \cite{Wang2025}. Further, techniques such as atomic frequency combs \cite{Afzelius2009} allow high-density, multi-mode storage of quantum information within a single REI crystal. 

Use of an ULC qubit as part of an STQM is viable provided the total accumulated error remains within the correction capability of a single QEC cycle. ULC systems typically exhibit biased noise with $T_1/T_2 \sim 55$ \cite{Wang2025}. By applying Hadamard rotations prior to transfer, the effective noise channel during storage is transformed to one consistent with the XZZX code, which has a threshold exceeding $20\%$. Upon retrieval and reversal of these rotations, the QPU corrects accumulated errors, provided that storage errors remain below the physical error threshold.

\section{Q-CHESS: Quantum Compiler for Heterogeneous Execution, Scheduling, and Synthesis}\label{sec:compiler}

The heterogeneous architecture proposed here introduces compilation and scheduling complexities absent in contemporary quantum platforms. Existing compilation heuristics, such as those implemented in Qiskit \cite{Javadiabhari2024} or PyTKET \cite{Sivarajah2021}, optimize primarily for circuit depth under the assumption that all idle windows are equally damaging and that SWAP operations incur similar cost. When applied to heterogeneous architectures, these assumptions lead to catastrophic error accumulation. 
As a result, qubit placement, scheduling, and data movement become first-order determinants of total error rather than secondary optimization choices.

Compilation pipelines include many components, such as logical-to-physical qubit assignment, unitary synthesis to a given gate set, scheduling and parallelizing operations, and information routing. When orchestrating a program with a modular architecture, all components must take into account the finite capacity of each module, the penalties of data transfer between modules, the non-universality of specific modules, and the tradeoffs associated with executing a task in one of a diversity of modules available.
Naively executing a program across multiple modules may degrade the overall performance, unless a program orchestrator balances the different tradeoffs; the ability to optimize individual modules for narrow functionalities is not sufficient.

\begin{figure}[t]
    \centering
    \includegraphics[width=\linewidth]{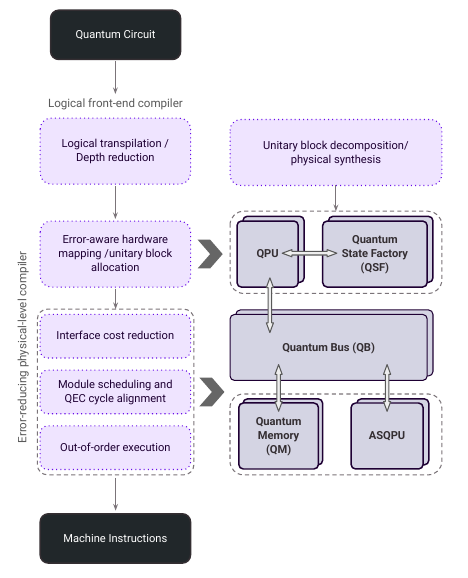}
    \caption{Visual representation of the Q-CHESS error-aware heterogeneous quantum compilation pipeline. The diagram illustrates the mapping of software compilation passes to a heterogeneous hardware architecture. Color coding and shapes distinguish the different components: dark gray rounded rectangles denote the pipeline's input (Quantum Circuit) and final output (Machine Instructions); light purple boxes with dotted borders represent the software compilation passes, divided into logical front-end and error-reducing physical-level stages; and grayish-purple boxes with solid borders designate the distinct physical hardware modules (QPU, QSF, QB, QM, ASQPU). The pathways between software and hardware are defined by three arrow styles. Vertical down arrows indicate the sequential control flow of the compilation pipeline, advancing the input circuit through successive logical and physical optimization passes. Large chevrons denote the mapping from specific software compilation stages to their corresponding hardware targets. Finally, double arrows signify bi-directional physical interconnects and communication channels between the disparate hardware components.} 
    \label{fig:compiler}
\end{figure}

We address these challenges through the introduction of Q-CHESS (Quantum Compiler for Heterogeneous Execution, Scheduling, and Synthesis), an error-aware compilation framework designed explicitly for modular heterogeneous architectures, as shown in Fig~\ref{fig:compiler}. Q-CHESS treats the target architecture not as a single device graph, but as a distributed system composed of interacting modules. The compiler explicitly models module-specific clock rates, information transfer latencies, and error accumulation. It performs coordinated logical and physical compilation passes to optimize logical depth, execution time, and resource utilization, while synchronizing mismatched clocks and reducing accumulated error. 

At the logical level, Q-CHESS performs transpilation and depth reduction before grouping operations into multi-qubit unitary blocks, with block sizes constrained by instantaneous QPU capacity. The preservation of a unitary block structure enables scheduling and mapping decisions to be made with granularity aligned with architectural constraints rather than individual gates. The compiler then performs an error-aware register allocation and hardware mapping, assigning logical qubits and unitary blocks to QPUs, QMs, and QSFs subject to capacity, connectivity, and interface constraints. Mapping decisions explicitly compare accumulated idle error on the QPU against the latency and error introduced by state transfers.

Physical synthesis is deferred until after logical scheduling and mapping. Unitary blocks assigned to QPUs are efficiently decomposed into hardware-native instruction sequences in a context that reflects their actual execution environment, including expected idle intervals and interface timing. Throughout this process, Q-CHESS maintains an explicit representation of execution timing across modules with disparate clock rates, enabling physical-level passes that align QEC cycles and exploit limited out-of-order execution to suppress idle error and reduce state transfers. 

By virtue of these operational features, Q-CHESS allows the analysis of different architectural choices. The type and size of modules, choice of QEC codes, qubit modalities, error profiles and cycle times are all inputs to Q-CHESS that can be easily adjusted by a user. Given these inputs, Q-CHESS outputs a fully scheduled and compiled program, accounting for the full count of operations, program durations, scheduling inefficiencies and penalties from idling and state transfers. 

Next, we provide a detailed description of the compilation pipeline components. 

\subsection{Clock Synchronization}
\label{sec:clock_Synchronization}
Different modules may have different clock cycles, a circumstance which poses synchronization challenges. If a logical gate to be executed in the QPU depends on a state stored in memory, the QPU must wait for the fault-tolerant retrieval of that state. Given the orders-of-magnitude differences in clock speeds, a naive wait results in thousands of accumulated idle cycles on the fast, decoherence-prone QPU qubits.

Q-CHESS addresses this challenge via specialized optimization steps. The compiler optimizes for maximal operational yield, by leveraging circuit manipulations, commutation relations, and out-of-order compilation techniques to avoid unnecessary state transfers between modules.  

RAQM employs active QEC in memory, typically using a slow, long-coherence qubit modality. We exploit the physical asymmetry of such qubit modalities, where control errors are much larger than coherence errors, and therefore, the logical error per QEC cycle, in such modalities, is dominated by physical operations (such as 2Q gates and measurements) rather than idling \cite{Wang2025}. This allows the compiler to treat the QEC cycle time, in memory, as a continuous quantity, $T_\textrm{QM} \in [T_\textrm{min},\, T_\textrm{max}]$. The minimal time, $T_\textrm{min}$, is set by the duration of physical gates and measurements, while $T_\textrm{max}$ is set such that $T_\textrm{max}/T_1 < p$, where $p$ is the physical error per cycle, see \cref{app: coherence} for more details.

When synchronization is required, the compiler inserts ``idling buffers'' into the QM schedule, effectively stretching its cycle time to align with an integer multiple of the QPU's cycle without incurring a logical error penalty \cite{Marton2025}. 

STQM by contrast does not include active error correction, and therefore, does not typically pose additional timing constraints.
In architectures where both types of memory exist, the compiler routes information as needed to each of the memory types and includes buffering as needed to guarantee the QPU operates at maximum capacity without incurring long wait times.

\subsection{Module-Aware Compilation Passes}

Q-CHESS implements a pipeline of specialized passes designed to minimize the global space-time volume of the computation.

\csubsubsec{Logical-depth and gate-count reduction:} 
Q-CHESS starts with a family of compiler passes designed to eliminate redundant computation, such as replacing predefined subroutines with efficient forms and applying symbolic rewrites that cancel redundant logic, exploit commutation relations, and collapse nested expressions. Reducing logical operations early prevents the proliferation of costly T gates, minimizing state transfers, routing complexity, and idle windows.

\csubsubsec{Unitary block consolidation:} 
To minimize expensive transfers between the QPU and QM, Q-CHESS decomposes the input algorithm into high-density logic blocks sized to fit the QPU's instantaneous capacity. Increasing the size of the QPUs allows more parallelization and reduction of state transfers at the price of increasing ``in-QPU" routing, control burden, and fabrication complexity. Future optimized architectures are likely to include bigger QPUs compared to near term architectures, however, their sizes are expected to remain small (e.g., $\lesssim$10 logical qubits). In the demonstrations below, we use small QPUs with only 3 logical qubits. Unlike standard approaches which synthesize unitaries to Clifford+T gate sequences directly, we maintain these blocks at intermediate representations to allocate registers and insert routing.

\csubsubsec{Register allocation:} 
To account for multiple modules, the compiler performs multi-objective optimization to enforce hardware constraints and maximize performance. Specifically, Q-CHESS accounts for the disparate capacity limits of the QPU and QM while maintaining operational locality and hardware connectivity. By strategically minimizing state transfers and idling durations, the pipeline reduces total execution latency. Furthermore, the compiler avoids high-error interfaces and efficiently leverages QM to maximize global execution fidelity.

\csubsubsec{Routing cost function:} 
Deciding when to send a qubit state to memory is a non-trivial optimization problem. Q-CHESS employs a cost-function router that weighs the \textit{Transfer Error} accounting for both the transfer time and the transfer quality, $\epsilon_{\text{ST}}$, against the integrated idling error in the QPU, $\epsilon_{\text{idle}}(t)$, or in memory, $\epsilon_{_\text{storage}}(t')$. 
\begin{equation}
    \text{Cost} = \begin{cases} 
    \epsilon_{\text{idle}}(t) & \text{, if kept in QPU} \\
    2 \cdot \epsilon_{\text{ST}} + \epsilon_{_\text{storage}}(t') & \text{, if moved to QM}
    \end{cases}
\end{equation}
For short idle windows, the compiler keeps the state in the QPU. However, as $t$ exceeds the ``breakeven'' duration defined by the QB fidelity, the compiler automatically generates the microcode to teleport the state to the QM, effectively using the QM as a high-fidelity delay line. 
Each memory module has its own cost function. Memories with low $\epsilon_{\text{ST}}$ and moderate $\epsilon_{_\text{storage}}(t')$ may serve as fast memories (cache), where the opposite regime serves as an example for a slower memory, where the storage time is long enough to justify the higher cost of transfer. 

\csubsubsec{Optimal Unitary block synthesis:}
Once high-density unitary blocks are allocated to QPUs and scheduled for execution, the compiler applies numerical unitary synthesis and T-gate decomposition passes, such as GridSynth \cite{Ross2016} and specialized Toffoli decompositions \cite{Selinger2013}, to decompose these multi-qubit unitaries into hardware-native instruction sequences. Deferring synthesis until execution preserves block structure and ensures that, once logical state is transferred to a QPU, it is processed with maximal computational density before being returned to memory. When a noise model is available, the compiler incorporates device-specific error characteristics and performs synthesis directly over the model-defined noisy operations, enabling error-aware realization of target unitaries.

\csubsubsec{Cross-module scheduling:} 
Finally, Q-CHESS orchestrates the operation of the different modules. It simultaneously: offloads long QPU idle periods to memory and leverages continuous-cycle timing to collapse these waits into fewer effective QEC cycles, schedules the operations of quantum state factories to ensure continuous supply of magic states, schedules the purification of Bell pairs on the QPU to prepare for upcoming state transfers, and routes LQs in the RAQM toward the transfer patch as their transfer is due.
This joint scheduling, ensures low latency, allowing the system to perform at the speed of the QPU while retaining the capacity and efficiency gains of the QM.


\section{Mid-scale analysis: scaling, error mechanisms, and resource trade-offs}
\label{sec:midnalysis}
\begin{figure*}[t]
    \centering
    \includegraphics[width=\linewidth]{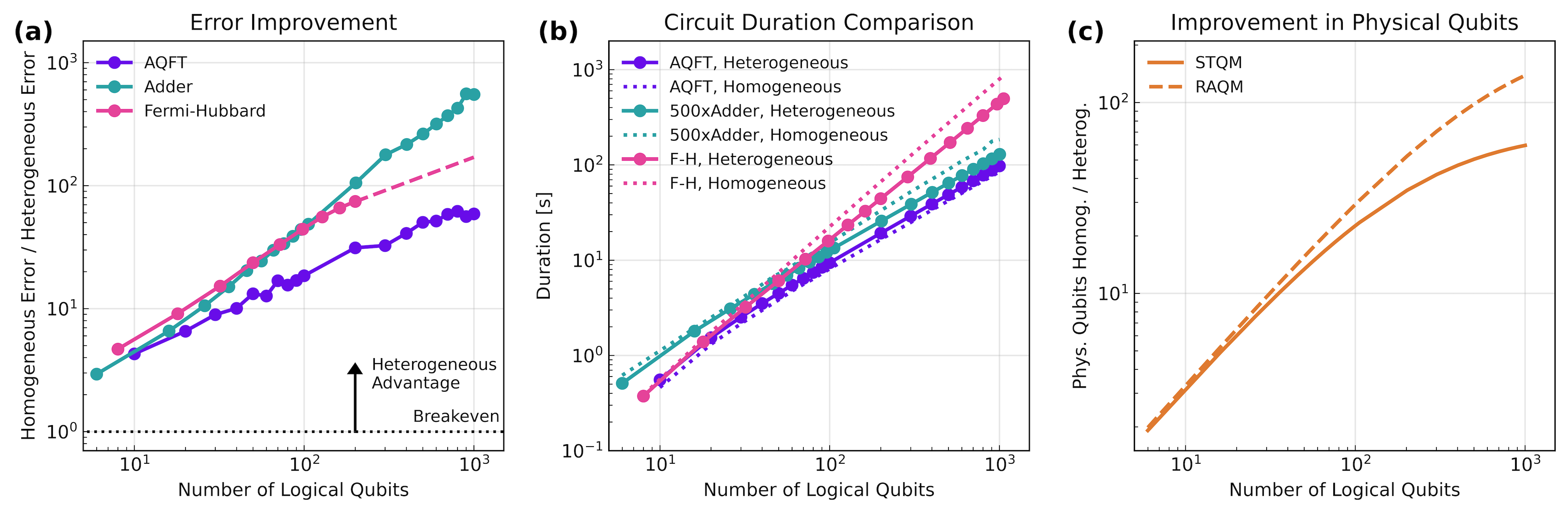}
    \caption{Comparison between two architectures. In (a) and (b) STQM is used as a memory, whereas in (c) values for both STQM and RAQM are plotted. (a) Fidelity improvement, compared to baseline, for three different quantum algorithms: AQFT, Quantum Adder, and Dynamics Simulation of the 2D Fermi-Hubbard model. For Fermi-Hubbard the baseline error saturates at 1 starting 200 LQ (where the error improvement factor is $74\times$); beyond that point the baseline algorithmic error reaches unity (zero fidelity), and the plotted error ratio is no longer meaningful (denoted by dashed line). (b) Circuit duration versus algorithm size. At 1000 LQ, duration factors are $0.9\times$ for AQFT, $1.7\times$ for Fermi-Hubbard and $1.4\times$ for Adder (scaled $500\times$ for visibility). (c) Improvement in the number of physical qubits. The estimated total number of physical qubits depends on the architecture, and is independent of the algorithm type. At 1000 LQ, the improvement factor reaches $60\times$ for STQM and $138\times$ for RAQM.}
    \label{fig:error_improvement_algorithms}
\end{figure*}

To validate the architectural framework and compilation strategies proposed above, we perform a mid-scale resource analysis up to 1,000 logical qubits (LQ). We benchmark the performance and resource requirements of a baseline homogeneous (monolithic) superconducting architecture against heterogeneous Q-NEXUS architectures, orchestrated by the Q-CHESS compiler, using three algorithmic workloads to ensure that observed trends are not specific to a single circuit structure: (i) the Approximate Quantum Fourier Transform (AQFT)\footnote{We utilize an approximate version of the QFT circuit by truncating all controlled-phase gates with angles below $\pi/2^{k_{\rm th}}$ with $k_{\rm th}$ chosen such that the error due to the operation omission is lower than its implementation error. All benchmarks are done with identical circuits.}; (ii) a Quantum Adder; and (iii) a dynamical simulation of the 2D Fermi–Hubbard model. These algorithms exhibit distinct depth, connectivity, and non-Clifford-resource profiles, allowing us to isolate how heterogeneous modularity impacts both performance metrics and resource overheads across different computational regimes. 

\subsection{Methodology}
The baseline reference architecture against which we compare Q-NEXUS is a 1000-LQ homogeneous superconducting grid with nearest-neighbor connectivity, using a distance-15 surface code, with a cycle time of $1\,\upmu$s \cite{GoogleFaultTolerant2025}. We assume physical error rates of $5 \times 10^{-4}$ and code threshold of $6\times 10^{-3}$, leading to a logical error rate of $7\times 10^{-11}$ per QEC cycle. The surface code supports a universal set of logical operations, where single-qubit Clifford operations are implemented within a single QEC cycle, CNOTs are implemented via lattice surgery and consume $d$ QEC cycles, and logical T gates are implemented via an integrated source of T-states~\cite{Gidney2024} and consume $2d$ QEC cycles.

The heterogeneous architecture we employ is outlined in Fig.~\ref{fig:rsa_architecture}, with state-of-the-art module-specific parameters described in Appendix \cref{table:compiler input}. In all cases explored, the QPU is a 3-LQ device assumed to be realized using superconducting qubits with identical hardware parameters as those used above for the homogeneous architecture. Again, we employ a surface code with $d=15$ and a $1\,\upmu$s cycle time, and operations are performed in a manner identical to that of the baseline architecture. We also add a photonic QSF as an accelerator.

In our comparisons we vary the assumptions about the employed QMs in the heterogeneous architectures, comparing use of the passive STQM and actively corrected RAQM across several cases:
\begin{itemize}
    \item \textbf{STQM:} A 1,000-LQ static memory based on a ULC-qubit modality with no active QEC (Fig.~\ref{fig:rsa_architecture}c). For this demonstration, we use REI parameters of \mbox{$T_1\sim23\,\text{days}$} and \mbox{$T_2>10\text{hr}$} \cite{Wang2025}. We denote this architectural choice as A1. 
    \item \textbf{RAQM:} A 1,000-LQ memory based on an LC-qubit modality with active QEC and lattice surgery as a transfer protocol. Memory uses a surface code with a lower distance ($d=9$) than the QPU. Assuming physical error rates of $1 \times 10^{-4}$ \cite{Hughes2025}, we have a logical error rate of $3.8\times 10^{-11}$ per QEC cycle. We analyze the impact of different cycle times on the quality and runtime of the execution: architectural choices A2 and A3 have QM cycle times of $50\,\upmu$s and $1000\,\upmu$s, respectively. 
\end{itemize}

Compilation for the heterogeneous architecture is executed using Q-CHESS. In order to ensure a fair comparison that isolates the architectural impact of heterogeneity from other performance differences, we compile and orchestrate the input program for the baseline architecture using a custom-designed transpiler that mimics the heterogeneous pipeline, but uses SWAP gates for routing. 

\begin{figure*}[t]
    \centering
    \includegraphics[width=\linewidth]{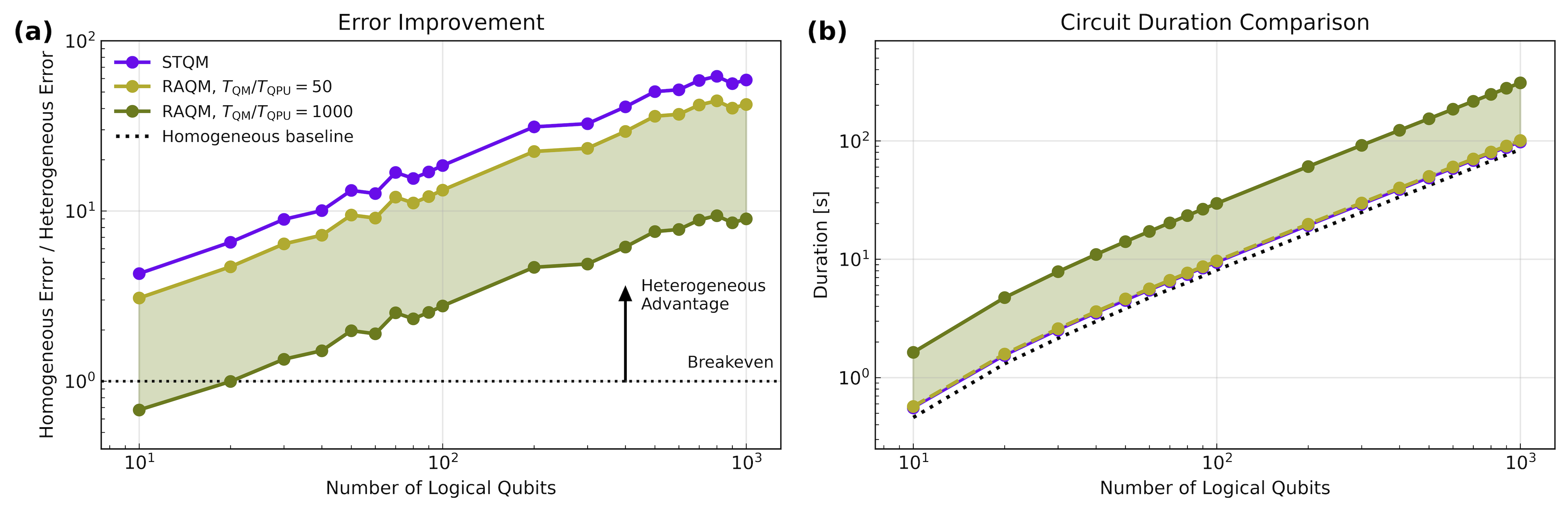}
    \caption{Comparison between STQM with transversal teleportation and RAQM with lattice-surgery transfer protocol for the AQFT algorithm in terms of execution (a) error and (b) duration. For RAQM, we present the results for two values of the QEC cycle ratio $T_{\rm QM}/T_{\rm QPU}$ between RAQM and QPU. $T_{\rm QM}/T_{\rm QPU} = 1000$ (green) corresponds to the current state of the art QM gate and measurement durations, whereas $T_{\rm QM}/T_{\rm QPU} = 50$ (yellow) is a look-ahead value. The shaded region marks the regime between these two scenarios.}
    \label{fig:error_improvement_interfaces}
\end{figure*}

\subsection{Break-even scaling and algorithm-dependent performance}
Performance benchmarking proceeds by calculating the algorithmic error, duration, and number of physical qubits required for each architecture, varying the overall number of \emph{logical qubits} defining the problem size. Scaling the logical qubit count from small instances of six up to 1,000 LQ allows us to directly observe when improvements in fidelity and resource efficiency overcome the overheads introduced by heterogeneity.

We estimate the total algorithmic error by accounting for the error contributions of every operation (logical gates, idling, and state transfers) in the scheduled circuit. We typically present the ratio of the homogeneous-architecture error over the heterogeneous-architecture error such that values greater than unity correspond to a net advantage for the heterogeneous architecture for fixed logical-qubit resourcing. The total number of physical qubits required for a specific algorithmic benchmark is calculated according to estimates derived in \cref{app:resource estimate}, and the total execution time is extracted from the scheduled circuit output by the compiler.

Fig.~\ref{fig:error_improvement_algorithms} shows a comparison between the baseline and the A1 (STQM) architectures for various benchmarking algorithms. Improvements derived from the heterogeneous architecture surpass the break-even point on the scale as small as 5 LQ, and increase with the size of the algorithm. For 1,000-LQ AQFT, the heterogeneous A1 architecture demonstrates a \textbf{$\bm{59\times}$} reduction in the total algorithmic error compared to the baseline architecture. At the same scale, we observe an error-reduction factor of up to $\bm{551\times}$ for the Quantum Adder subroutine and similar benefits for the Fermi-Hubbard model before the homogeneous-architecture baseline error reaches unity and the comparisons no longer become meaningful. Across all problem instances and scales, the algorithmic runtime shown in Fig.~\ref{fig:error_improvement_algorithms}b is comparable between the two architectures (within $2\times$), demonstrating that optimal scheduling can largely mask the latencies typically associated with modular architectures.

Fig.~\ref{fig:error_improvement_algorithms}c shows the improvement factor in the count of total physical qubits relative to the baseline architecture. The estimated number of physical qubits is independent of the algorithmic benchmark, but varies based on architecture type. At 1,000 LQ, the improvement factor reaches $60\times$ (from $\sim 49$ million to $\sim 0.8$ million physical qubits) for architecture A1 (STQM). 
\begin{figure*}[t]
    \centering
    \includegraphics[width=\linewidth]{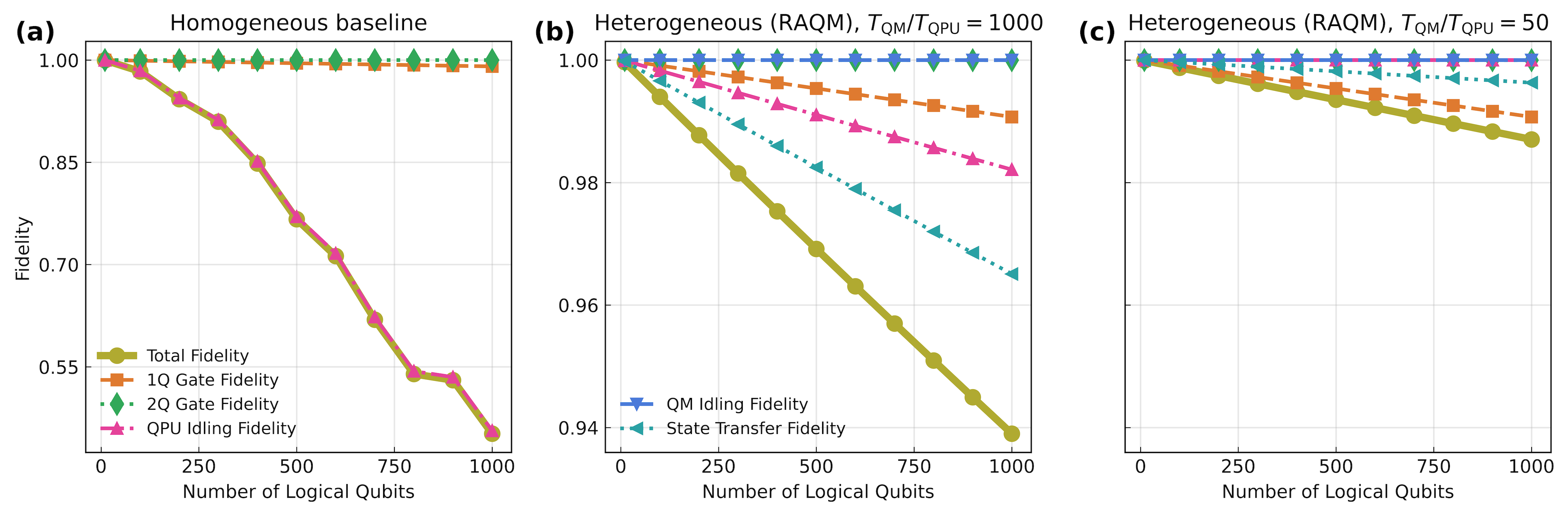}
    \caption{Error budget for 1000-qubit AQFT algorithm for (a) homogeneous baseline (b,c) heterogeneous (RAQM) with QEC cycle ratios: (b) $T_{\rm QM}/T_{\rm QPU} = 1000$ and (c) $T_{\rm QM}/T_{\rm QPU} = 50$. In (a), the error is dominated by idling in the QPU (pink). In the heterogeneous case with $T_{\rm QM}/T_{\rm QPU} = 1000$, state transfer is the main contributor to the error, followed by other substantial contributions from QPU idling and single-qubit gates (whose error is dominated by T-gates). As we decrease the cycle duration ratio to $T_{\rm QM}/T_{\rm QPU} = 50$, the state transfer and QPU idling errors are considerably lower, decreasing the overall error and leaving T-gates as the primary source of error in the circuit.}
    \label{fig:fidelity_contributions}
\end{figure*}

Next, we shift our analysis to architectures A2 and A3 (RAQM). The lattice-surgery-based transfer protocol used by RAQM does not require transversal teleportation, leading to a further reduction in the number of physical qubits required for the protocol. Fig.~\ref{fig:error_improvement_algorithms}c shows that at 1,000 LQ, the improvement factor in the number of physical qubits required reaches $138\times$ (from $\sim 49$ million to $\sim 0.4$ million physical qubits) using RAQM -- a significant improvement beyond the $60\times$ advantage achieved for STQM.

In this case, executing active error correction in memory, with a slow qubit modality, introduces execution bottlenecks that reduce the overall algorithmic execution speed. In Fig.~\ref{fig:error_improvement_interfaces}, we compare the performance advantages achieved using both STQM and RAQM, and explicitly show the impact of varying the RAQM QEC cycle time on these benefits. First, for modest QEC-cycle-time mismatches, the Q-CHESS compiler manages to approximately match the performance observed with architecture A1 (STQM). As the cycle-time mismatch between RAQM and QPU increases, overall performance degrades and the execution time increases. Nonetheless, even with an extreme QEC-cycle mismatch of 1,000$\times$, the modular architecture crosses the performance break-even threshold with respect to the baseline architecture at the scale of 20 LQ, growing to $\sim~10\times$ improvement at the 1,000-LQ scale. The circuit duration, across the range of problem sizes, the execution time increases only by $\sim 3\times$. Importantly, varying the QEC-cycle-time mismatch shifts the break-even point but does not eliminate it; heterogeneous architectures consistently surpass the baseline beyond a finite problem size, with larger mismatches delaying—but not preventing—the onset of advantage for heterogeneous architectures. 

Here we have treated the STQM and RAQM approaches to quantum memory independently in order to isolate effects of QEC cycle times on the performance of heterogeneous architectures. A natural extension combines both memory types in one architecture, using STQM as a fast ``cache'' and RAQM as a slower long-duration memory. We treat this case in the context of RSA factorization in Sec.~\ref{sec:rsa}. 

\subsection{Dominant error mechanism: Idling}

Q-CHESS provides a full decomposition of the error budget, allowing us to attribute algorithmic failure probability to specific sources and identify the limiting factors in each architectural design. Fig.~\ref{fig:fidelity_contributions} shows the error budget for the baseline and the RAQM architectures (A2, A3) for 1,000-LQ AQFT. 

The most significant finding is the decoupling of algorithmic logical error and runtime in the heterogeneous architectures, arising from a shift in the dominant error mechanism. In the monolithic baseline, Fig.~\ref{fig:fidelity_contributions}a, logical error is dominated by idling across a large qubit array, leading to linear error accumulation with problem size. In contrast, in the heterogeneous case, Fig.~\ref{fig:fidelity_contributions}b-c, the Q-CHESS compiler offloads idle states to RAQM, where error accumulation is negligible. This effectively suppresses idle-induced errors and decouples logical error from total runtime, as reflected in the difference in the vertical range (fidelity scale) between panel a and panels b,c.

For heterogeneous architectures, the dominant error sources shift to the state transfer (ST) and T-gate operations.  
Optimizing the QEC-cycle-time mismatch (the ratio of $T_\textrm{QM}/T_\textrm{QPU}$) allows the compiler to minimize the impact of transfer errors. For a large cycle mismatch, Fig.~\ref{fig:fidelity_contributions}b, state-transfer errors dominate; as this mismatch is lowered, Fig.~\ref{fig:fidelity_contributions}c, we approach the regime where the overall performance is dominated by non-Clifford (T) gates~\footnote{Optimal unitary synthesis in terms of T-gates, efficient T-states cultivation, additional non-Clifford factories, and modules that can implement T-gates natively (such as high-dimension color codes) may all further improve performance.}.

\begin{table*}[t!]
\centering
\renewcommand{\arraystretch}{1.3}
\begin{tblr}{
colspec={ll|c|c|cccc||c|cc|ccc|cc|c}, 
colsep=2pt,
column{5} = {leftsep=1pt, rightsep=1pt},
column{6} = {leftsep=1pt, rightsep=1pt},
column{7} = {leftsep=1pt, rightsep=1pt},
column{8} = {leftsep=1pt, rightsep=1pt},
width=\textwidth
}
\hline\hline
\SetCell[r=2, c=2]{c} \textbf{Arc.} && %
\SetCell[r=2]{c} \textbf{QM} & %
\SetCell[r=2]{c}\textbf{$\frac{T_\text{QM}}{T_\text{QPU}}$} & %
\SetCell[c=4]{c} \textbf{Opt. Req.} &&&& %
\textbf{Total} &%
\SetCell[c=2]{c}\textbf{L. Ops. (K)} && %
\SetCell[c=3]{c} \textbf{Qub. (M)} &&& %
\SetCell[c=2]{c} \textbf{Con. (M)} && %
\textbf{RT} \\
&&&& %
\textbf{LNG} & \textbf{STC} & \textbf{RND} & \textbf{QBT} & %
\textbf{error} & %
\textbf{CNOT}  & %
\textbf{ST}  & %
\textbf{Act.}  & %
\textbf{Stat.}  & %
\textbf{Tot.}  & %
\textbf{Loc.}  & %
\textbf{Int.}  & %
\textbf{(sec)}  \\ 
\hline
\SetCell[c=2]{r} Baseline &
    & -
    & --
    & \tN No & \tN No & \tN No & \tN No
    & 55\%
    & 46.69 & --
    & 49.14 & 0 & 49.14
    & 98.28 & 0
    & 85
    \\ \hline 
\SetCell[r=3]{c}Q-NEXUS & A1
    & STQM
    & --
    & \tY Yes & \tY Yes & \tN No & \tY Yes
    & 0.93\%
    & 15.93 & 34.84 
    & 0.602 & 0.223 & 0.825
    & 0.229 & 0.453
    & 98
    \\
 & A2 
    & RAQM 
    & 50 
    &\tY Yes & \tN No & \tY Yes & \tY Yes   
    & 1.3\%
    & 15.93 & 34.84
    & 0.354 & 0 & 0.354
    & 0.336 & 0.045
    & 100 \\
& A3
    & RAQM
    & 1000
    & \tY Yes & \tN No & \tY Yes & \tY Yes
    & 6.1\%
    & 15.93 & 16.92
    & 0.355 & 0 & 0.355    
    & 0.336 & 0.045
    & 309\\
\hline\hline
\end{tblr}

\caption{Resource estimation for a 1000-qubit AQFT circuit, comparing performance metrics between heterogeneous baseline and homogeneous architectures, such as total circuit error and runtime (RT). The metrics also quantify CNOT and state transfer (ST) logical operation counts (L. Ops.); active (Act.), static (Stat.) and total (Tot.) qubit counts (Qub.); and local coupler (Loc.) and interconnect (Int.) connection counts (Con.), all of which are shown in millions/thousands (M/K). Three variations of the Q-NEXUS heterogeneous architecture are compared, highlighting how different Optional Requirements (Opt. Req.) can impact performance. Unlike the homogeneous baseline, all Q-NEXUS architectures encode quantum data in multiple Qubit (QBT) modalities and utilize Long-range (LNG) routing of quantum data via the QB to minimize logical SWAP gates. In addition, we explore the benefit of Static (STC) short-term storage and Random-access (RND) long-term storage. A1 utilizes STQM memory only and transversal teleportation. A2 and A3 both utilize long-term RAQM, enabled by the lattice-surgery transfer protocol, and operate at different QEC cycles times in the QM: (A2) $T_\textrm{QM}=50\,\upmu$s ($T_\textrm{QM}/T_\textrm{QPU}=50$) and (A3) $T_\textrm{QM}=1000\,\upmu$s ($T_\textrm{QM}/T_\textrm{QPU}=1000$).
\label{table:qubit_resources_QFT}}
\end{table*}

\subsection{Requirement-driven resource analysis}
\label{sec:midoptreq}

The dominant performance advantage identified in the previous section arises from the suppression of idling errors by offloading quantum states into memory. Here, we examine how this structural shift translates into improvements in resource overheads. Crucially, the benefits of heterogeneous architectures extend beyond reducing total qubit count, encompassing simplification of control complexity and efficient bus-mediated routing. To make this explicit, we analyze a representative workload -- AQFT at the scale of 1,000 logical qubits -- and decompose the required resources into the components reported in Table~\ref{table:qubit_resources_QFT} for each architecture under study.

We are able to examine measures of logical operations -- specifically the CNOT gates and state transfers (ST) -- as well as calculated runtime. Our resource analysis also includes counts over the following distinct physical resources:

\begin{itemize}
    \item \textbf{Active qubits (Act.)}: physical qubits participating in active QEC. These require high connectivity (e.g., four couplers for surface code stabilizers), continuous calibration, and real-time decoding. For example, a 1,000-LQ system at $d=25$ and $1\,\upmu$s cycle time generates $\sim6.25\times10^{11}$ bits/s of decoding data.
    \item \textbf{Static qubits (Stat.)}: physical qubits that do not participate in active QEC (e.g., STQM). These avoid the control, connectivity, and decoding overheads associated with active qubits.
    \item \textbf{Local couplers (Loc.)}: Nearest-neighbor connections required to support QEC operations.
    \item \textbf{Interconnects (Int.)}: Interconnects required for Bell-pair generation (i.e. part of the quantum bus). 
\end{itemize}

We isolate the structural impact of heterogeneous modularity by examining how individual optional requirements quantitatively affect compiled-resource metrics.

\csubsubsec{Long-range routing optional requirement:} 
In the monolithic baseline, long-range routing is implemented through SWAP networks, which decompose into sequences of CNOT gates and therefore dominate the two-qubit gate count. In contrast, the heterogeneous architecture offloads long-range communication to state-transfer (ST) operations via the quantum bus. This shift is directly reflected in the compiled metrics shown in \cref{table:qubit_resources_QFT}: we achieve a ~$3\times$ reduction in the two-qubit gate count by replacing routing with a roughly equivalent $\sim35,000$ state transfer operations. Despite this substitution, circuit error drops from $55\%$ to $\sim1-6\%$, indicating a net benefit from shifting routing to bus-mediated transfer. Runtime increases moderately (up to $3.6\times$ at 309\,s). Notably, when the QM cycle time $T_\textrm{QM}$ is extended in A3, the number of transfer operations decreases compared to A1/A2, reflecting the compiler cost function (\cref{sec:compiler}.B) which trades off transfer overhead against idling cost as the QM-to-QPU cycle-time ratio increases.

\csubsubsec{Static memory optional requirement:} 
Introducing static qubits in A1 reduces the number of active qubits $\sim82\times$ relative to the baseline. Low-compexity static qubits account for 27\% of the total qubit count. Local couplers drop $>400\times$, directly reflecting the reduction in actively controlled hardware. This restructuring is accompanied by the introduction of $\sim10^5$ interconnects, which replace local connectivity with bus-mediated communication. While this increases reliance on interconnect infrastructure, these links do not require the same dense, error-corrected connectivity as local couplers, shifting complexity away from tightly coupled QEC hardware toward modular communication. Runtime remains relatively constant (15\% increase).

\csubsubsec{Random-access memory optional requirement:} 
By utilizing long-term RAQM in A2 and A3, active qubits are reduced a further $\sim2\times$ ($138\times$ total relative to the baseline), indicating increased storage density, while local couplers remain low ($50\%$ higher than STQM, but still $\sim300\times$ lower than the baseline). Unlike STQM, all qubits are actively corrected, and runtime increases by up to $3.6\times$, reflecting the cost of fault-tolerant access with finite memory cycle times. Notably, the number of interconnects is significantly reduced, by an order-of-magnitude compared to STQM. This reduction arises from the lattice-surgery-based transfer protocol used in RAQM, which enables structured, fault-tolerant communication between modules without requiring dedicated point-to-point interconnects for each transfer. As a result, RAQM achieves a more efficient use of interconnect resources while preserving the benefits of modular communication.

\csubsubsec{Mixed qubit modalities optional requirement:} 
Across A1--A3, total qubits and local couplers are both reduced by approximately two orders of magnitude. These reductions arise from modality-specific advantages: ULC qubits enabling effectively static storage in STQM (A1), and LC qubits enabling reduced encoding overhead in RAQM (A2/A3). This demonstrates that mixed qubit modalities directly enable system-level resource savings.

\section{Large scale resource analysis: RSA-2048 factorization} \label{sec:rsa}

The resource requirements of RSA factorization have been extensively analyzed in prior work, particularly by Gidney \emph{et al.} \cite{GidneyEkera2021RSA,Gidney2025RSAUpdate}. These studies provide well-defined circuit constructions, resource estimates, and performance models that make RSA an ideal workload for evaluating architectural trade-offs at scale. Most importantly, analyzing the structure of the algorithm reveals that only $0.5\%$ of cycles are devoted to pure logical operations, while the remainder correspond to idling or routing. This structure makes RSA a particularly informative workload for evaluating heterogeneous architectures that explicitly separate processing and storage.

In this section, we compile and schedule Gidney’s RSA-2048 algorithm~\cite{Gidney2025RSAUpdate} using the Q-CHESS toolchain. Our objective is to systematically evaluate how architectural heterogeneity transforms resource requirements and execution time without modification of the algorithm\footnote{Further algorithmic innovations, e.g.,~\cite{Webster2026Iceberg}, may provide additional gains.}. We structure the resource estimation in three stages. First, we introduce the metrics we will calculate for each architecture via application of the Q-CHESS compiler to fully compiling and scheduling the dominant subroutines of the RSA algorithm. Second, we define specific reference architectures for our analysis, including a homogeneous baseline with fixed QPU and QSF configurations and a common heterogeneous architecture with QPU, QSF and QM modules shared across all Q-NEXUS designs. Third, we aggregate the specific system-level time and space-cost estimates for a range of specific implementations of the Q-NEXUS architecture. 

We adopt a requirement-driven analysis. Starting from a monolithic baseline, we progressively introduce architectural capabilities corresponding to the optional requirements defined in Sec.~\ref{sec:intro-qnexus}, and quantify their impact on compiled-resource metrics in isolation:
Multi-core QPUs; Hierarchical memory; Mixed QEC codes; Application-specific accelerators. 

\csubsubsec{Subroutine-level resource estimation:}
Direct compilation of the full RSA circuit is computationally intensive at this scale. Instead, we decompose the $1{,}399$-logical-qubit program into its dominant subroutines, fully compile and schedule each one, and reconstruct total resource usage from their repetition counts. This allows us to extract both qubit/coupler overheads and execution times directly from compiled schedules. The three subroutines -- an Adder, Lookup and Phaseup -- are described in \cref{table:subroutines}, including the number of times they occur in the full program ($N_\textrm{adder}, N_\textrm{lookup}, N_\textrm{phaseup}$). 


\begin{table}[t]
    \centering
    \begin{tblr}{ccc}
        \hline\hline
         Subroutine & Logical qubits & Occurrences \\ \hline
         33-bit Adder & 68~\cite{Cuccaro2004} & $N_\textrm{adder}=10,621,207$~\cite{Gidney2025RSAUpdate} \\
         6-bit Lookup & 70~\cite{Babbush2018} & $N_\textrm{lookup}=7,646,081$~\cite{Gidney2025RSAUpdate} \\
         6-bit Phaseup & 14~\cite{Gidney2025RSAUpdate} & $N_\textrm{phaseup}=1,581,186$~\cite{Gidney2025RSAUpdate} \\
         \hline\hline
    \end{tblr}
    \caption{The RSA circuit can be decomposed into three dominant subroutines: a $33$-bit Adder, a “Lookup” that acts on a $6$-bit address register, and a “Phaseup” that acts on a $6$-bit target register. The number of logical qubits required for each module's implementation and the number of occurrences in the full program are detailed.}
    \label{table:subroutines}
\end{table}

Q-CHESS produces an optimized schedule for each subroutine, along with execution metadata including runtime ($\tau_\textrm{adder}, \tau_\textrm{lookup}, \tau_\textrm{phaseup}$), error accumulation, and hardware utilization. Qubit and coupler overheads are derived directly from these compiled schedules.

To estimate total execution time, we follow the methodology of~\cite{Gidney2025RSAUpdate}, combining subroutine runtimes with repetition counts and accounting for fidelity and sampling overhead. The total time per shot is
\begin{align*}
\text{Total time per shot} = &N_\textrm{adder}\tau_\textrm{adder} + N_\textrm{lookup}\tau_\textrm{lookup} \\
&+ N_\textrm{phaseup}\tau_\textrm{phaseup},
\end{align*}
and the total runtime (RT) is
\begin{align*}
\text{RT} = \frac{\text{(Total time per shot)}}{F_\text{RSA}}\times 9.2\times 1.14,
\end{align*}
where $F_\text{RSA}$ is the total program fidelity, $9.2$ is the average number of shots required for success, and $1.14$ accounts for additional overhead~\cite{Gidney2025}. 

\csubsubsec{Space-time cost metrics:}
Estimating space cost requires distinguishing between different hardware resources. Accordingly we explicitly track: Active (Act.) qubits, $N_{q\text{-act}}$, Static (Stat.) qubits, $N_{q\text{-st}}$, Local (Loc.) couplers, $N_\textrm{lcp}$, non-local (N-L) couplers, $N_\textrm{nlcp}$, and interconnects (int.), $N_\textrm{int}$ (see \cref{sec:midoptreq}), capturing physical qubit, connectivity and communication overheads. 

These quantities enable a generalized space-cost model:
\begin{align}\label{eq:space_cost}
\text{cost}_\textrm{space} = &W_\textrm{ac}N_{q\text{-act}} + W_\textrm{st}N_{q\textrm{-st}}+\\ \nonumber
&W_\textrm{lcp}N_\textrm{lcp} + W_\textrm{nlcp}N_\textrm{nlcp} + W_\textrm{int}N_\textrm{int}  
\end{align}

\noindent The weights $W_i = (W_\textrm{ac},\, W_\textrm{st},\, W_\textrm{lcp},\, W_\textrm{nlcp},\, W_\textrm{int})$ should be chosen according to hardware realities and constraints. In the following, we present the complete space-cost of each architectural choice we make (all $N_i$ are explicitly extracted), along with two explicit examples: one that coincides with the common total number of physical qubits and another that uses a complexity weighting $W_\textrm{st} = W_\textrm{ac}/2$ and $W_\textrm{lcp} = W_\textrm{nlcp}/4 = 2W_\textrm{int}$.   

\begin{table*}[t]
\centering
\renewcommand{\arraystretch}{1.3}
\begin{tblr}{
  width=\textwidth,
  colsep=2pt, 
  colspec={cc|c|cccc||ccc|ccc|c|cc}, 
}
\hline\hline
\SetCell[r=2, c=2]{c}\textbf{Arch.} &
& \SetCell[r=2]{c}{\textbf{RAQM/Cache/}\textbf{ASQPU} }
& \SetCell[c=4]{c} \textbf{Opt. Req.} &&&
& \SetCell[c=3]{c} \textbf{Qub. (M)}&&
& \SetCell[c=3]{c} \textbf{Con. (M)} && 
& \SetCell[c=1]{c} \textbf{RT}
& \SetCell[c=2]{c} \textbf{Cost (M-day)} &
\\
&& 
& \textbf{MLT} & \textbf{APL} & \textbf{HRC} & \textbf{CDE}   %
& \textbf{Act.} & \textbf{Stat.} & \textbf{Tot.}
& \textbf{Loc.} & \textbf{N-L.} & \textbf{Int.}
& \textbf{(days)}
& \textbf{Qb.}
& \textbf{Cn.}
\\ \hline
\SetCell[c=2]{c} Mono.~\cite{Gidney2025RSAUpdate}&
    & --
    & \tN No & \tN No & \tN No & \tN No
    & 0.9  & 0 & 0.9
    & 1.8 & 0 & 0
    & 5 
    & 4.5 & 9 \\ \hline
\SetCell[r=6]{c} \rotatebox[origin=c]{90}{Q-NEXUS}
    & B1
    & -- / STQM / --
    & \tY Yes & \tN No & \tN No & \tN No 
    & 0.53 & 0.51 & 1.04
    & 0.51 & 0 & 0.51
    & 9.2 
    & 9.54 & 7.04 \\
& B2
    & Surface / STQM / --
    & \tY Yes & \tN No & \tY Yes & \tN No
    & 0.33  & 0.05 & 0.38
    & 0.50 & 0 & 0.06
    & 9.2  
    & 3.5 & 4.88 \\
& B3 
    & Gross / STQM / --
    & \tY Yes & \tN No & \tY Yes & \tY Yes
    & 0.14 & 0.05 & 0.19
    & 0.16 & 0.03 & 0.07
    & 9.2 
    & 1.74 & 2.90 \\
& B4
    & -- / STQM / Adder
    & \tY Yes & \tY Yes & \tN No & \tN No 
    & 0.56 & 0.54 & 1.10
    & 0.57 & 0 & 0.52
    & 4.9 
    & 5.37 & 4.07 \\
& B5
    & Surface / STQM / Adder
    & \tY Yes & \tY Yes & \tY Yes & \tN No 
    & 0.39 & 0.05 & 0.44
    & 0.55 & 0 & 0.08
    & 4.9
    & 2.15 & 2.89 \\
& B6
    & Gross / STQM / Adder
    & \tY Yes & \tY Yes & \tY Yes & \tY Yes 
    & 0.20 & 0.05  & 0.25
    & 0.21 & 0.03 & 0.08
    & 4.9 
    & 1.22 & 1.81 \\
\hline\hline
\end{tblr}
\caption{Resource estimation for factoring a 2048-bit RSA integer (1,399 logical qubits), comparing space-time trade-offs between homogeneous baseline and an increasing heterogeneous architectures. In addition to the metrics shown in \cref{table:qubit_resources_QFT}, we quantify the unweighted qubit (Qb.) and weighted connection (Cn.) cost in millions of component-days (M-days). 
The monolithic architecture proposed by Gidney \cite{Gidney2025RSAUpdate} serves as the state-of-the-art baseline. The Q-NEXUS architecture examples systematically increase the heterogeneity in the system, by increasing the number of requirements they meet. 
Processing requirements: MLT utilizes MuLTi-core units, here with two QPUs (activated in B1-B6); APL employs an APpLication-specific QPU (ASQPU) accelerator for the Adder subroutine (B4-B6). 
Memory requirements: all Q-NEXUS architectures include an STQM cache tier (B1-B6); HRC employs HieRarChical memory with both STQM and RAQM tiers (B2, B3, B5, B6). 
Representation requirements: CDE supports multiple error-correcting CoDEs, using a qLDPC Gross code in RAMQ (B3, B6). 
Non-local couplers are considered to be twice as costly as local couplers while calculating the coupler Space-time volume. Further details are provided in \cref{app:rsa} and \cref{sec:rsa}.
\label{table:qubit_resources_RSA}}
\end{table*}

\csubsubsec{Reference architectures:}
We begin by defining a monolithic baseline architecture comprising two key components: a superconducting QPU and dedicated embedded QSF modules supplying high-rate CCZ states. 

For the QPU, Gidney's algorithm for factoring 2048-bit integers requires a total of $\sim6.9\times10^{13}$ QEC logical cycles with a cycle depth of $\sim4.3\times10^{10}$. This scale demands an exceptionally low error rate per cycle ($<10^{-15}$), necessitating a resource-intensive $d=25$ code. The depth of the program further biases toward fast qubit modalities as runtimes quickly become impractical. For example, with a 1~ms QEC cycle, a single shot requires 500 days and a full solution ($\sim10$ shots) requires 13.7 years, compared to approximately 5 days with a $1\,\upmu$s cycle time. Extending beyond 2048-bit integers to 8192-bit integers requires code distances $d>30$ and exceeds one year of execution time even assuming $1\,\upmu$s cycle times. The QPU is augmented by a QSF optimized for Toffoli gates. We therefore employ CCZ-factories based on T-cultivation \cite{Gidney2019}, which significantly reduce resource overheads. The physical error budget is kept consistent with~\cref{app:rsa}.

\begin{figure*}[t]
    \centering
    \includegraphics[width=\linewidth]{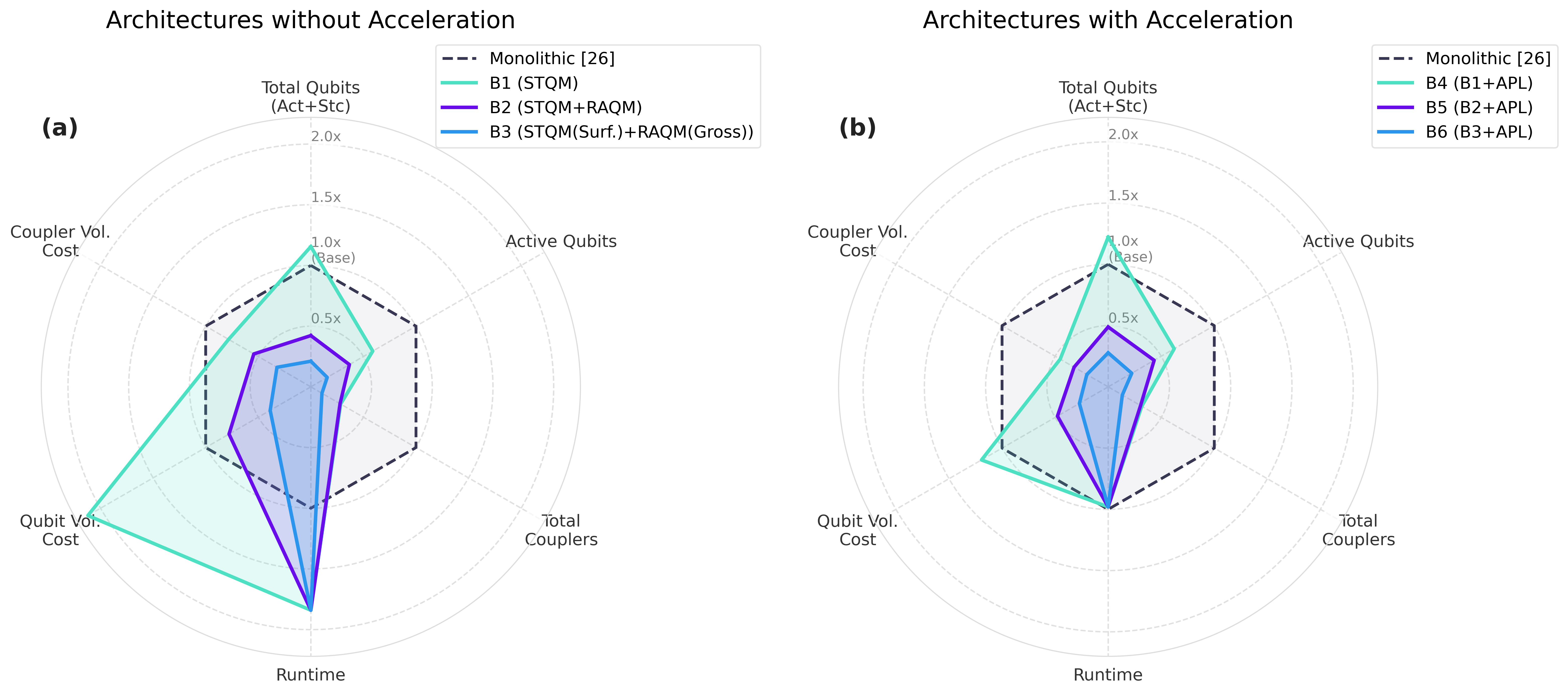}
    \caption{Normalized performance of Q-NEXUS modular architectures (solid polygons) compared to a state-of-the-art monolithic surface-code baseline (dashed black hexagon, normalized to a value of $1.0\times$), highlighting the role of algorithmic acceleration in driving net performance improvements. Radial axes represent physical hardware requirements (Total Qubits, Active Qubits, Total Couplers) and execution metrics (Runtime, Qubit Vol. Cost, Coupler Vol. Cost), with values closer to the origin indicating higher efficiency. See Table~\ref{table:qubit_resources_RSA} for full architectural details.(a) Performance of candidate architectures without algorithmic acceleration; B1 relies only on a short term STQM, B2 adds RAQM using surface codes, B3 adds RQAM using high-density qLDPC codes. (b) Performance of architectures with algorithmic acceleration; architectures B4, B5, B6 add an ASQPU to accelerate the Adder subroutine to architectures B1, B2, B3.}
    \label{fig:rsa_spider}
\end{figure*}

For the Q-NEXUS heterogeneous architectures, we utilize the processing and state-generation capabilities from the monolithic baseline, and additionally introduce quantum memory. The common modules for Q-NEXUS are:
\begin{itemize}
    \item \textbf{QPU}: Q-NEXUS employs small QPU cores, with only three logical qubits per core. We utilize a lower distance \mbox{$d=19$} surface code for the QPU because the overhead from idling has been offloaded from the QPU into the cache. Hardware design maintains use of an existing experimentally demonstrated grid-coupling topology.
    \item \textbf{QSF}: Similarly to the monolithic baseline, we employ CCZ-factories based on T-cultivation~\cite{Gidney2019} (see \cref{app:rsa} for physical error budgets).
    \item \textbf{QM}: As an initial configuration, we consider a single QPU coupled to a STQM with a one-to-one logical capacity. The cache size will depend on the specific implementation.
\end{itemize}
In the following subsections, we expand this common configuration by systematically introducing additional heterogeneity reflecting instantiations of the optional recommendations made in Sec.~\ref{sec:intro-qnexus}: multi-core processing, hierarchical memory, mixed codes, and application-specific accelerators. The impacts of the various architectural choices we explore on physical resources, runtime, and overall space-time cost, compared against the homogeneous baseline, are presented in summary form in Table~\ref{table:qubit_resources_RSA}. A comprehensive visualization of Pareto improvements over physical space-time resources, across all evaluated architectures, is provided in \cref{fig:rsa_spider}, demonstrating that two heterogeneous configurations combining hierarchical memory and acceleration (B5, B6) achieve strict Pareto dominance across all evaluated metrics.

\subsection{Impact of Multi-core QPUs}

We first evaluate the impact of introducing multi-core processing units. Allowing multiple QPUs enables parallel execution of logical operations when the algorithm exposes sufficient concurrency. We compare its performance metrics in \cref{table:qubit_resources_RSA} (B1) to the monolithic baseline (Mono.).

The specific realization we evaluate consists of two 3Q-QPUs (with QSF modules) alongside an STQM, the latter comprising sufficient logical qubits to support the algorithm size (here, 1,399). A parallelization analysis suggests that two QPU units provide runtime speedup, but we do not observe further speedup with an increased number of QPUs unless the algorithm itself is modified to support parallelization. Specifically, for all architectures with MLT enabled, we use two 3Q-QPUs with $d=19$ surface codes, $1\,\upmu$s cycle time, and $p/p_{th} = 1/12$. Compared to a single-core Q-NEXUS architecture, this achieves a $15\%$ runtime speed-up at the cost of a $5\%$ increase in qubit resources. 

As highlighted in \cref{table:qubit_resources_RSA}, this architecture achieves factorization of 2048-bit RSA integers with 1.04M physical qubits and in under 9.2 days. Furthermore, nearly half the qubits are found in the lower-complexity cache. Introducing multi-core processing (B1 vs. Monolithic in \cref{table:qubit_resources_RSA}) reduces local couplers $\sim3.5\times$ to $0.51$M, reflecting the distribution of compute across modules. This comes with an increase in runtime from $5$ to $9.2$ days and a rise in ``qubit-time cost" from $4.5$ to $9.54$ million qubit days (M-days), establishing a clear space–time trade-off where reduced connectivity is achieved at the expense of longer execution.

\subsection{Impact of hierarchical memory}

We next analyze the impact of including hierarchical quantum memory. For this purpose, we reduce the cache to a fixed-sized (145Q) STQM directly connected to the QPU via transversal teleportation, together with a 1,300Q RAQM for long-duration storage. 
The RAQM uses lattice surgery modules with $d=9$ surface codes, $p/p_{th} = 1/60$, and $d=19$ transfer patches (matching the QPU code distance) with a maximum swap-distance of $k=4$.

A scheduler for such architectures works at two levels of abstraction. At a high level, the scheduler guarantees that active qubits reside in the STQM cache. In the example of Gidney's algorithm, as the subroutines are sequential, only a small number of qubits are active at any moment. The average number of active logical qubits is 64 at any given moment, hence by having a cache of size 145 we ensure that the active qubits and those expected to participate in upcoming subroutines are already in cache.Meanwhile, the scheduler can locally swap or route qubits in the RAQM to guarantee that qubits reside at the appropriate transfer location when a transfer to the cache is due. At a lower level, the scheduler acts, as before, between the QPUs and the cache, ensuring that transfer times remain short and the QPU does not wait idle.

With this approach, we show that 2048-bit RSA integers can be factored successfully with 381k physical qubits and in 9.2 days. Adding hierarchical memory (B2 vs. B1 in \cref{table:qubit_resources_RSA}) reduces the total physical-qubit count by $\sim2.7\times$ to $0.38$M,and qubit-time cost to $3.5$ M-days. This is accompanied by a reduction in the fraction of static qubits from $49\%$ to $13\%$, indicating a more efficient storage utilization. Coupler requirements remain essentially unchanged, showing that these gains arise from improved memory structure rather than changes in connectivity.

\subsection{Impact of increased memory density through mixed QEC codes}

Both architectural choices above relied on experimentally demonstrated grid topology. To further illustrate the advantage provided by heterogeneity in Q-NEXUS, we show that code heterogeneity provides a straightforward path to leverage long-range couplings if and when such devices mature. With hypothetical long-range coupling, the implementation of quantum memory using qLDPC codes can reduce the total physical qubit resources required for the implementation of Gidney's algorithm. In the following we assume the device connectivity envisioned by IBM in \cite{Yoder2025}.

We analyze the hierarchical memory architecture described above, changing the RAQM implementation from a $d=9$ surface code to a $d=12$ Gross (qLDPC) code. We envision a long-time storage composed of 105 Gross code ($[[144,12,12]]$) modules, each comprising 12 logical qubits and implemented with long coherence modalities to allow the use of low-distance codes for long-term storage. This code also presents an opportunity to reduce the density of interconnects beyond the transfer-patch construction used with the surface code. Using the intrinsic shift automorphism property of the Gross code, information can be shuffled efficiently within the memory module. This allows information transfer between the QPU and memory modules to be narrowed to a small region of the memory.In this example only 26 interconnects are needed for lattice surgery between a surface-code qubit in the Q-cache and Gross-code qubits in the RAQM \cite{stein2024architecturesheterogeneousquantumerror}. 

With these architectural choices we find that 2048-bit integers can be factored with 190k qubits (see \cref{app:rsa} for details) and in 9.2 days.
The use of mixed error-correcting codes (B3 vs. B2 in \cref{table:qubit_resources_RSA}) reduces physical qubits $\sim 2\times$ to $0.19$M, and local couplers $\sim 3.1 times$ to $0.16$M. This leads to a reduction in qubit-time cost from $1.74$ M-days and coupler-time cost from $2.9$ M-days. These results show that code heterogeneity directly improves encoding efficiency and hardware footprint without introducing additional time overhead.

\subsection{Impact of algorithm-specific accelerators}

Finally, we identify that $\sim70\%$ of factoring runtime is spent executing the Adder subroutine. As its implementation is fundamentally serial, adding QPU units cannot speed up this subroutine execution, motivating the introduction of application-specific accelerators (ASQPUs) in order to address recurring computational bottlenecks.

In this case, we analyze the impact of adding an arithmetic accelerator in the form of a dedicated $37$-logical-qubit ASQPU for the Adder subroutine. We find that with an additional Adder ASQPU, 2048-bit RSA integers can be factored in 4.9 days, a $1.87\times$ runtime speedup, at the cost of an increased physical qubit count between $5\%$-$20\%$, depending on the architecture. Application-specific acceleration (B4 vs. B1) reduces runtime $\sim1.9\times$ to $4.9$ days ($\sim1.9\times$) without increasing overall hardware complexity: the local coupler count increased only slightly ($0.51$M $\rightarrow 0.57$M). The reduced runtime directly lowers both qubit-time cost ($9.54 \rightarrow 5.37$ M-days) and coupler-time cost ($7.04 \rightarrow 4.07$ M-days), demonstrating that targeted architectural specialization converts the time overhead introduced by modularity into a net system-level gain. This behavior is also observed with HRC (B5 vs. B2) and CDE (B6 vs. B3) enabled.

\section{Discussion}\label{sec:discussion}

The central result of this work is that quantum computer \emph{architecture} provides an under-explored and high-potential pathway to accelerate the development of large-scale quantum computers. Discussions obviating inclusion of this element in favor of a narrow focus on QEC code development or physical-device scaling miss a profound opportunity to achieve more with hardware being developed in near-term industry roadmaps. 

The impact of embracing heterogeneous architectural design concepts in quantum computer development is most clearly demonstrated in the RSA-2048 resource analysis (Table~\ref{table:qubit_resources_RSA} and~\cref{fig:rsa_spider}). Starting from the state-of-the-art monolithic baseline requiring $0.9$M actively controlled qubits and $1.8$M local couplers, Q-NEXUS reduces the total physical qubit count to $0.19$M (B3) and $0.25$M (B6), a $\sim4.7\times$ reduction, while simultaneously collapsing the volume of actively controlled hardware. In B3, only $0.14$M qubits remain under active QEC, with the remainder offloaded to memory, and local couplers are reduced from $1.8$M to $0.16$M ($\sim11\times$), with only $0.03$M non-local couplers and $0.07$M interconnects required. This represents a qualitative shift in system complexity: rather than scaling millions of densely connected, actively corrected qubits, the architecture concentrates control into a small computational core while the majority of qubits reside in a narrowly applied, simplified storage. Crucially, these reductions are achieved without runtime penalty, and in the fully heterogeneous configuration (B6) with application-specific acceleration, both qubit and coupler space-time costs are reduced to $1.22$ M-days and $1.81$ M-days.

Throughout the manuscript, we introduced heterogeneity through explicitly defined functional requirements, each corresponding to a concrete architectural capability with directly measurable impact on compiled resource metrics: LNG, STC, RND, and QBT at mid-scale (Sec.~\ref{sec:midoptreq}), and MLT, APL, HRC, and CDE at large scale (Sec.~\ref{sec:rsa}). The impact of introducing modularity and heterogeneity was robust to changes in algorithm (AQFT, Quantum Adder and Fermi–Hubbard workloads), and, crucially, hardware constraints. Even under memory–processor cycle mismatches of $10^3\times$, heterogeneous architectures retain a performance advantage at total system sizes as small as $\sim20$ logical qubits, and achieve $\sim11\times$ error improvement at 1,000 LQ. These results demonstrate that by structuring the system according to function, rather than enforcing uniformity, one can suppress dominant error mechanisms and reduce resource overheads in a verifiable and scalable manner.

These conclusions are not derived from a single algorithmic case study, but arise from a structured and quantitative sampling of an extremely broad architectural-design trade space across algorithms with diverse characteristics. All specific implementations embraced in our exploration were derived from realistic hardware capabilities demonstrated in the literature and/or appearing on industry roadmaps. We do not assume unphysical connectivity, biased error rates detached from physical demonstrations, or abstracted performance of QEC widgets, favoring instead detailed resource estimation enabled by the Q-CHESS compiler for heterogeneous architectures.

Importantly, these results show that achieving large-scale fault-tolerant quantum computing is not merely a race to find a single ``Goldilocks'' code, but instead create an opportunity for practically motivated \emph{functional specialization} of QEC codes within a heterogeneous architecture. This is most clearly illustrated in the RSA-2048 analysis (Table~\ref{table:qubit_resources_RSA}). Incorporating mixed QEC codes (B2$\rightarrow$B3) yields a $\sim2\times$ reduction in total physical-qubit count $0.38$M to $0.19$M ($\sim2\times$). Crucially, the additional code is deployed exclusively within the memory subsystem, where the requirements are fundamentally different as long-term storage does not demand the full overhead of universal fault-tolerant gate synthesis. Matching the strengths and weaknesses of QEC codes to target functional specifications for heterogeneous modules provides clear wins in overall system performance.

\vspace{0.5em}
Having established the opportunity presented by adopting a heterogeneous architecture, we pivot the discussion to \emph{realizing} a Q-NEXUS fault-tolerant quantum computer.

Q-NEXUS provides a concrete framework for developing specialized hardware components within a fault-tolerant quantum architecture. First, it defines a clear micro-architecture for all modules, including interconnect structure, fault-tolerant transfer protocols, and detailed specifications of the quantum memory subsystems (Sec.~\ref{sec:intro-qnexus}). Second, our RSA-2048 analysis (resource estimates in Table~\ref{table:qubit_resources_RSA}) translates these architectural requirements into concrete performance targets for hardware developers built from bottom-up detailed accounting:
\begin{enumerate}
\item \emph{Computation:} QPUs operating with just three logical qubits per core, two cores, fast cycle times of $\sim1\,\upmu$s, and two CCZ QSFs per core are sufficient, breaking previous trends towards large monolithic quantum processors.
\item \emph{Communication:} The QB (interconnects) must sustain a logical state transfer rate $0.1$ MHz.
\item \emph{Memory:} Quantum memory systems comprising $1260$ logical qubits in RAQM and $145$ unencoded ultra-long-coherence qubits in STQM.
\end{enumerate}

All results presented in this work are underpinned by the Q-CHESS compiler, which provides end-to-end orchestration from high-level algorithms to machine-level instructions across heterogeneous modules. By fully compiling and scheduling realistic workloads, Q-CHESS exposes the true resource costs and system bottlenecks (Secs.~\ref{sec:midnalysis} and \ref{sec:rsa}) that are not visible in abstract scaling analyses. We view Q-CHESS as a development tool for the broader community, enabling algorithm designers, hardware developers, and QEC researchers to quantitatively evaluate when and how their innovations impact fault-tolerant quantum computing, through architecture-aware resource estimation.

Our results demonstrate that the pathway to large-scale deployable systems is not defined by a monolithic scaling ``race'' among individual qubit modalities. In Q-NEXUS, the scaling challenge is fundamentally redefined: the growth of the system is no longer tied to expanding a single, fully fault-tolerant processor. Instead the QPU size is fixed and scaling is shunted to a technologically simplified quantum memory with limited functionality. We identify that both unencoded ultra-long-coherence quantum memory (STQM) and QEC-protected random access quantum memory (RAQM) can deliver multiple orders of magnitude in increased efficiency. Compared to the homogeneous baseline STQM reduces total physical-qubit overhead by $\sim 60\times$ and local couplers by $\sim 400\times$ through control simplification, while using RAQM reduces physical qubits by $\sim138\times$ via storage density and logical efficiency. These two approaches are not mutually exclusive, as demonstrated via our analyses of hierarchical memory architectures in which one can achieve control simplification, high-density storage, and low-latency read/write access (Sec.~\ref{sec:rsa}). Accordingly, new emphasis on high-density and high-performance quantum memories should be prioritized going forward

The central implementation challenge in Q-NEXUS is the quantum bus (QB) that facilitates communication between modules and enables their specialized functions. We have aimed to ground this proposal in technologies that are already demonstrated or under current development, but the QB remains technically demanding. The difficulty of realizing a high-performance QB should not be interpreted as a weakness of the architectural approach as the development of interconnect technologies has already become a necessity even in homogeneous architectures \cite{Heya2025,Pasqal2025,IonQ2026,NuSite,AliroSite,QunnectSite}. In this sense, the present work acts to reinforce existing trends while opening new architectural approaches in which the development of interconnects and fault-tolerant transfer protocols can reduce overall resource requirements by orders of magnitude, dramatically accelerating progress in the field.

Taken together, the innovations captured in Q-NEXUS and Q-CHESS open significant opportunities across the quantum computing stack. For QEC code developers,
Q-NEXUS provides a natural framework for \emph{code specialization}, where different error-correcting codes can be matched to the functional role of each module (e.g., codes with mature fault-tolerant gate sets, high encoding rate codes for memory), enabling system-level optimization beyond a single-code paradigm. Hardware developers can leverage Q-NEXUS to define clear targets for modular hardware components, including fixed-size QPUs, scalable memory subsystems, and quantum interconnects. It enables each qubit modality to be deployed in the role best aligned with its physical characteristics, rather than ``competing'' to be the single winner in a homogeneous architecture. Lastly, Q-CHESS provides algorithmic developers a compilation framework that translates high-level algorithms into machine-level execution on heterogeneous systems, enabling realistic resource estimation and revealing pathways to substantial reductions in fault-tolerant overheads.

\vspace{0.5em}

\section{Conclusion}\label{sec:conclusion}

We have introduced Q-NEXUS, a heterogeneous quantum computing architecture that integrates specialized modules for computation, communication, and storage, together with a micro-architecture-aware compiler (Q-CHESS) for end-to-end orchestration. A defining feature of this architecture is that the size of the quantum processing unit (QPU) is \emph{fixed}, fundamentally shifting the responsibility for system scaling to the quantum memory (QM). Q-NEXUS provides two complementary pathways to achieve this scaling: increasing storage density through random-access quantum memory (RAQM), and simplifying control through static transversal quantum memory (STQM). Combined within a hierarchical memory system, these approaches enable low-latency access to quantum data while dramatically reducing the hardware and control overhead required for large-scale fault-tolerant computation.

This architectural shift yields substantial and verifiable gains. In the RSA-2048 analysis, a fully heterogeneous implementation reduces the total physical qubit requirement from $0.9$M in a monolithic baseline to $0.19$M, while simultaneously reducing actively controlled qubits to $0.14$M and local couplers from $1.8$M to $0.16$M. These reductions reflect not only a decrease in total hardware, but a qualitative simplification of system complexity: the majority of quantum data is stored in modules with reduced control and connectivity requirements.

Nonetheless we believe the greatest strength of Q-NEXUS and Q-CHESS is derived from the new opportunities for innovation enabled across hardware device engineering, architectures, and QEC code development. We outline several ongoing areas of development and continued research.

\csubsubsec{Heterogeneity within a qubit modality:} 
The results presented here are fundamentally \emph{architectural}, and similar advantages are expected from heterogeneity within a single qubit modality. For example, mixed-species ion traps, already standard for sympathetic cooling \cite{Ransford2025}, have been proposed for memory–compute separation \cite{Monroe2014,Brandl2017}. Likewise, in superconducting systems, combining qubits for computation with resonators for memory has demonstrated strong potential \cite{Milul2023,Mariantoni2011,Stein2023}. Novel analyses seeking to further diversify subcomponents provide a promising pathway.

\csubsubsec{Defining sharp module interfaces:} 
In this work, we deliberately avoided specifying fully detailed module interfaces, as doing so introduces substantial consideration of the \emph{classical} control infrastructure. Our focus here was on quantum resource estimation, deferring the equally important classical infrastructure to future work. We believe that the full development of memory interfaces is a critical area of continued research, for instance addressing the operation of the classical address bus that assists with information load/store \cite{Patterson2017}.

\csubsubsec{Augmenting the QB micro-architecture:}
The QB considered in this work is based on teleportation-enabled quantum interconnects. Other viable realizations of this communication layer are possible, as Table~\ref{table:architecture_comparison} shows that several existing architectures instead employ physical qubit shuttling \cite{Brandl2017,Ransford2025}. We believe it could be advantage to incorporate shuttling-based transport into the Q-NEXUS micro-architecture, providing a hierarchical bus akin to our hierarchical approach to memory..

\csubsubsec{Further optimizing Q-CHESS:} 
Two promising directions exist for further reducing resource overheads in Q-CHESS. First, it may be possible to better leverage classical computer architecture techniques, such as pipelining and advanced scheduling heuristics, in order to improve parallelism and hardware utilization \cite{Patterson2017}. Second, we identify the ``Clifford+T decomposition" as a key bottleneck in heterogeneous execution. Significant improvements are possible: synthesis is naturally parallelizable (e.g., in cost search and candidate evaluation), numerical routines can be offloaded to GPUs or HPC systems, and emerging approaches—including hybrid GPU-assisted search and direct decomposition of arbitrary unitaries \cite{Hao2025,Tan2025}—offer pathways toward real-time compilation of fault-tolerant workloads.

Taken together, Q-NEXUS and Q-CHESS establish a pathway toward scalable quantum computing in which architectural design, rather than pure device or code optimization, serves as the primary lever for achieving practical fault tolerance. We believe this represents an exciting transition opening many new opportunities for technical development across the community and acceleration of progress in the development of large-scale quantum computers

\section*{Acknowledgments}
We are grateful to all other colleagues at Q-CTRL whose technical work has supported the results presented in this paper. We also thank Rose Ahlefeldt, John Bartholomew, and Matthew Sellars for insightful discussions on quantum memories.

\bibliographystyle{apsrev4-2}
\bibliography{bibliography}

@article{Andres2024,
	title = {Distributing circuits over heterogeneous, modular quantum computing network architectures},
	volume = {9},
	issn = {2058-9565},
	url = {https://doi.org/10.1088/2058-9565/ad6734},
	doi = {10.1088/2058-9565/ad6734},
	language = {en},
	number = {4},
	urldate = {2026-03-30},
	journal = {Quantum Science and Technology},
	publisher = {IOP Publishing},
	author = {Andres-Martinez, Pablo and Forrer, Tim and Mills, Daniel and Wu, Jun-Yi and Henaut, Luciana and Yamamoto, Kentaro and Murao, Mio and Duncan, Ross},
	month = aug,
	year = {2024},
	pages = {045021},
}

@article{Khodjasteh_LowLatencyMemory,
	author = {Khodjasteh, Kaveh and Sastrawan, Jarrah and Hayes, David and Green, Todd J. and Biercuk, Michael J. and Viola, Lorenza},
	da = {2013/06/19},
	doi = {10.1038/ncomms3045},
	id = {Khodjasteh2013},
	isbn = {2041-1723},
	journal = {Nature Communications},
	number = {1},
	pages = {2045},
	title = {Designing a practical high-fidelity long-time quantum memory},
	ty = {JOUR},
	url = {https://doi.org/10.1038/ncomms3045},
	volume = {4},
	year = {2013}}

@article{Magnard2020,
  title = {Microwave Quantum Link between Superconducting Circuits Housed in Spatially Separated Cryogenic Systems},
  author = {Magnard, P. and Storz, S. and Kurpiers, P. and Sch\"ar, J. and Marxer, F. and L\"utolf, J. and Walter, T. and Besse, J.-C. and Gabureac, M. and Reuer, K. and Akin, A. and Royer, B. and Blais, A. and Wallraff, A.},
  journal = {Phys. Rev. Lett.},
  volume = {125},
  issue = {26},
  pages = {260502},
  numpages = {7},
  year = {2020},
  month = {Dec},
  publisher = {American Physical Society},
  doi = {10.1103/PhysRevLett.125.260502},
  url = {https://link.aps.org/doi/10.1103/PhysRevLett.125.260502}
}

@article{Milul2023,
	title = {Superconducting {Cavity} {Qubit} with {Tens} of {Milliseconds} {Single}-{Photon} {Coherence} {Time}},
	volume = {4},
	issn = {2691-3399},
	url = {https://link.aps.org/doi/10.1103/PRXQuantum.4.030336},
	doi = {10.1103/PRXQuantum.4.030336},
	language = {en},
	number = {3},
	urldate = {2026-03-27},
	journal = {PRX Quantum},
	author = {Milul, Ofir and Guttel, Barkay and Goldblatt, Uri and Hazanov, Sergey and Joshi, Lalit M. and Chausovsky, Daniel and Kahn, Nitzan and Çiftyürek, Engin and Lafont, Fabien and Rosenblum, Serge},
	month = sep,
	year = {2023},
	pages = {030336},
}

@article{Giovannetti2008,
	title = {Quantum random access memory},
	volume = {100},
	issn = {0031-9007, 1079-7114},
	url = {http://arxiv.org/abs/0708.1879},
	doi = {10.1103/PhysRevLett.100.160501},
	number = {16},
	urldate = {2026-03-20},
	journal = {Physical Review Letters},
	author = {Giovannetti, Vittorio and Lloyd, Seth and Maccone, Lorenzo},
	month = apr,
	year = {2008},
	note = {arXiv:0708.1879 [quant-ph]},
	pages = {160501},
}

@misc{Jiao2025,
	title = {Transversal {Toffoli}-gate in {Hybrid}-code {System}},
	url = {http://arxiv.org/abs/2511.09265},
	doi = {10.48550/arXiv.2511.09265},
	urldate = {2026-03-20},
	publisher = {arXiv},
	author = {Jiao, Dawei and Bayanifar, Mahdi and Ashikhmin, Alexei and Tirkkonen, Olav},
	month = nov,
	year = {2025},
	note = {arXiv:2511.09265 [quant-ph]},
}

@article{Haah2018,
	title = {Codes and {Protocols} for {Distilling} \${T}\$, controlled-\${S}\$, and {Toffoli} {Gates}},
	volume = {2},
	issn = {2521-327X},
	url = {http://arxiv.org/abs/1709.02832},
	doi = {10.22331/q-2018-06-07-71},
	language = {en},
	urldate = {2026-03-20},
	journal = {Quantum},
	author = {Haah, Jeongwan and Hastings, Matthew B.},
	month = jun,
	year = {2018},
	note = {arXiv:1709.02832 [quant-ph]},
	pages = {71},
}

@article{Pham2013,
	title = {A {2D} nearest-neighbor quantum architecture for factoring in polylogarithmic depth},
	volume = {13},
	issn = {15337146, 15337146},
	url = {http://www.rintonpress.com/journals/doi/QIC13.11-12-3.html},
	doi = {10.26421/QIC13.11-12-3},
	language = {en},
	number = {11\&12},
	urldate = {2026-03-20},
	journal = {Quantum Information and Computation},
	author = {Pham, Paul and Svore, Krysta M.},
	month = nov,
	year = {2013},
	pages = {937--962},
}

@article{Afzelius2009,
  title = {Multimode quantum memory based on atomic frequency combs},
  author = {Afzelius, Mikael and Simon, Christoph and de Riedmatten, Hugues and Gisin, Nicolas},
  journal = {Phys. Rev. A},
  volume = {79},
  issue = {5},
  pages = {052329},
  numpages = {9},
  year = {2009},
  month = {May},
  publisher = {American Physical Society},
  doi = {10.1103/PhysRevA.79.052329},
  url = {https://link.aps.org/doi/10.1103/PhysRevA.79.052329}
}

@article{Ahlefeldt2020,
  title = {Quantum processing with ensembles of rare-earth ions in a stoichiometric crystal},
  author = {Ahlefeldt, R. L. and Pearce, M. J. and Hush, M. R. and Sellars, M. J.},
  journal = {Phys. Rev. A},
  volume = {101},
  issue = {1},
  pages = {012309},
  numpages = {11},
  year = {2020},
  month = {Jan},
  publisher = {American Physical Society},
  doi = {10.1103/PhysRevA.101.012309},
  url = {https://link.aps.org/doi/10.1103/PhysRevA.101.012309}
}

@article{Bombin2006,
	title = {Topological {Quantum} {Distillation}},
	volume = {97},
	url = {https://link.aps.org/doi/10.1103/PhysRevLett.97.180501},
	doi = {10.1103/PhysRevLett.97.180501},
	number = {18},
	urldate = {2026-03-19},
	journal = {Physical Review Letters},
	publisher = {American Physical Society},
	author = {Bombin, H. and Martin-Delgado, M. A.},
	month = oct,
	year = {2006},
	pages = {180501},
}

@misc{Zhu2025,
	title = {Computing with error-corrected quantum computers {\textbar} {IBM} {Quantum} {Computing} {Blog}},
	url = {https://www.ibm.com/quantum/blog/qldpc-codes},
	urldate = {2024-02-22},
    author = {Zhu, Guanyu and Cross, Andrew and Brown, Ben and  Mandelbaum, Ryan},
	year = {2024},
}

@article{Jacinto_2026,
   title={Network requirements for distributed quantum computation},
   volume={8},
   ISSN={2643-1564},
   url={http://dx.doi.org/10.1103/v9ln-c4v2},
   DOI={10.1103/v9ln-c4v2},
   number={1},
   journal={Physical Review Research},
   publisher={American Physical Society (APS)},
   author={Jacinto, Hugo and Gouzien, Elie and Sangouard, Nicolas},
   year={2026},
   month=feb }

@article{Singh2025modular,
  title = {Modular architectures and entanglement schemes for error-corrected distributed quantum computation},
  author = {Siddhant Singh and Fenglei Gu and Sébastian de Bone and Eduardo Villaseñor and David Elkouss and Johannes Borregaard},
  year = {2025},
  publisher = {Springer Science and Business Media LLC},
  journal = {npj Quantum Information},
  volume = {12},
  doi = {10.1038/s41534-025-01146-2},
  url = {https://doi.org/10.1038/s41534-025-01146-2}
}

@article{RyanAnderson2024,
    author = {C. Ryan-Anderson  and others},
    title = {High-fidelity teleportation of a logical qubit using transversal gates and lattice surgery},
    journal = {Science},
    volume = {385},
    number = {6715},
    pages = {1327-1331},
    year = {2024},
    doi = {10.1126/science.adp6016},
    URL = {https://www.science.org/doi/abs/10.1126/science.adp6016},
    eprint = {https://www.science.org/doi/pdf/10.1126/science.adp6016},
}

@article{Ye2025,
	title = {Quantum error correction for long chains of trapped ions},
	volume = {9},
	issn = {2521-327X},
	url = {http://arxiv.org/abs/2503.22071},
	doi = {10.22331/q-2025-11-27-1920},
	language = {en},
	urldate = {2026-03-19},
	journal = {Quantum},
	author = {Ye, Min and Delfosse, Nicolas},
	month = nov,
	year = {2025},
	note = {arXiv:2503.22071 [quant-ph]},
	pages = {1920},
}

@article{RigettiMultiChip2024,
   title={Modular superconducting-qubit architecture with a multichip tunable coupler},
   volume={21},
   ISSN={2331-7019},
   url={http://dx.doi.org/10.1103/PhysRevApplied.21.054063},
   DOI={10.1103/physrevapplied.21.054063},
   number={5},
   journal={Physical Review Applied},
   publisher={American Physical Society (APS)},
   author={Field, Mark and Chen, Angela Q. and Scharmann, Ben and Sete, Eyob A. and Oruc, Feyza and Vu, Kim and Kosenko, Valentin and Mutus, Joshua Y. and Poletto, Stefano and Bestwick, Andrew},
   year={2024},
   month=may }

@article{Gottesman1999,
  author = {D. Gottesman and I. L. Chuang},
  title = {Demonstrating the viability of universal quantum computation using teleportation and single-qubit operations},
  journal = {Nature},
  volume = {402},
  pages = {390--393},
  year = {1999}
}

@article{Kimble2008,
	title = {The quantum internet},
	volume = {453},
	issn = {1476-4687},
	doi = {10.1038/nature07127},
	language = {eng},
	number = {7198},
	journal = {Nature},
	author = {Kimble, H. J.},
	month = jun,
	year = {2008},
	pages = {1023--1030},
}

@misc{Yamamoto2025,
	title = {Quantum {Error}-{Corrected} {Computation} of {Molecular} {Energies}},
	url = {http://arxiv.org/abs/2505.09133},
	doi = {10.48550/arXiv.2505.09133},
	urldate = {2026-03-04},
	publisher = {arXiv},
	author = {{Yamamoto \emph{et al.} (Quantinuum)}, Kentaro},
	month = may,
	year = {2025},
	note = {arXiv:2505.09133 [quant-ph]
version: 1},
}

@article{Wang2024,
	title = {Fault-tolerant one-bit addition with the smallest interesting color code},
	volume = {10},
	url = {https://www.science.org/doi/10.1126/sciadv.ado9024},
	doi = {10.1126/sciadv.ado9024},
	number = {29},
	urldate = {2026-03-04},
	journal = {Science Advances},
	publisher = {American Association for the Advancement of Science},
	author = {{Wang \emph{et al.} (Quantinuum)}, Yang},
	month = jul,
	year = {2024},
	pages = {eado9024},
}

@misc{Fang2026,
	title = {Bridging {Superconducting} and {Neutral}-{Atom} {Platforms} for {Efficient} {Fault}-{Tolerant} {Quantum} {Architectures}},
	url = {http://arxiv.org/abs/2601.10144},
	doi = {10.48550/arXiv.2601.10144},
	urldate = {2026-02-27},
	publisher = {arXiv},
	author = {Fang, Xiang and Ruan, Jixuan and Prabhu, Sharanya and Li, Ang and Humble, Travis and Tullsen, Dean and Ding, Yufei},
	month = jan,
	year = {2026},
	note = {arXiv:2601.10144 [quant-ph]},
}

@article{Gouzien2021,
	title = {Factoring 2048-bit {RSA} {Integers} in 177 {Days} with 13 436 {Qubits} and a {Multimode} {Memory}},
	volume = {127},
	url = {https://link.aps.org/doi/10.1103/PhysRevLett.127.140503},
	doi = {10.1103/PhysRevLett.127.140503},
	number = {14},
	urldate = {2026-02-27},
	journal = {Physical Review Letters},
	publisher = {American Physical Society},
	author = {Gouzien, Elie and Sangouard, Nicolas},
	month = sep,
	year = {2021},
	pages = {140503},
}

@article{Mariantoni2011,
	title = {Implementing the {Quantum} von {Neumann} {Architecture} with {Superconducting} {Circuits}},
	volume = {334},
	url = {https://www.science.org/doi/10.1126/science.1208517},
	doi = {10.1126/science.1208517},
	number = {6052},
	urldate = {2026-02-25},
	journal = {Science},
	publisher = {American Association for the Advancement of Science},
	author = {Mariantoni, Matteo and Wang, H. and Yamamoto, T. and Neeley, M. and Bialczak, Radoslaw C. and Chen, Y. and Lenander, M. and Lucero, Erik and O’Connell, A. D. and Sank, D. and Weides, M. and Wenner, J. and Yin, Y. and Zhao, J. and Korotkov, A. N. and Cleland, A. N. and Martinis, John M.},
	month = oct,
	year = {2011},
	pages = {61--65},
}

@misc{Brandl2017,
	title = {A {Quantum} von {Neumann} {Architecture} for {Large}-{Scale} {Quantum} {Computing}},
	url = {http://arxiv.org/abs/1702.02583},
	doi = {10.48550/arXiv.1702.02583},
	urldate = {2026-02-25},
	publisher = {arXiv},
	author = {Brandl, Matthias F.},
	month = nov,
	year = {2017},
	note = {arXiv:1702.02583 [quant-ph]},
}

@misc{Ball2024,
	title = {The {Best} {Qubits} for {Quantum} {Computing} {Might} {Just} {Be} {Atoms}},
	url = {https://www.quantamagazine.org/the-best-qubits-for-quantum-computing-might-just-be-atoms-20240325/},
	urldate = {2026-02-25},
	journal = {Quanta Magazine},
	author = {Ball, Philip},
	month = mar,
	year = {2024},
}

@misc{Frackiewicz2025,
	title = {Quantum {Showdown}: {Superconducting} vs {Trapped} {Ion} vs {Photonic} – {Who} {Will} {Rule} {Quantum} {Computing}?},
	shorttitle = {Quantum {Showdown}},
	url = {https://ts2.tech/en/quantum-showdown-superconducting-vs-trapped-ion-vs-photonic-who-will-rule-quantum-computing/},
	urldate = {2026-02-25},
	journal = {TechStock},
	author = {Frackiewicz, Marcin},
	month = aug,
	year = {2025},
}

@book{Patterson2017,
	address = {Amsterdam ; Boston},
	edition = {ARM® edition},
	title = {Computer organization and design: the hardware/software interface},
	isbn = {978-0-12-801733-3},
	shorttitle = {Computer organization and design},
	publisher = {Elsevier/Morgan Kaufmann},
	author = {Patterson, David A. and Hennessy, John L. and Alexander, Perry},
	year = {2017},
}

@book{Hennessy2012,
	address = {Waltham, MA},
	edition = {5th ed},
	title = {Computer architecture: a quantitative approach},
	isbn = {978-0-12-383872-8},
	shorttitle = {Computer architecture},
	publisher = {Morgan Kaufmann/Elsevier},
	author = {Hennessy, John L. and Patterson, David A. and Asanović, Krste},
	year = {2012},
}

@techreport{vonNeumann1945,
  author       = { {von Neumann, (J. P. Eckert and J. Mauchly are seen by many to have made equal intellectual contributions but their names were controversially omitted \cite{Patterson2017}.) }, John },
  title        = {First Draft of a Report on the {EDVAC}},
  institution  = {Moore School of Electrical Engineering, University of Pennsylvania},
  year         = {1945},
  month        = {June},
  note         = {Contract No. W-670-ORD-4926, U.S. Army Ordnance Department}
}

@misc{Yoder2025,
      title={{Tour de gross: A modular quantum computer based on bivariate bicycle codes}}, 
      author={Theodore J. Yoder and Eddie Schoute and Patrick Rall and Emily Pritchett and Jay M. Gambetta and Andrew W. Cross and Malcolm Carroll and Michael E. Beverland},
      year={2025},
      eprint={2506.03094},
      archivePrefix={arXiv},
      primaryClass={quant-ph},
      url={https://arxiv.org/abs/2506.03094}, 
}

@article{Bravyi2024,
  title = {High-threshold and low-overhead fault-tolerant quantum memory},
  volume = {627},
  ISSN = {1476-4687},
  url = {http://dx.doi.org/10.1038/s41586-024-07107-7},
  DOI = {10.1038/s41586-024-07107-7},
  number = {8005},
  journal = {Nature},
  publisher = {Springer Science and Business Media LLC},
  author = {Bravyi,  Sergey and Cross,  Andrew W. and Gambetta,  Jay M. and Maslov,  Dmitri and Rall,  Patrick and Yoder,  Theodore J.},
  year = {2024},
  month = mar,
  pages = {778–782}
}

@inproceedings{Sutcliffe2025,
	title = {Distributed {Quantum} {Error} {Correction} {Based} on {Hyperbolic} {Floquet} {Codes}},
	volume = {01},
	url = {https://ieeexplore.ieee.org/document/11250341},
	doi = {10.1109/QCE65121.2025.00076},
	urldate = {2026-02-20},
	booktitle = {2025 {IEEE} {International} {Conference} on {Quantum} {Computing} and {Engineering} ({QCE})},
	author = {Sutcliffe, Evan and Jonnadula, Bhargavi and Le Gall, Claire and Moylett, Alexandra E. and Westoby, Coral M.},
	month = aug,
	year = {2025},
	pages = {649--657},
}

@article{Kaiser2022,
	title = {Cavity-driven {Rabi} oscillations between {Rydberg} states of atoms trapped on a superconducting atom chip},
	volume = {4},
	url = {https://link.aps.org/doi/10.1103/PhysRevResearch.4.013207},
	doi = {10.1103/PhysRevResearch.4.013207},
	number = {1},
	urldate = {2026-02-25},
	journal = {Physical Review Research},
	publisher = {American Physical Society},
	author = {Kaiser, Manuel and Glaser, Conny and Ley, Li Yuan and Grimmel, Jens and Hattermann, Helge and Bothner, Daniel and Koelle, Dieter and Kleiner, Reinhold and Petrosyan, David and Günther, Andreas and Fortágh, József},
	month = mar,
	year = {2022},
	pages = {013207},
}

@article{Kumar2023,
	title = {Quantum-enabled millimetre wave to optical transduction using neutral atoms},
	volume = {615},
	copyright = {2023 The Author(s), under exclusive licence to Springer Nature Limited},
	issn = {1476-4687},
	url = {https://www.nature.com/articles/s41586-023-05740-2},
	doi = {10.1038/s41586-023-05740-2},
	language = {en},
	number = {7953},
	urldate = {2026-02-25},
	journal = {Nature},
	publisher = {Nature Publishing Group},
	author = {Kumar, Aishwarya and Suleymanzade, Aziza and Stone, Mark and Taneja, Lavanya and Anferov, Alexander and Schuster, David I. and Simon, Jonathan},
	month = mar,
	year = {2023},
	pages = {614--619},
}

@misc{Liu2023,
	title={{Quantum Memory: A Missing Piece in Quantum Computing Units}}, 
	url = {https://arxiv.org/abs/2309.14432v2},
	urldate = {2026-02-20},
	journal = {arXiv.org},
	author = {Liu, Chenxu and Wang, Meng and Stein, Samuel A. and Ding, Yufei and Li, Ang},
	month = sep,
	year={2023},
    eprint={2309.14432},
    archivePrefix={arXiv},
    primaryClass={quant-ph},
}

@article{Xu2024,
  title = {Constant-overhead fault-tolerant quantum computation with reconfigurable atom arrays},
  volume = {20},
  ISSN = {1745-2481},
  url = {http://dx.doi.org/10.1038/s41567-024-02479-z},
  DOI = {10.1038/s41567-024-02479-z},
  number = {7},
  journal = {Nature Physics},
  publisher = {Springer Science and Business Media LLC},
  author = {Xu,  Qian and Bonilla Ataides,  J. Pablo and Pattison,  Christopher A. and Raveendran,  Nithin and Bluvstein,  Dolev and Wurtz,  Jonathan and Vasić,  Bane and Lukin,  Mikhail D. and Jiang,  Liang and Zhou,  Hengyun},
  year = {2024},
  month = apr,
  pages = {1084–1090}
}

@article{Zhong2015,
  title = {Optically addressable nuclear spins in a solid with a six-hour coherence time},
  volume = {517},
  ISSN = {1476-4687},
  url = {http://dx.doi.org/10.1038/nature14025},
  DOI = {10.1038/nature14025},
  number = {7533},
  journal = {Nature},
  publisher = {Springer Science and Business Media LLC},
  author = {Zhong,  Manjin and Hedges,  Morgan P. and Ahlefeldt,  Rose L. and Bartholomew,  John G. and Beavan,  Sarah E. and Wittig,  Sven M. and Longdell,  Jevon J. and Sellars,  Matthew J.},
  year = {2015},
  month = jan,
  pages = {177–180}
}

@article{Konz2003,
  title = {Temperature and concentration dependence of optical dephasing, spectral-hole lifetime, and anisotropic absorption in ${\mathrm{Eu}}^{3+}{:\mathrm{Y}}_{2}{\mathrm{SiO}}_{5}$},
  author = {K\"onz, Flurin and Sun, Y. and Thiel, C. W. and Cone, R. L. and Equall, R. W. and Hutcheson, R. L. and Macfarlane, R. M.},
  journal = {Phys. Rev. B},
  volume = {68},
  issue = {8},
  pages = {085109},
  numpages = {9},
  year = {2003},
  month = {Aug},
  publisher = {American Physical Society},
  doi = {10.1103/PhysRevB.68.085109},
  url = {https://link.aps.org/doi/10.1103/PhysRevB.68.085109}
}

@article{Bluvstein2024,
  title={Logical quantum processor based on reconfigurable atom arrays},
  author={Bluvstein, Dolev and Evered, Simon J and Geim, Alexandra A and Li, Sophie H and Zhou, Hengyun and Manovitz, Tom and Ebadi, Sepehr and Cain, Madelyn and Kalinowski, Marcin and Hangleiter, Dominik and others},
  journal={Nature},
  volume={626},
  number={7997},
  pages={58--65},
  year={2024},
  publisher={Nature Publishing Group UK London},
  url={https://www.nature.com/articles/s41586-023-06927-3}
}

@misc{Ransford2025,
	title = {Helios: {A} 98-qubit trapped-ion quantum computer},
	shorttitle = {Helios},
	url = {https://arxiv.org/abs/2511.05465v1},
	urldate = {2026-02-20},
	journal = {arXiv.org},
	author = {{Ransford \emph{et al.} (Quantinuum)}, Anthony },
	month = nov,
	year={2025},
    eprint={2511.05465},
    archivePrefix={arXiv},
    primaryClass={quant-ph},
}

@misc{cain2026,
      title={Shor's algorithm is possible with as few as 10,000 reconfigurable atomic qubits}, 
      author={Madelyn Cain and Qian Xu and Robbie King and Lewis R. B. Picard and Harry Levine and Manuel Endres and John Preskill and Hsin-Yuan Huang and Dolev Bluvstein},
      year={2026},
      eprint={2603.28627},
      archivePrefix={arXiv},
      primaryClass={quant-ph},
      url={https://arxiv.org/abs/2603.28627}, 
}

@misc{stein2024architecturesheterogeneousquantumerror,
      title={Architectures for Heterogeneous Quantum Error Correction Codes}, 
      author={Samuel Stein and Shifan Xu and Andrew W. Cross and Theodore J. Yoder and Ali Javadi-Abhari and Chenxu Liu and Kun Liu and Zeyuan Zhou and Charles Guinn and Yufei Ding and Yongshan Ding and Ang Li},
      year={2024},
      eprint={2411.03202},
      archivePrefix={arXiv},
      primaryClass={quant-ph},
      url={https://arxiv.org/abs/2411.03202}, 
}

@misc{Stein2023,
	title={Microarchitectures for Heterogeneous Superconducting Quantum Computers}, 
	url = {https://arxiv.org/abs/2305.03243v1},
	urldate = {2026-02-20},
	journal = {arXiv.org},
	author = {Stein, Samuel and Sussman, Sara and Tomesh, Teague and Guinn, Charles and Tureci, Esin and Lin, Sophia Fuhui and Tang, Wei and Ang, James and Chakram, Srivatsan and Li, Ang and Martonosi, Margaret and Chong, Fred T. and Houck, Andrew A. and Chuang, Isaac L. and DeMarco, Michael Austin},
	month = may,
	year = {2023},
    eprint={2305.03243},
    archivePrefix={arXiv},
    primaryClass={quant-ph},
}

@misc{Chen2025,
	title={{Adaptive Resource Orchestration for Distributed Quantum Computing Systems}}, 
	url = {http://arxiv.org/abs/2512.24902},
	doi = {10.48550/arXiv.2512.24902},
	urldate = {2026-02-20},
	publisher = {arXiv},
	author = {Chen, Kuan-Cheng and Burt, Felix and Panigrahy, Nitish K. and Leung, Kin K.},
	month = dec,
    year = {2025},
	eprint={2512.24902},
    archivePrefix={arXiv},
    primaryClass={quant-ph},
}

@misc{NuSite,
	title = {Nu {Quantum}},
	url = {https://www.nu-quantum.com/},
	author = {Nu Quantum},
	urldate = {2026-02-19},
    howpublished = {\url{https://www.nu-quantum.com/}},
    note = {Accessed: 2026-02-19}
}

@book{KletteRosenfeld2004,
  title     = "Digital Geometry",
  author    = "Reinhard Klette and Azriel Rosenfeld",
  year      = 2004,
  publisher = "Elsevier",
}

@misc{AliroSite,
	title = {Aliro - {Quantum} {Powered} {Security}},
	url = {https://www.aliroquantum.com},
    howpublished = {\url{https://www.aliroquantum.com}},
    author = {Aliro},
	urldate = {2026-02-19},
    note = {Accessed: 2026-02-19}
}

@misc{QunnectSite,
	title = {Enabling the {Quantum} {Internet}},
	url = {https://qunnect.inc/},
    howpublished = {\url{https://qunnect.inc/}},
	author = {Qunnect},
    urldate = {2026-02-19},
    note = {Accessed: 2026-02-19}
}

@misc{Kawabata2026,
	title = {Integration and {Resource} {Estimation} of {Cryoelectronics} for {Superconducting} {Fault}-{Tolerant} {Quantum} {Computers}},
	url = {http://arxiv.org/abs/2601.03922},
	doi = {10.48550/arXiv.2601.03922},
	urldate = {2026-02-19},
	publisher = {arXiv},
	author = {Kawabata, Shiro},
	month = jan,
	year={2026},
    eprint={2601.03922},
    archivePrefix={arXiv},
    primaryClass={quant-ph},
}

@article{Monroe2014,
	title = {Large-scale modular quantum-computer architecture with atomic memory and photonic interconnects},
	volume = {89},
	url = {https://link.aps.org/doi/10.1103/PhysRevA.89.022317},
	doi = {10.1103/PhysRevA.89.022317},
	number = {2},
	urldate = {2026-02-19},
	journal = {Physical Review A},
	publisher = {American Physical Society},
	author = {Monroe, C. and Raussendorf, R. and Ruthven, A. and Brown, K. R. and Maunz, P. and Duan, L.-M. and Kim, J.},
	month = feb,
	year = {2014},
	pages = {022317},
}

@misc{Abughanem2024,
      title={Photonic Quantum Computers}, 
      author={M. AbuGhanem},
      year={2024},
      eprint={2409.08229},
      archivePrefix={arXiv},
      primaryClass={quant-ph},
      url={https://arxiv.org/abs/2409.08229}, 
}

@article{Gu2017,
	title = {Microwave photonics with superconducting quantum circuits},
	volume = {718-719},
	issn = {03701573},
	url = {http://arxiv.org/abs/1707.02046},
	doi = {10.1016/j.physrep.2017.10.002},
	urldate = {2026-02-19},
	journal = {Physics Reports},
	author = {Gu, Xiu and Kockum, Anton Frisk and Miranowicz, Adam and Liu, Yu-xi and Nori, Franco},
	month = nov,
	year = {2017},
	pages = {1--102},
}

@misc{Robertson2025,
	title = {{A Resource Allocating Compiler for Lattice Surgery}},
	url = {http://arxiv.org/abs/2506.04620},
	doi = {10.48550/arXiv.2506.04620},
	urldate = {2026-02-19},
	publisher = {arXiv},
	author = {Robertson, Alan and Gao, Haowen and Sanders, Yuval R.},
	month = jun,
    year={2025},
    eprint={2506.04620},
    archivePrefix={arXiv},
    primaryClass={quant-ph},
}

@misc{Kan2025,
	title = {{SPARO}: {Surface}-code {Pauli}-based {Architectural} {Resource} {Optimization} for {Fault}-tolerant {Quantum} {Computing}},
	shorttitle = {{SPARO}},
	url = {http://arxiv.org/abs/2504.21854},
	doi = {10.48550/arXiv.2504.21854},
	urldate = {2026-02-19},
	publisher = {arXiv},
	author = {Kan, Shuwen and Du, Zefan and Liu, Chenxu and Wang, Meng and Ding, Yufei and Li, Ang and Mao, Ying and Stein, Samuel},
	month = may,
	year={2025},
    eprint={2504.21854},
    archivePrefix={arXiv},
    primaryClass={quant-ph},
}

@misc{Marqversen2025,
	title = {Fault-tolerant interfaces for modular quantum computing on diverse qubit platforms},
	url = {http://arxiv.org/abs/2510.05221},
	doi = {10.48550/arXiv.2510.05221},
	urldate = {2026-02-19},
	publisher = {arXiv},
	author = {Marqversen, Frederik K. and Baranes, Gefen and Sirotin, Maxim and Borregaard, Johannes},
	month = oct,
	year = {2025},
	eprint={2510.05221},
    archivePrefix={arXiv},
    primaryClass={quant-ph},
}

@article{Strikis2023,
	title = {Quantum {Low}-{Density} {Parity}-{Check} {Codes} for {Modular} {Architectures}},
	volume = {4},
	url = {https://link.aps.org/doi/10.1103/PRXQuantum.4.020321},
	doi = {10.1103/PRXQuantum.4.020321},
	number = {2},
	urldate = {2026-02-19},
	journal = {PRX Quantum},
	publisher = {American Physical Society},
	author = {Strikis, Armands and Berent, Lucas},
	month = may,
	year = {2023},
	pages = {020321},
}

@misc{Guinn2023,
	title = {Co-{Designed} {Superconducting} {Architecture} for {Lattice} {Surgery} of {Surface} {Codes} with {Quantum} {Interface} {Routing} {Card}},
	url = {http://arxiv.org/abs/2312.01246},
	doi = {10.48550/arXiv.2312.01246},
	urldate = {2026-02-19},
	publisher = {arXiv},
	author = {Guinn, Charles and Stein, Samuel and Tureci, Esin and Avis, Guus and Liu, Chenxu and Krastanov, Stefan and Houck, Andrew A. and Li, Ang},
	month = dec,
	year={2023},
    eprint={2312.01246},
    archivePrefix={arXiv},
    primaryClass={quant-ph},
}

@article{Zhang2025,
	title = {Demonstrating a universal logical gate set in error-detecting surface codes on a superconducting quantum processor},
	volume = {11},
	issn = {2056-6387},
	url = {https://www.nature.com/articles/s41534-025-01118-6},
	doi = {10.1038/s41534-025-01118-6},
	language = {en},
	number = {1},
	urldate = {2026-02-19},
	journal = {npj Quantum Information},
	author = {Zhang, Jiaxuan and Chen, Zhao-Yun and Wang, Yun-Jie and Lu, Bin-Han and Zhang, Hai-Feng and Li, Jia-Ning and Duan, Peng and Wu, Yu-Chun and Guo, Guo-Ping},
	month = nov,
	year = {2025},
	pages = {177},
}

@Article{Wang2021,
author={Wang, Pengfei
and Luan, Chun-Yang
and Qiao, Mu
and Um, Mark
and Zhang, Junhua
and Wang, Ye
and Yuan, Xiao
and Gu, Mile
and Zhang, Jingning
and Kim, Kihwan},
title={Single ion qubit with estimated coherence time exceeding one hour},
journal={Nature Communications},
year={2021},
month={Jan},
day={11},
volume={12},
number={1},
pages={233},
issn={2041-1723},
doi={10.1038/s41467-020-20330-w},
url={https://doi.org/10.1038/s41467-020-20330-w}
}

@article{Pogorelov2025,
  title = {Experimental fault-tolerant code switching},
  volume = {21},
  ISSN = {1745-2481},
  url = {http://dx.doi.org/10.1038/s41567-024-02727-2},
  DOI = {10.1038/s41567-024-02727-2},
  number = {2},
  journal = {Nature Physics},
  publisher = {Springer Science and Business Media LLC},
  author = {Pogorelov,  Ivan and Butt,  Friederike and Postler,  Lukas and Marciniak,  Christian D. and Schindler,  Philipp and M\"{u}ller,  Markus and Monz,  Thomas},
  year = {2025},
  month = jan,
  pages = {298–303}
}

@article{Harper2019,
	title = {Fault-{Tolerant} {Logical} {Gates} in the {IBM} {Quantum} {Experience}},
	volume = {122},
	issn = {0031-9007, 1079-7114},
	url = {http://arxiv.org/abs/1806.02359},
	doi = {10.1103/PhysRevLett.122.080504},
	number = {8},
	urldate = {2026-02-19},
	journal = {Physical Review Letters},
	author = {Harper, Robin and Flammia, Steven T.},
	month = feb,
	year = {2019},
	pages = {080504},
}

@article{Rodriguez2025,
	title = {Experimental demonstration of logical magic state distillation},
	volume = {645},
	issn = {0028-0836, 1476-4687},
	url = {http://arxiv.org/abs/2412.15165},
	doi = {10.1038/s41586-025-09367-3},
	number = {8081},
	urldate = {2026-02-19},
	journal = {Nature},
	author = { {Rodriguez \emph{et al.} (QuEra)}, Pedro Sales },
	month = sep,
	year = {2025},
	pages = {620--625},
}

@misc{Xanadu2026,
	title = {Xanadu {\textbar} {Xanadu} introduces {Aurora}: world's first scalable, networked and modular quantum computer},
	shorttitle = {Xanadu {\textbar} {Xanadu} introduces {Aurora}},
    year = {2025},
	url = {https://www.xanadu.ai/press/xanadu-introduces-aurora-worlds-first-scalable-networked-and-modular-quantum-computer},
	urldate = {2026-02-18},
	author = {Xanadu},
    howpublished = {\url{https://www.xanadu.ai/press/xanadu-introduces-aurora-worlds-first-scalable-networked-and-modular-quantum-computer}},
    note = {Accessed: 2026-02-18}
}

@misc{Pasqal2025,
	title = {Pasqal {Releases} 2025 {Roadmap} {Showcasing} {Upgradable} {Platform} from {Today}’s {Quantum} {Solutions} to {Tomorrow}’s {Fault}-{Tolerant} {Systems}},
	url = {https://www.pasqal.com/newsroom/pasqal-releases-2025-roadmap/},
	urldate = {2026-02-18},
	journal = {Pasqal},
	month = jun,
	year = {2025},
	author = {Pasqal},
    howpublished = {\url{https://www.pasqal.com/newsroom/pasqal-releases-2025-roadmap/}},
    note = {Accessed: 2026-02-18}
}

@misc{IonQ2026,
	title = {{IonQ} {\textbar} {Roadmap}},
	url = {https://www.ionq.com/roadmap},
	urldate = {2026-02-18},
	author = {IonQ},
    year = {2025},
    howpublished = {\url{https://www.ionq.com/roadmap}},
    note = {Accessed: 2026-02-18}
}

@article{Babbush2018,
  title = {Encoding Electronic Spectra in Quantum Circuits with Linear T Complexity},
  author = {Babbush, Ryan and Gidney, Craig and Berry, Dominic W. and Wiebe, Nathan and McClean, Jarrod and Paler, Alexandru and Fowler, Austin and Neven, Hartmut},
  journal = {Phys. Rev. X},
  volume = {8},
  issue = {4},
  pages = {041015},
  numpages = {36},
  year = {2018},
  month = {Oct},
  publisher = {American Physical Society},
  doi = {10.1103/PhysRevX.8.041015},
  url = {https://link.aps.org/doi/10.1103/PhysRevX.8.041015}
}

@article{GidneyEkera2021RSA,
	title = {How to factor 2048 bit {RSA} integers in 8 hours using 20 million noisy qubits},
	volume = {5},
	url = {https://quantum-journal.org/papers/q-2021-04-15-433/},
	doi = {10.22331/q-2021-04-15-433},
	language = {en-GB},
	urldate = {2026-02-17},
	journal = {Quantum},
	publisher = {Verein zur Forderung des Open Access Publizierens in den Quantenwissenschaften},
	author = {Gidney, Craig and Ekera, Martin},
	month = apr,
	year = {2021},
	pages = {433},
}

@misc{Gidney2025RSAUpdate,
	title = {How to factor 2048 bit {RSA} integers with less than a million noisy qubits},
	url = {http://arxiv.org/abs/2505.15917},
	doi = {10.48550/arXiv.2505.15917},
	urldate = {2026-02-17},
	publisher = {arXiv},
	author = {Gidney, Craig},
	month = may,
	year={2025},
    eprint={2505.15917},
    archivePrefix={arXiv},
    primaryClass={quant-ph},
}

@misc{Webster2026Iceberg,
	title = {The {Pinnacle} {Architecture}: {Reducing} the cost of breaking {RSA}-2048 to 100 000 physical qubits using quantum {LDPC} codes},
	shorttitle = {The {Pinnacle} {Architecture}},
	url = {http://arxiv.org/abs/2602.11457},
	doi = {10.48550/arXiv.2602.11457},
	urldate = {2026-02-17},
	publisher = {arXiv},
	author = {Webster, Paul and Berent, Lucas and Chandra, Omprakash and Hockings, Evan T. and Baspin, Nouedyn and Thomsen, Felix and Smith, Samuel C. and Cohen, Lawrence Z.},
	month = feb,
	year = {2026},
    eprint={2602.11457},
    archivePrefix={arXiv},
    primaryClass={quant-ph},
}

@INPROCEEDINGS{Kobori2025,
  author={Kobori, Takumi and Suzuki, Yasunari and Ueno, Yosuke and Tanimoto, Teruo and Todo, Synge and Tokunaga, Yuuki},
  booktitle={2025 IEEE International Symposium on High Performance Computer Architecture (HPCA)}, 
  title={LSQCA: Resource-Efficient Load/Store Architecture for Limited-Scale Fault-Tolerant Quantum Computing}, 
  year={2025},
  volume={},
  number={},
  pages={304-320},
  doi={10.1109/HPCA61900.2025.00033}}

@misc{IBMOldRoadmap,
	title = {{IBM Quantum System Two: the era of quantum utility is here}},
    author = {IBM},
    year = {2023},
	url = {https://www.ibm.com/quantum/blog/quantum-roadmap-2033},
	urldate = {2026-02-17},
    howpublished = {\url{https://www.ibm.com/quantum/blog/quantum-roadmap-2033}},
    note = {Accessed: 2026-02-17}
}

@misc{IBMCurrentRoadmap,
	title = {{IBM Quantum Roadmap}},
    author = {IBM},
    year = {2025},
	url = {https://www.ibm.com/downloads/documents/us-en/1443d5cda24021e4},
	urldate = {2026-02-17},
    note = {Accessed: 2026-02-17},
    howpublished = {\url{https://www.ibm.com/downloads/documents/us-en/1443d5cda24021e4}}
}

@article{Krinner2019,
	title = {Engineering cryogenic setups for 100-qubit scale superconducting circuit systems},
	volume = {6},
	issn = {2196-0763},
	url = {https://doi.org/10.1140/epjqt/s40507-019-0072-0},
	doi = {10.1140/epjqt/s40507-019-0072-0},
	language = {en},
	number = {1},
	urldate = {2026-02-17},
	journal = {EPJ Quantum Technology},
	author = {Krinner, S. and Storz, S. and Kurpiers, P. and Magnard, P. and Heinsoo, J. and Keller, R. and Lütolf, J. and Eichler, C. and Wallraff, A.},
	month = may,
	year = {2019},
	pages = {2},
}

@article{Franke2018,
	title = {Rent’s rule and extensibility in quantum computing},
	volume = {67},
	issn = {01419331},
	url = {https://linkinghub.elsevier.com/retrieve/pii/S014193311830293X},
	doi = {10.1016/j.micpro.2019.02.006},
	language = {en},
	urldate = {2026-02-17},
	journal = {Microprocessors and Microsystems},
	author = {Franke, D.P. and Clarke, J.S. and Vandersypen, L.M.K. and Veldhorst, M.},
	month = jun,
	year = {2019},
	pages = {1--7},
}

@article{Malinowski2023,
	title = {How to {Wire} a \$1000\$-{Qubit} {Trapped}-{Ion} {Quantum} {Computer}},
	volume = {4},
	url = {https://link.aps.org/doi/10.1103/PRXQuantum.4.040313},
	doi = {10.1103/PRXQuantum.4.040313},
	number = {4},
	urldate = {2026-02-17},
	journal = {PRX Quantum},
	publisher = {American Physical Society},
	author = {Malinowski, M. and Allcock, D.T.C. and Ballance, C.J.},
	month = oct,
	year = {2023},
	pages = {040313},
}

@article{Reilly2015,
	title = {Cryogenic {Control} {Architecture} for {Large}-{Scale} {Quantum} {Computing}},
	volume = {3},
	url = {https://link.aps.org/doi/10.1103/PhysRevApplied.3.024010},
	doi = {10.1103/PhysRevApplied.3.024010},
	number = {2},
	urldate = {2026-02-17},
	journal = {Physical Review Applied},
	publisher = {American Physical Society},
	author = {Hornibrook, J.M. and Colless, J.I. and Conway Lamb, I.D. and Pauka, S.J. and Lu, H. and Gossard, A.C. and Watson, J.D. and Gardner, G.C. and Fallahi, S. and Manfra, M.J. and Reilly, D.J.},
	month = feb,
	year = {2015},
	pages = {024010},
}

@article{QueraFaultTolerant2025,
	title = {Low-overhead transversal fault tolerance for universal quantum computation},
	volume = {646},
	copyright = {2025 The Author(s), under exclusive licence to Springer Nature Limited},
	issn = {1476-4687},
	url = {https://preview-www.nature.com/articles/s41586-025-09543-5},
	doi = {10.1038/s41586-025-09543-5},
	language = {en},
	number = {8084},
	urldate = {2026-02-17},
	journal = {Nature},
	publisher = {Nature Publishing Group},
	author = {Zhou, Hengyun and Zhao, Chen and Cain, Madelyn and Bluvstein, Dolev and Maskara, Nishad and Duckering, Casey and Hu, Hong-Ye and Wang, Sheng-Tao and Kubica, Aleksander and Lukin, Mikhail D.},
	month = oct,
	year = {2025},
	pages = {303--308},
}

@misc{AtomFaultTolerant2025,
	title = {Fault-tolerant quantum computation with a neutral atom processor},
	url = {http://arxiv.org/abs/2411.11822},
	doi = {10.48550/arXiv.2411.11822},
	author = {{Reichardt \emph{et al.} (Atom Computing)}, Ben W.  },
	month = jun,
	year = {2025},
    eprint={2411.11822},
    archivePrefix={arXiv},
    primaryClass={quant-ph},
}

@article{GoogleFaultTolerant2025,
	title = {Quantum error correction below the surface code threshold},
	volume = {638},
	issn = {0028-0836, 1476-4687},
	url = {https://www.nature.com/articles/s41586-024-08449-y},
	doi = {10.1038/s41586-024-08449-y},
	language = {en},
	number = {8052},
	urldate = {2026-02-17},
	journal = {Nature},
	author = { {Acharya \emph{et al.} (Google Quantum AI and Collaborators)}, Rajeev },
	month = feb,
	year = {2025},
	pages = {920--926},
}

@misc{WhiteHouse2022,
  author       = {{The White House}},
  title        = {National security memorandum on promoting United States leadership in quantum computing while mitigating risks to vulnerable cryptographic systems},
  howpublished = {National Archives},
  month        = {May},
  year         = {2022}
}

@article{Castelvecchi2023,
  title = {IBM releases first-ever 1, 000-qubit quantum chip},
  volume = {624},
  ISSN = {1476-4687},
  url = {http://dx.doi.org/10.1038/d41586-023-03854-1},
  DOI = {10.1038/d41586-023-03854-1},
  number = {7991},
  journal = {Nature},
  publisher = {Springer Science and Business Media LLC},
  author = {Castelvecchi,  Davide},
  year = {2023},
  month = dec,
  pages = {238–238}
}

@article{Heya2025,
  title = {Randomized Benchmarking of a Remote cnot Gate Via a Meter-Scale Microwave Link},
  author = {Heya, Kentaro and Phung, Timothy and Malekakhlagh, Moein and Steiner, Rachel and Turchetti, Marco and Shanks, William and Mamin, John and Lu, Wen-Sen and Kandel, Yadav Prasad and Sundaresan, Neereja and Orcutt, Jason},
  journal = {Phys. Rev. Lett.},
  volume = {135},
  issue = {20},
  pages = {200801},
  numpages = {8},
  year = {2025},
  month = {Nov},
  publisher = {American Physical Society},
  doi = {10.1103/xx24-r7q6},
  url = {https://link.aps.org/doi/10.1103/xx24-r7q6}
}

@article{Caleffi2024,
  title = {Distributed quantum computing: A survey},
  volume = {254},
  ISSN = {1389-1286},
  url = {http://dx.doi.org/10.1016/j.comnet.2024.110672},
  DOI = {10.1016/j.comnet.2024.110672},
  journal = {Computer Networks},
  publisher = {Elsevier BV},
  author = {Caleffi,  Marcello and Amoretti,  Michele and Ferrari,  Davide and Illiano,  Jessica and Manzalini,  Antonio and Cacciapuoti,  Angela Sara},
  year = {2024},
  month = dec,
  pages = {110672}
}

@article{Wang2025,
  title = {Nuclear Spins in a Solid Exceeding 10-Hour Coherence Times for Ultra-Long-Term Quantum Storage},
  author = {Wang, Fudong and Ren, Miaomiao and Sun, Weiye and Guo, Mucheng and Sellars, Matthew J. and Ahlefeldt, Rose L. and Bartholomew, John G. and Yao, Juan and Liu, Shuping and Zhong, Manjin},
  journal = {PRX Quantum},
  volume = {6},
  issue = {1},
  pages = {010302},
  numpages = {12},
  year = {2025},
  month = {Jan},
  publisher = {American Physical Society},
  doi = {10.1103/PRXQuantum.6.010302},
  url = {https://link.aps.org/doi/10.1103/PRXQuantum.6.010302}
}

@misc{Hughes2025,
      title={Trapped-ion two-qubit gates with $>99.99\%$ fidelity without ground-state cooling}, 
      author={A. C. Hughes and R. Srinivas and C. M. Löschnauer and H. M. Knaack and R. Matt and C. J. Ballance and M. Malinowski and T. P. Harty and R. T. Sutherland},
      year={2025},
      eprint={2510.17286},
      archivePrefix={arXiv},
      primaryClass={quant-ph},
      url={https://arxiv.org/abs/2510.17286}
}

@misc{Gidney2024,
      title={{Magic state cultivation: growing T states as cheap as CNOT gates}}, 
      author={Craig Gidney and Noah Shutty and Cody Jones},
      year={2024},
      eprint={2409.17595},
      archivePrefix={arXiv},
      primaryClass={quant-ph},
      url={https://arxiv.org/abs/2409.17595}, 
}

@misc{Kutin2006,
  author        = {Kutin, S. A.},
  title         = {Shor's algorithm on a nearest-neighbor machine},
  year          = {2006},
  eprint        = {quant-ph/0609001},
  archivePrefix = {arXiv},
  url={https://arxiv.org/abs/quant-ph/0609001},
  primaryClass  = {quant-ph}
}

@article{Marton2025,
  title = {Optimal number of stabilizer measurement rounds in an idling surface code patch},
  volume = {9},
  ISSN = {2521-327X},
  url = {http://dx.doi.org/10.22331/q-2025-06-12-1767},
  DOI = {10.22331/q-2025-06-12-1767},
  journal = {Quantum},
  publisher = {Verein zur Forderung des Open Access Publizierens in den Quantenwissenschaften},
  author = {M{\'a}rton,  {\'A}ron and Asb{\'o}th,  J{\'a}nos K.},
  year = {2025},
  month = jun,
  pages = {1767}
}

@article{Ross2016,
  author  = {Ross, N. J. and Selinger, P.},
  title   = {{Optimal ancilla-free Clifford+T approximation of z-rotations}},
  journal = {Quantum Info. Comput.},
  volume  = {16},
  number  = {11-12},
  pages   = {901--953},
  year    = {2016},
  url = {https://dl.acm.org/doi/abs/10.5555/3179330.3179331}
}

@misc{rudolph2023logical,
  author = {Rudolph, Terry},
  title = {What is the logical gate speed of a photonic quantum computer?},
  howpublished = {Quantum Frontiers},
  month = {June 21},
  year = {2023},
  url = {https://quantumfrontiers.com/2023/06/21/what-is-the-logical-gate-speed-of-a-photonic-quantum-computer/},
  note = {Accessed: 2024-05-22}
}

@article{Litinski2019,
  title = {A Game of Surface Codes: Large-Scale Quantum Computing with Lattice Surgery},
  volume = {3},
  ISSN = {2521-327X},
  url = {http://dx.doi.org/10.22331/q-2019-03-05-128},
  DOI = {10.22331/q-2019-03-05-128},
  journal = {Quantum},
  publisher = {Verein zur Forderung des Open Access Publizierens in den Quantenwissenschaften},
  author = {Litinski,  Daniel},
  year = {2019},
  month = mar,
  pages = {128}
}

@article{Gidney2025,
  title = {Yoked surface codes},
  volume = {16},
  url = {http://dx.doi.org/10.1038/s41467-025-59714-1},
  journal = {Nature Communications},
  number = {1},
  publisher = {Springer Science and Business Media LLC},
  author = {Gidney,  Craig and Newman,  Michael and Brooks,  Peter and Jones,  Cody},
  year = {2025},
  month = may 
}

@misc{Hao2025,
  author        = {Hao, T. and Xu, A. and Tannu, S.},
  title         = {{Reducing T gates with unitary synthesis}},
  year          = {2025},
  url = {https://arxiv.org/abs/2503.15843},
  eprint        = {2503.15843},
  archivePrefix = {arXiv},
  primaryClass  = {quant-ph}
}

@misc{Tan2025,
  author        = {Tan, X.},
  title         = {{Unitary synthesis with fewer T gates}},
  year          = {2025},
  eprint        = {2509.25702},
  archivePrefix = {arXiv},
  primaryClass  = {quant-ph},
  url={https://arxiv.org/abs/2509.25702}
}

@misc{Javadiabhari2024,
      title={Quantum computing with Qiskit}, 
      author={Ali Javadi-Abhari and Matthew Treinish and Kevin Krsulich and Christopher J. Wood and Jake Lishman and Julien Gacon and Simon Martiel and Paul D. Nation and Lev S. Bishop and Andrew W. Cross and Blake R. Johnson and Jay M. Gambetta},
      year={2024},
      eprint={2405.08810},
      archivePrefix={arXiv},
      primaryClass={quant-ph},
      url={https://arxiv.org/abs/2405.08810}
}

@article{Sivarajah2021,
doi = {10.1088/2058-9565/ab8e92},
url = {https://doi.org/10.1088/2058-9565/ab8e92},
year = {2020},
month = {nov},
publisher = {IOP Publishing},
volume = {6},
number = {1},
pages = {014003},
author = {Sivarajah, Seyon and Dilkes, Silas and Cowtan, Alexander and Simmons, Will and Edgington, Alec and Duncan, Ross},
title = {t|ket⟩: a retargetable compiler for NISQ devices},
journal = {Quantum Science and Technology}
}

@article{Nickerson2014,
  title = {Freely Scalable Quantum Technologies Using Cells of 5-to-50 Qubits with Very Lossy and Noisy Photonic Links},
  author = {Nickerson, Naomi H. and Fitzsimons, Joseph F. and Benjamin, Simon C.},
  journal = {Phys. Rev. X},
  volume = {4},
  issue = {4},
  pages = {041041},
  numpages = {17},
  year = {2014},
  month = {Dec},
  publisher = {American Physical Society},
  doi = {10.1103/PhysRevX.4.041041},
  url = {https://link.aps.org/doi/10.1103/PhysRevX.4.041041}
}

@article{Manetsch2025,
  title = {A tweezer array with 6, 100 highly coherent atomic qubits},
  volume = {647},
  ISSN = {1476-4687},
  url = {http://dx.doi.org/10.1038/s41586-025-09641-4},
  DOI = {10.1038/s41586-025-09641-4},
  number = {8088},
  journal = {Nature},
  publisher = {Springer Science and Business Media LLC},
  author = {Manetsch,  Hannah J. and Nomura,  Gyohei and Bataille,  Elie and Lv,  Xudong and Leung,  Kon H. and Endres,  Manuel},
  year = {2025},
  month = sep,
  pages = {60–67}
}

@article{Barnes2022,
  title = {Assembly and coherent control of a register of nuclear spin qubits},
  volume = {13},
  url = {http://dx.doi.org/10.1038/s41467-022-29977-z},
  number = {1},
  journal = {Nature Communications},
  publisher = {Springer Science and Business Media LLC},
  author = {Barnes,  Katrina and Battaglino,  Peter and Bloom,  Benjamin J. and Cassella,  Kayleigh and Coxe,  Robin and Crisosto,  Nicole and King,  Jonathan P. and Kondov,  Stanimir S. and Kotru,  Krish and Larsen,  Stuart C. and others},
  year = {2022},
  month = may 
}

@article{Bluvstein2022,
  title = {A quantum processor based on coherent transport of entangled atom arrays},
  volume = {604},
  ISSN = {1476-4687},
  url = {http://dx.doi.org/10.1038/s41586-022-04592-6},
  DOI = {10.1038/s41586-022-04592-6},
  number = {7906},
  journal = {Nature},
  publisher = {Springer Science and Business Media LLC},
  author = {Bluvstein,  Dolev and Levine,  Harry and Semeghini,  Giulia and Wang,  Tout T. and Ebadi,  Sepehr and Kalinowski,  Marcin and Keesling,  Alexander and Maskara,  Nishad and Pichler,  Hannes and Greiner,  Markus and Vuletić,  Vladan and Lukin,  Mikhail D.},
  year = {2022},
  month = apr,
  pages = {451–456}
}

@misc{Haug2025,
      title={Lattice surgery with Bell measurements: Modular fault-tolerant quantum computation at low entanglement cost}, 
      author={Trond Hjerpekjøn Haug and Timo Hillmann and Anton Frisk Kockum and Raphaël Van Laer},
      year={2025},
      eprint={2510.13541},
      archivePrefix={arXiv},
      primaryClass={quant-ph},
      url={https://arxiv.org/abs/2510.13541}, 
}

@misc{DARPA2025HARQ,
  author      = {{Defense Advanced Research Projects Agency}},
  title       = {Program Solicitation: {Heterogeneous Architectures for Quantum (HARQ)}},
  number      = {DARPA-PS-25-31},
  institution = {U.S. Department of Defense},
  year        = {2025},
  month       = {August},
  url         = {https://sam.gov/opp/944007d554364a1aad811469028a7e73/view},
  howpublished = {\url{https://sam.gov/opp/944007d554364a1aad811469028a7e73/view}},
  note        = {Accessed: 2026-02-20}
}

@article{Horsman2012,
doi = {10.1088/1367-2630/14/12/123011},
url = {https://doi.org/10.1088/1367-2630/14/12/123011},
year = {2012},
month = {dec},
publisher = {IOP Publishing},
volume = {14},
number = {12},
pages = {123011},
author = {Horsman, Dominic and Fowler, Austin G and Devitt, Simon and Meter, Rodney Van},
title = {Surface code quantum computing by lattice surgery},
journal = {New Journal of Physics}
}

@article{Gidney2019,
  doi = {10.22331/q-2019-04-30-135},
  url = {https://doi.org/10.22331/q-2019-04-30-135},
  title = {Efficient magic state factories with a catalyzed {$|CCZ\rangle$} to {$2|T\rangle$} transformation},
  author = {Gidney, Craig and Fowler, Austin G.},
  journal = {{Quantum}},
  issn = {2521-327X},
  publisher = {{Verein zur F{\"{o}}rderung des Open Access Publizierens in den Quantenwissenschaften}},
  volume = {3},
  pages = {135},
  month = apr,
  year = {2019}
}

@article{Fowler2012,
  title = {Surface codes: Towards practical large-scale quantum computation},
  author = {Fowler, Austin G. and Mariantoni, Matteo and Martinis, John M. and Cleland, Andrew N.},
  journal = {Phys. Rev. A},
  volume = {86},
  issue = {3},
  pages = {032324},
  numpages = {48},
  year = {2012},
  month = {Sep},
  publisher = {American Physical Society},
  doi = {10.1103/PhysRevA.86.032324},
  url = {https://link.aps.org/doi/10.1103/PhysRevA.86.032324}
}

@misc{Cuccaro2004,
      title={A new quantum ripple-carry addition circuit}, 
      author={Steven A. Cuccaro and Thomas G. Draper and Samuel A. Kutin and David Petrie Moulton},
      year={2004},
      eprint={quant-ph/0410184},
      archivePrefix={arXiv},
      primaryClass={quant-ph},
      url={https://arxiv.org/abs/quant-ph/0410184}, 
}

@article{Selinger2013,
  title = {Quantum circuits of $T$-depth one},
  author = {Selinger, Peter},
  journal = {Phys. Rev. A},
  volume = {87},
  issue = {4},
  pages = {042302},
  numpages = {4},
  year = {2013},
  month = {Apr},
  publisher = {American Physical Society},
  doi = {10.1103/PhysRevA.87.042302},
  url = {https://link.aps.org/doi/10.1103/PhysRevA.87.042302}
}

\clearpage

\appendix

\section{Nomenclature}\label{app:nomenclature}
\cref{table:definition} summarizes the notations and numerical assumptions used throughout our heterogeneous resource and error modeling. The number of logical qubits in the QPU remains fixed at $N_\textrm{QPU}=3$, with the burden of system growth shifted to quantum memory, $N_\textrm{QM}$, satisfying the requirements FXD and SCL. 
In both QPU and RAQM, we choose the minimal code distances to achieve the target logical error per QEC cycle under standard surface-code scaling. Given our physical error assumptions (summarized later in \cref{table:compiler input}), these correspond to \mbox{$d_{\rm QPU}=15$} and \mbox{$d_{\rm QM}=9$} for the QPU and RAQM, respectively. 
The ancilla overhead, $c_{\rm anc}$, required for stabilizer measurements follows the standard surface-code construction~\cite{Fowler2012,Litinski2019}. 

The number of interconnects between the QPU and QM logical qubits, \mbox{$N_{\text{bdry}}=2N_\text{QPU}+N_\text{QM}$}, assumes two physical qubit interconnects per logical qubit in the QPU and one physical qubit per logical in the QM. The additional interconnect qubit required by the QPU enables a buffer to ensure smooth state transfer operations given the QPU's fast cycle time. 
The parameters $n_{\text{buf}}$ and $n_{\text{anc,pump}}$ quantify the physical-qubit overhead for Bell-pair buffering and one-sided entanglement purification on each QB rail respectively, and are taken from~\cite{Nickerson2014}. 
We assume $N_{\text{MF/QPU}}=3$ T-state factories per logical qubit in QPU, to ensure that a T-gate can be injected at any clock cycle. Reducing this number will reduce the physical qubit overhead of T-gates at the expense of the increased runtime and hence algorithmic error, as the QPU would idle while waiting for the T-states to be distilled. 
The QM cycle time $T_{\rm QM}$ is relevant for any interface utilizing a RAQM that implements full QEC. We consider two values intended to represent near-term capabilities ($1000\times T_\textrm{QPU}$) and future hardware ($50\times T_\textrm{QPU}$). 

For all quantities, \cref{table:poc_values_AQFT} and \cref{table:poc_values_RSA} contain the explicit proof-of-concept values used in this work for AQFT and RSA, respectively. These vary across two heterogeneous interfaces for AQFT and six interfaces for an RSA decryption algorithm.

\begin{table}[t!]
\centering
\renewcommand{\arraystretch}{1.3}
\begin{tblr}{colspec={X[1, l]X[3.5, l]}, width=\columnwidth}
\hline\hline
Symbol & Definition \\
\hline
$N_{\text{QPU}}$ & \# of logical qubits in QPU\\
$N_{\text{QM}}$ & \# of logical qubits in QM\\
$N_{\text{cache}}$ & \# of logical qubits in cache\\
$N_{\text{transfer}}$ & \# of transfer patches\\
$N_{\text{dist}}$ & Overhead for magic state distillation \\
$N_{\text{QPU,edges}}$ & \# of connections between QPU logical qubits \\
$N_{\text{bdry}}$ & \# of interconnects between QM and QPU logical qubits \\ 
$d_{\text{QPU}}$ & Distance of the QPU code \\
$d_{\text{QM}}$ & Distance of the QM code \\
$d_{\text{bdry}}$ & $\min(d_{\text{QPU}}, d_{\text{QM}})$ \\
$d_{\text{time}}$ & $\max(d_{\text{QPU}}, d_{\text{QM}})$ \\  
$c_{\text{anc}}$ & Ancilla qubit overhead for syndrome detection \\
$n_{\text{buf}}$ & \# of stored Bell-pair halves per link rail on the \textit{fast} module\\
$n_{\text{anc, pump}}$ & \# of ancilla qubits per rail for one-sided entanglement pumping \\
$N_{\text{MF/QPU}}$ & Number of magic state factories per logical qubit in QPU \\
$T_{\text{QM}}$ & Minimum QEC cycle time in QM \\
$T_{\text{QPU}}$ & QEC cycle time in QPU \\
$N_{\text{ASQPU}}$ & Number of logical qubits in ASQPU\\
\hline\hline
\end{tblr}
\caption{Notation for modeling heterogeneous execution with a fixed-size QPU, memory tier(s), and fault-tolerant QB.}\label{table:definition}
\end{table}

\begin{table}[t!]
\centering
\renewcommand{\arraystretch}{1.3}
\begin{tblr}{
colspec={X[1.8, l]||X[1.2, c]|X[1, c]|X[1, c]|X[1, c]},
width=\columnwidth,
colsep=0pt}
\hline\hline
\SetCell[r=2]{c} \textbf{POC value (AQFT)}
    &\SetCell[r=2]{c} {Baseline\\Arc.}
    & \SetCell[c=3]{c} Q-NEXUS Architectures &&
    \\ \hline
&& A1 & A2 & A3
    \\ \hline\hline
$N_{\text{QPU}}$
    & 1000
    & \SetCell[c=3]{c} 3 &&
    \\ \hline
$N_{\text{QM}}$
    & - 
    & \SetCell[c=3]{c} Algorithm size (1000) &&
    \\ \hline
$N_{\text{dist}}$
    & \SetCell[c=4]{c} 72 (T-state)~\cite{Gidney2019}&&&
    \\ \hline
$N_{\text{QPU,edges}}$
    & $2N_{\text{QPU}}$
    & \SetCell[c=3]{c} 3 &&
    \\ \hline
$N_{\text{bdry}}$
    & \SetCell[c=1]{c} - 
    & \SetCell[c=3]{c} $N_\text{QM}+2N_\text{QPU}$ &&
    \\  \hline
$d_{\text{QPU}}$
    & \SetCell[c=4]{c} 15 &&&
    \\ \hline
$d_{\text{QM}}$
    & \SetCell[c=1]{c} - 
    & \SetCell[c=1]{c} $d_{\text{QPU}}$ 
    & \SetCell[c=2]{c} 9 &
    \\ \hline
$d_{\text{bdry}}$
    & \SetCell[c=1]{c} - 
    & \SetCell[c=1]{c} - 
    & \SetCell[c=2]{c} 9 &
    \\ \hline
$d_{\text{time}}$
    & \SetCell[c=1]{c} - 
    & \SetCell[c=1]{c} - 
    & \SetCell[c=2]{c} 15 &
    \\ \hline
$c_{\text{anc}}$
    && \SetCell[c=3]{c}1 \cite{Litinski2019} &&
    \\ \hline
$n_{\text{buf}}$
    && \SetCell[c=3]{c}2 \cite{Nickerson2014} &&
    \\ \hline
$n_{\text{anc, pump}}$
    && \SetCell[c=3]{c}1-2 \cite{Nickerson2014} &&
    \\ \hline
$N_{\text{MF/QPU}}$
    & \SetCell[c=4]{c} 3~\cite{Gidney2019} &&&
    \\ \hline
$T_{\text{QM}}$
    & \SetCell[c=1]{c} - 
    & \SetCell[c=1]{c} - 
    & $1\,$ms \cite{Bluvstein2024}
    & $50\upmu$s \cite{Wang2025, Hughes2025}
    \\ \hline
$T_{\text{QPU}}$
    & \SetCell[c=4]{c} 1\,$\upmu$s~\cite{GoogleFaultTolerant2025, Castelvecchi2023} &&&
    \\
\hline\hline
\end{tblr}
\caption{Proof-of-concept baseline numerical assumptions for modeling heterogeneous execution with a fixed-size QPU, memory tier(s), and fault-tolerant QB.}\label{table:poc_values_AQFT}
\end{table}

\begin{table*}[t!]
\centering
\renewcommand{\arraystretch}{1.3}
\begin{tblr}{colspec={
    @{\hspace{1pt}}X[1.1, l]@{\hspace{1pt}}|| 
    @{\hspace{1pt}}X[0.9, c]@{\hspace{1pt}}|
    @{\hspace{1pt}}X[1.2, c]@{\hspace{1pt}}|
    @{\hspace{1pt}}X[1.1, c]@{\hspace{1pt}}|
    @{\hspace{1pt}}X[0.9, c]@{\hspace{1pt}}|
    @{\hspace{1pt}}X[1.2, c]@{\hspace{1pt}}|
    @{\hspace{1pt}}X[1.1, c]@{\hspace{1pt}}}, width=\textwidth}
\hline\hline
\SetCell[r=2]{c} \textbf{POC value (RSA)}
    & \SetCell[c=6]{c} Q-NEXUS Architectures &&&&&\\ \hline
& B1 & B2 & B3 & B4 & B5 & B6\\ \hline
{Storage (Code)/ \\Cache/ \\ASQPU}
    & \SetCell{c}{-- /\\STQM/  \\--}
    & \SetCell{c}{RAQM (Surface)/\\STQM/\\--}
    & \SetCell{c}{RAQM (Gross)/ \\STQM/\\--}
    & \SetCell{c}{-- /\\STQM/\\Adder}
    & \SetCell{c}{RAQM (Surface)/ \\STQM/\\Adder}
    & \SetCell{c}{RAQM (Gross)/ \\STQM/ \\Adder} \\ \hline\hline
$N_{\text{QPU}}$ 
    & \SetCell[c=6]{c} 6 &&&&&\\ \hline
$N_{\text{RAQM}}$ 
    & 0
    & 1254
    & 1260
    & 0
    & 1254
    & 1260\\ \hline
$N_{\text{cache}}$ 
    & 1399
    & 145
    & 145
    & 1399
    & 145
    & 145 \\ \hline
$N_{\text{transfer}}$ 
    & - 
    & 22
    & 26
    & - 
    & 22
    & 26\\ \hline
$d_{\text{QPU}}$ 
    & \SetCell[c=6]{c} 19 &&&&&
    \\ \hline
$d_{\text{LTS}}$ 
    & \SetCell[c=1]{c} - 
    & \SetCell[c=1]{c} 9 
    & \SetCell[c=1]{c} 12
    & \SetCell[c=1]{c} - 
    & \SetCell[c=1]{c} 9 
    & \SetCell[c=1]{c} 12
    \\ \hline
$N_{\text{dist}}$ 
    & \SetCell[c=6]{c} 12 (CCZ-state)~\cite{Gidney2025RSAUpdate} &&&&&\\ \hline
$N_{\text{MF/QPU}}$ 
    & \SetCell[c=6]{c} 0.67 &&&&&
    \\ \hline
$N_{\text{ASQPU}}$ 
    & \SetCell[c=6]{c} 37 &&&&&
    \\ \hline
$T_{\text{QM}}$ 
    & \SetCell[c=6]{c} 1ms &&&&& \\ \hline
$T_{\text{QPU}}$
    & \SetCell[c=6]{c} 1\,$\upmu$s &&&&&\\
\hline\hline
\end{tblr}
\caption{Proof-of-concept baseline numerical assumptions for modeling heterogeneous execution with a fixed-size QPU, memory tier(s), and fault-tolerant QB. Note, here the $N_\textrm{QPU}=6$ QPU consists of two 3Q-modules.}\label{table:poc_values_RSA}
\end{table*}

\begin{table*}[t] 
\centering
\renewcommand{\arraystretch}{1.5}
\begin{tblr}{
    width=\linewidth,
    colspec={
    X[1.3, l]X[4, l]X[1.6, l]},
    }
\hline\hline
\textbf{Interconnect}
    & \textbf{State transfer error budget (including idling)}
    & \textbf{State transfer duration} \\ \hline
Transversal teleportation
    & $2\epsilon_\text{QPU}+(\frac{\epsilon_\text{eff,idle}+\epsilon_\text{tele}}{\epsilon_{\text{th}}})^{(d_\text{QPU}+1)/2}$
    & $2T_\text{QPU}$ \\
Lattice surgery swap 
    & $2d_\text{time} [\epsilon_\text{QM} + \epsilon_\text{QPU}(T_\text{QM}/T_\text{QPU})]+N_\text{idle}\epsilon_\text{QM}$ 
    & $2d_\text{time}T_\text{QM}$ \\
\hline\hline
\end{tblr}
\caption{State transfer error and duration for the two heterogeneous interconnects. Notation is defined in \cref{table:definition}. \label{table:SWAP analysis}}
\end{table*}

\begin{table*}[t]
\centering
\renewcommand{\arraystretch}{2.2} 
\begin{tblr}{
    width=\linewidth,
    colspec={X[1.05, l]X[1.7, l]X[2.15, l]X[2.6, l]},
    }
\hline\hline
    & \textbf{Homogeneous} 
    & \textbf{Heterogeneous type 1} 
    & \textbf{Heterogeneous type 2} \\ \hline
\makecell[lc]{\textbf{Logical qubits}\\\textbf{\& stabilizers}}
    & $N_{\text{QPU}}(1+c_{\text{anc}})d_{\text{QPU}}^2$
    & $N_{\text{QM}}d_{\text{QPU}}^2 + N_{\text{QPU}}(1+c_{\text{anc}})d_{\text{QPU}}^2$
    & $N_{\text{QM}}(1+c_{\text{anc}})d_{\text{QM}}^2+N_{\text{QPU}}(1+c_{\text{anc}})d_{\text{QPU}}^2$ \\
\makecell[lc]{\textbf{Local}\\\textbf{Couplers}}
    &$2N_{\text{QPU}}(1+c_{\text{anc}})d_{\text{QPU}}^2$
    &$2N_{\text{QPU}}(1+c_{\text{anc}})d_{\text{QPU}}^2+N_{\text{bdry}}d_{\text{QPU}}^2$
    &$2N_{\text{QM}}(1+c_{\text{anc}})d_{\text{QM}}^2+2N_{\text{QPU}}(1+c_{\text{anc}})d_{\text{QPU}}^2 + N_{\text{bdry}}d_{\text{bdry}}$\\
\makecell[lc]{\textbf{QM - QPU}\\\textbf{interconnect}}
    & \makecell[cc]{---}
    & $N_{\text{bdry}}d_{\text{QPU}}^2(1+n_{\text{anc, pump}})$
    & $N_{\text{bdry}}d_{\text{bdry}}(2+n_{\text{buf}}+n_{\text{anc, pump}})$ \\
\makecell[lc]{\textbf{QPU lattice}\\\textbf{surgery}}
    & $2N_{\text{QPU,edges}}d_{\text{QPU}}$ 
    & $2N_{\text{QPU,edges}}d_{\text{QPU}}$ 
    & $2N_{\text{QPU,edges}}d_{\text{QPU}}$ \\ 
\makecell[lc]{\textbf{T-gate}\\\textbf{injection}}
    & $2d_{\text{QPU}}N_{\text{QPU}}$ 
    & $2d_{\text{QPU}}N_{\text{QPU}}$ 
    & $2d_{\text{QPU}}N_{\text{QPU}}$ \\
\makecell[lc]{\textbf{QSF T-state}\\\textbf{distillation}}
    & $N_{\text{MF/QPU}}N_{\text{dist}}N_{\text{QPU}}d_{\text{QPU}}^2$ 
    & $N_{\text{MF/QPU}}N_{\text{dist}}N_{\text{QPU}}d_{\text{QPU}}^2$ 
    & $N_{\text{MF/QPU}}N_{\text{dist}}N_{\text{QPU}}d_{\text{QPU}}^2$ \\
\hline\hline
\end{tblr}
\caption{Qubit resource estimation for the different architectures. Heterogeneous type 1 refers to architectures using short-term STQM without active QEC, while Heterogeneous type 2 refers to architectures with error-corrected RAQM. Notation is defined in \cref{table:definition}. \label{table:qubit_resources}}
\end{table*}

\section{Resource estimation}
\label{app:resource estimate}
We provide analytical expressions of the resource estimates used in \cref{sec:midnalysis} and \cref{sec:rsa}. \cref{table:SWAP analysis} details the expressions used to estimate the logical state transfer error (including memory idling) and duration for the two different interconnect protocols shown in \cref{fig:protocol}. We proceed by describing the protocol for transversal teleportation and lattice surgery swap, which are used to transfer data between the logical QPU qubits and the QM qubits, either in the short-term ULC-qubit STQM or long-term error-corrected RAQM.

\subsection{Transversal teleportation}

Transversal teleportation requires one-to-one matching between physical qubits in the QPU and QM. The QPU logical qubit is encoded by $d^2$ physical qubits and therefore the QM must contain a corresponding set of $d^2$ physical qubits (see \cref{fig:rsa_architecture}). This holds for both the STQM and the RAQM (where the $d^2$ qubits make up the ``transfer patch''). To allow fast transfer time, transversal teleportation is the only interconnect used for the STQM. The general transfer protocol is as follows.
\begin{enumerate}
    \item Prepare rails: To establish a quantum connection, we store Bell pairs. Assuming a 100~MHz pair-generation rate with error rate $10^{-3}$~\cite{DARPA2025HARQ}, we generate $n_\text{buf}$ long-range physical Bell pairs between the $d^2$ physical qubits in the QPU and their corresponding qubits in the QM. We then run multiple rounds of one-sided entanglement purification \cite{Nickerson2014} on the QPU logical qubit, leveraging its fast cycle time. The purified halves are stored in advance in small buffers on each rail, and can be accessed when required.
    \item \textit{(Optional)} If transferring to a STQM made up of ultra-long-coherence (ULC) qubits, we can optionally exploit their biased idling noise, with bias factors up to $T_1/T_2 \sim 55$ \cite{Wang2025}. We leverage this bias by applying Hadamard rotations on the data qubits in the QPU surface code prior to the state teleportation. Thus, the idling errors accrued in the STQM follow the favorable, high-threshold ($>20\%$) XZZX code.
    \item Transfer on demand: With a high-fidelity EPR connection established, transversal teleportation can be achieved within a single code cycle.
    \item QM storage: The storage behavior varies between the short-term STQM and long-term RAQM.

    \textbf{Short-term STQM:} once the data are transferred to ULC storage qubits defining the STQM, the qubits idle freely with no active error correction. As long as the physical errors accrued during this idling period are smaller than the \textit{physical} error assumed in the QPU, then, after the state is transferred back to the QPU, all errors will be corrected to a level set by the QPU error per cycle. 
    
    \textbf{Long-term RAQM:} information transfer via transversal teleportation between QPU and RAQM qubits is enabled via large ``transfer'' patches, which are sparsely and strategically placed in the QM. Information is then shuttled to storage patches protected by the QM code. During idling, the memory qubits are then protected from all error sources and have less stringent requirements on the physical error rates compared to the QPU.

    \item State retrieval: Prepare rails as per step 1, and transfer information back from the $d^2$ memory qubits to the QPU. For the STQM, any necessary Hadamard rotations are applied once the data are transferred back to the QPU.
\end{enumerate}

If information is transferred transversally to the RAQM via its transfer patches, it must subsequently be shuttled within the memory. In \cref{fig:rsa_architecture}d, we show an example of a RAQM setup: surface code patches are tiled in a QM, with each patch representing a logical qubit. The large transfer patches use a distance $d_\text{QPU}$ code, while the small patches use a distance $d_\text{QM}<d_\text{QPU}$ code. Shuttling information within the RAQM is accomplished via lattice-surgery-based \textit{local} swap operations. To satisfy requirement LNG, long-range routing is not permitted. The transfer patches are placed such that no logical qubit in memory is more than $k$ swaps away from a transfer patch. In \cref{fig:rsa_architecture}d, the colors of the small patches represent the swap-distance (Manhattan) from the closest transfer patch. For $N$ logical qubits in memory, placing $N/(2k^2 + 6k +1)$ transfer patches is sufficient to guarantee a maximum swap-distance of $k$. In order to leverage the long-coherence (LC) qubits composing the RAQM, the QEC cycle times are dynamically adjusted to minimize the number of QEC cycles. For more details on modular clock cycles, see \cref{sec:clock_Synchronization}.

The resource requirements for transversal teleportation are summarized in \cref{table:SWAP analysis}. The state transfer overhead is two QPU cycles, with transfer time $2T_\textrm{QPU}$. The total error accumulated from the effective memory idling error, $\epsilon_\text{eff,idle}$, and the teleportation error, $\epsilon_\text{tele}$, can subsequently be error-corrected in the QPU using its surface code, provided it remains below the threshold $\epsilon_\text{th}$. 

\subsection{Lattice-surgery transfer}

A different approach for RAQM relies on the lattice-surgery-based transfer protocol in \cref{fig:protocol}b. With this approach, the information transfer protocol is:
\begin{enumerate}
    \item Prepare rails: Similar to the transversal teleportation.
    \item Lattice surgery merge-split: Consume \mbox{$2d_\text{bdry} = 2\min(d_\text{QPU}, d_\text{QM})$} Bell pairs to perform lattice-surgery ``merge'' checks (ZZ and XX type) along a boundary of length $d_\text{bdry}$. The long-range parity checks are repeated \mbox{$d_\text{time}=\max(d_\text{QPU},d_\text{QM})$} to suppress time-like error chains; $d_\text{time}$ is chosen to equalize the effective 3D space-time code distance between the QPU and QM codes. The decoder utilizes anisotropic weights to account for differences in the code distance, and Pauli-frame updates are done classically.
    \item QM storage: qubits idle with protection from the QM code. 
    \item State retrieval: Prepare rails as per step 1, and execute lattice surgery as per step 2. Importantly, classical Pauli-frame corrections can be performed in later steps and do not block logical operations on the QPU.
\end{enumerate}

This approach does not require large transfer patches or information shuffling within the RAQM. On the other hand, it relies on a slower transfer mechanism that may introduce runtime bottlenecks.

For the lattice surgery swap, the state transfer overhead $2d_{\text{time}} = 2\max(d_{\text{QPU}}, d_{\text{QM}})$ represents the overhead necessitated by the lattice surgery \cite{Horsman2012}, and the memory error is determined by the logical error in memory, $\epsilon_\text{QM}$, as well as the idling time $N_\text{idle}$ (\cref{table:SWAP analysis}).

\subsection{Architecture resources}

\cref{table:qubit_resources} lists the expressions used to estimate the number of physical qubits required for three types of architecture: (1) a homogeneous QPU-only device; (2) a heterogeneous architecture using STQM and transversal teleportation interconnect (interface 1); and (3) a heterogeneous architecture using RAQM and lattice-surgery SWAP (interface 2). The number of physical qubits needed for a logical qubit, including the ancillae $c_\text{anc}$ needed for stabilizer measurements, follow the standard surface code construction \cite{Fowler2012, Litinski2019}. Note that for STQM, no ancillae are required in memory as there is no need to perform active error correction. The number of physical qubits acting as local couplers also follows the standard surface code construction. In addition, we require physical qubits for the $N_{\text{bdry}}$ interconnects in the two heterogeneous architectures. The estimates for QPU lattice surgery, as well as the factor of $2$ in the expression for the lattice-surgery swap interconnect utilized in interface 2, are taken from \cite{Litinski2019}. The factors of $n_{\text{buf}}$ and $n_{\text{anc, pump}}$ represent the physical qubit overhead associated with Bell pair buffer and purification, respectively~\cite{Nickerson2014}. Note that the transversal teleportation protocol requires no Bell pair buffer.
Physical qubit overheads for the T-state injection and distillation are taken from \cite{Litinski2019} and \cite{Gidney2019}, although the T-state cultivation protocol \cite{Gidney2024} may further reduce these numbers.

\section{Leveraging extreme physical coherence time in QEC context}\label{app: coherence}
While long physical-level coherence times are undeniably a desirable property, it is not straightforward to translate them to advantages in a QEC context.
For example, qubit modalities (such as REI or NA) have $T_1$ that ranges from tens of minutes to many days \cite{Konz2003,Zhong2015}. However, the typical physical error rates they allow are not much better than lower coherence modalities ($1 \times 10^{-4}$ vs.~$5 \times 10^{-4}$), and typical cycle times are in the range $1-5\,$ms, which seems at odds with the fact that $T_\textrm{cycle}/T_1\sim 10^{-7}$, three orders of magnitude lower than the reported physical error. In such modalities, the physical error is dominated by physical operations (single- and two-qubit gates, and measurements) and not by idling. 

This error asymmetry provides a knob that allows the extension of the cycle time without changing the physical error, and hence, the logical error per cycle. This is a somewhat counterintuitive knob, as it goes against the community effort to shorten cycle times. Indeed, when QEC is used to process data, the duration of \textbf{logical} operations scales with the QEC cycle time, rendering cycle times of $100\,$ms impractical for relevant large scale algorithms. 
However, if one can separate the rate of logical execution from long-time storage, long cycle times become an advantage rather than an obstacle.

For example, assuming a logical qubit idle for a full second. Code A, with a $1\,\upmu$s cycle time, must run $10^{6}$ QEC cycles, while code B, with a $250\,$ms cycle time run only four cycles during the idling period. If we assume identical logical error accumulation during the full (1 sec) time span, we find:
\begin{equation}
    (1 - e_A)^{10^{6}} = (1-e_B)^4 \to e_B \sim 2.5\times10^{5} e_A,
\end{equation}
where $e_A, e_B$ are the logical errors per cycle of each code. This means that a much lower distance code can be used for code B without sacrificing the overall idling error. 
Moreover, in modalities that are not $T_1$ dominated, increasing the cycle duration allows for slower but higher quality operations. 
A $d=25$ surface-code with physical error $p/p_{th} = 1/12$ and cycle time $1\upmu$s has a $1\,$s idling error of $3.33\times 10^{-10}$, while a $d=9$ surface-code with physical error $p/p_{th} = 1/60$ ($5\times$ better) and cycle time $250\,$ms has a $1\,$s idling error of $1.54\times 10^{-10}$, hence, $2.1\times$ better error with a much lower distance code.

More formally, if code A has a physical error to threshold ratio of $p$, cycle time $T$ and distance $d_A$, and code B has a physical error to threshold ratio of $p/\kappa$, and cycle time $R T$, the distance $d_B$ that provide similar long time idling to code A is given by:

\begin{equation}
    d_B = 2\frac{\log \left(p^{\frac{d_A+1}{2}} R\right)}{\log{\left(\frac{p}{\kappa}\right)}} - 1
\end{equation}

As before, assuming $p = 1/12$, $d_A=25$ and $\kappa=5$, we find that for any $RT > 137\,$ms a code with $d_B=9$ provides better idling protection than a $d_A=25$ code with cycle time $T$.

In general, the cycle time of code B can be thought of as a range. The lower end of the range is determined by the physical operation duration and the fastest possible syndrome detection. The upper bound is set such that the idling error of that range is still lower than the assumed physical error rate.

\section{Compiler input}
\cref{table:compiler input} summarizes the baseline physical and logical parameters provided to the proof-of-concept compiler for both the homogeneous superconducting architecture and the heterogeneous QPU+QM+QSF setting. For the baseline QPU (and the homogeneous reference), we assume a distance-$15$ surface code operated with a $1\,\upmu$s QEC cycle, consistent with recent superconducting fault-tolerance experiments and roadmap-scale devices~\cite{GoogleFaultTolerant2025,Castelvecchi2023}. For the heterogeneous memory tier, STQM is modeled as a REI repetition-code memory without active QEC, leveraging demonstrated multi-hour coherence~\cite{Wang2025}, while RAQM is modeled as a slower surface-code memory with a representative physical error rate of $10^{-4}$~\cite{Hughes2025} and a cycle time range $T_{\rm QM}\in[50,1000]\,\upmu$s to bracket near-term and look-ahead readout/gate assumptions. Non-Clifford resources are supplied by a photonic QSF, with a representative T-gate latency of $30\,\upmu$s consistent with fast magic-state generation/distillation and injection assumptions in photonic approaches~\cite{rudolph2023logical,Gidney2024}. Together, these inputs allow the compiler to model module-specific logical gate errors, idling, and state-transfer costs under heterogeneous timing, and to quantify how performance depends on interface latency and QPU--QM clock mismatch.

\begin{table*}[t!] 
\centering
\renewcommand{\arraystretch}{1.3} 
\begin{tblr}{width=\linewidth, colspec={X[1.8, l]|X[1, l]X[1, l]X[1, l]X[1, l]|X[1.1, l]}}
\hline\hline
\SetCell[r=2]{l}\textbf{Architecture} & \SetCell[c=4]{c} \textbf{Heterogeneous} &&&& \textbf{Homogeneous} \\
 & QPU & QSF & STQM & RAQM ($\frac{T_{\rm QM}}{T_{\rm QPU}} = 1000$) & \SetCell[c=1]{c}- \\ \hline
\textbf{Qubit type} & SC & Photonic & ULC (REI) & LC (NA) & SC \\
\textbf{Logical Connectivity} & 2D grid & - & None & None & 2D grid \\
\textbf{Logical qubits ($k$)} & 3 & - & $k$ & $k$ & $k$ \\
\textbf{QEC distance, ($d$)} & 15 & - & 15$^\dagger$ & 9 & 15 \\
\textbf{Logical gate set} & Clifford & T-state & - & - & Clifford + T \\
\textbf{Error: Physical} & $5 \times 10^{-4}$ & - & $1 \times 10^{-10 \ast}$ & $1 \times 10^{-4}$ & $5 \times 10^{-4}$ \\
\textbf{Error: Logical / QEC} & $7 \times 10^{-11}$ & - & - & $3.8 \times 10^{-11}$ & $7 \times 10^{-11}$ \\
\textbf{Error: Logical 2Q} & $1 \times 10^{-9}$ & - & - & - & $1 \times 10^{-9}$ \\
\textbf{Error: Logical T} & - & $2.1 \times 10^{-9}$ & - & - & $2.1 \times 10^{-9}$ \\
\textbf{Error: State transfer} & - & - & $1.4 \times 10^{-10}$ & $2.1 \times 10^{-6}$ & - \\
\textbf{Duration: QEC [s]} & $1 \times 10^{-6}$ & - & - & $1 \times 10^{-3}$ & $1 \times 10^{-6}$ \\
\textbf{Duration: 2Q [s]} & $1.5 \times 10^{-5}$ & - & - & - & $1.5 \times 10^{-5}$ \\
\textbf{Duration: T-gate [s]} & - & $3 \times 10^{-5}$ & - & - & $3 \times 10^{-5}$ \\
\textbf{Duration: Transfer [s]} & - & - & $2 \times 10^{-6}$ & $3 \times 10^{-2}$ & - \\
\hline\hline
\end{tblr}
\caption{Compiler input parameters for heterogeneous (QPU+QSF+QM) and homogeneous architectures. $^\dagger$For STQM, no active error correction is applied. The code distance represents the size of logical encoding. $^\ast$For STQM, physical error is due to idling only.\label{table:compiler input}}
\end{table*}

\section{RSA-2048 analysis} \label{app:rsa}

\csubsubsec{Parallelization} The parallelization of quantum algorithms in general is beyond the scope of this manuscript, but we benchmark with RSA-2048 factorization, so it is worth looking at the parallelism of Shor's algorithm in particular. Shor's algorithm is amenable to ``extreme'' parallelism with the use of ``fanout'' techniques resulting in a drastic reduction in algorithm depth $\mathcal{O}(\log^3 n)$, where $n$ is the binary size of the input \cite{Pham2013}. However, this comes at great cost increasing algorithm width to $\mathcal{O}(n^4)$ and volume to $\mathcal{O}(n^4 log^3 n)$ (in contrast to the best known width $\mathcal{O}(n)$ and $\mathcal{O}(n^3)$ volume scaling \cite{Pham2013,Gidney2025RSAUpdate}). Fault-tolerant quantum computers have significantly higher resource requirements to expanding logical width, compared to depth, hence the preference for optimal width and volume algorithms. The algorithm we implement has optimal scaling for both width and volume, and has limited parallelism, so there is rapid diminishing returns, resulting in only two cores being used in our RSA-2048 analysis.

\csubsubsec{Resource estimation} We use a larger distance code to ensure that the error accumulated in the RSA-2048 circuit is about 90\%. Physical qubit resources presented in \cref{table:expanded_qubit_resources_RSA} use the formulae provided in \cref{table:qubit_resources} with updated parameter values $d_\textrm{QPU}=19$ and replacing T-factory with a CCZ factory\cite{Gidney2025RSAUpdate}. Since the QPU only contains 3-qubits, there are two state transfer operations between subsequent CCZ operations. Thus we only need 2 CCZ factories per QPU module without slowing down the runtime.

\begin{table*}[t]
\centering
\renewcommand{\arraystretch}{1.3}
\begin{tblr}{colspec={llllll}, width=\textwidth}
\hline\hline
\textbf{Storage} & \textbf{STQM} & \textbf{ASQPU} & \textbf{\makecell[lc]{$\boldsymbol{\tau_\textrm{adder}}$(ms)}} & \textbf{\makecell[lc]{$\boldsymbol{\tau_\textrm{lookup}}$(ms)}} & \textbf{\makecell[lc]{$\boldsymbol{\tau_\textrm{phaseup}}$(ms)}} & \textbf{\makecell[lc]{Runtime (day)}} \\
\hline
Baseline\cite{Gidney2025RSAUpdate}& --- &---& 2 & 2 & 1 & 5.6\\
\hline
STQM & --- & --- & 5.2 & 2.2 & 0.15 & 9.2\\
STQM & --- & Adder & 2 & 2.2 & 0.15 &  4.9\\
RAQM (Surface code) & STQM & --- & 5.2 & 2.2 & 0.15 & 9.2\\
RAQM (Surface code) & STQM & Adder & 2 & 2.2 & 0.15 & 4.9\\
RAQM (qLDPC)& STQM & --- & 5.2 & 2.2 & 0.15 & 9.2\\
\hline\hline
\end{tblr}
\caption{Detailed resource estimation for a 2048-RSA factorization circuit with 1399 logical qubits, comparing space-time tradeoffs between different heterogeneous and homogeneous architectures. We consider the architecture proposed by Gidney, in \cite{Gidney2025RSAUpdate}, to the state-of-art baseline which holds under realistic assumptions.
\label{table:expanded_qubit_resources_RSA}}
\end{table*}

For estimating the qubit resources mentioned in \cref{table:qubit_resources_RSA}, we use the following parameter values from \cref{table:poc_values_RSA} in \cref{table:rsa_qubit_resources}.


\begin{table*}[t]
\centering
\renewcommand{\arraystretch}{2.2} 
\begin{tblr}{
    width=\linewidth,
    colspec={X[1.05, l]X[1.8, l]X[1.7, l]X[1.7, l]X[1.7, l]X[1.8, l]},
    }
\hline\hline
    & \makecell[lc]{\textbf{Logical qubits}\\\textbf{\& stabilizers}} 
    &\makecell[lc]{\textbf{Local}\\\textbf{couplers}} 
    &\makecell[lc]{\textbf{Non-local}\\\textbf{couplers}} 
    & \makecell[lc]{\textbf{Cache}\\\textbf{interconnect}} 
    & \makecell[lc]{\textbf{Clifford}\\\textbf{overhead}} 
    & \makecell[lc]{\textbf{CCZ-factory}\\\textbf{\& injection}}\\ \hline
\makecell[lc]{QPU \\(Surface)}
    & $2N_{\text{QPU}}d_{\text{QPU}}^2$
    & $4N_{\text{QPU}}d_{\text{QPU}}^2$
    & \makecell[cc]{---}
    & $(2N_{\text{QPU}}+N_{\text{cache}})d_\text{QPU}^2$
    & $2N_{\text{QPU}}d_{\text{QPU}}$ 
    & $2N_{\text{QPU}}(4d_{\text{QPU}}^2+2d_{\text{QPU}})$\\
Cache (STQM)
    & $N_{\text{cache}}d_\text{QPU}^2$
    & $N_{\text{cache}}d_\text{QPU}^2$
    & \makecell[cc]{---}
    & \makecell[cc]{---}
    & \makecell[cc]{---} 
    & \makecell[cc]{---}\\
\makecell[lc]{LTS \\(Surface)}
    & $2N_{\text{LTS}}d_{\text{LTS}}^2 + 2N_\text{transfer}d_{\text{QPU}}^2$ 
    & $4N_{\text{LTS}}d_{\text{LTS}}^2 + 4N_\text{transfer}d_{\text{QPU}}^2$
    & \makecell[cc]{---}
    & $N_{\text{transfer}}d_{\text{QPU}}^2$
    & $2N_{\text{LTS}}d_{\text{LTS}}$
    & \makecell[cc]{---}\\
\makecell[lc]{LTS \\(Gross)}
    & $\sim24N_\text{LTS}+ 2N_\text{transfer}d_{\text{QPU}}^2$
    & $\sim48N_\text{LTS} + 4N_\text{transfer}d_{\text{QPU}}^2$
    & $\sim24N_\text{LTS}$
    & $N_{\text{transfer}}d_{\text{QPU}}^2$
    & \makecell[cc]{0$^a$}
    & \makecell[cc]{---}\\
Adder ASQPU
    & $2N_{\text{ASQPU}}d_{\text{QPU}}^2$ 
    & $4N_{\text{ASQPU}}d_{\text{QPU}}^2$
    & \makecell[cc]{---}
    & $N_{\text{ASQPU}}d_{\text{QPU}}^2$
    & \makecell[cc]{---}
    & $12(4d_{\text{QPU}}^2+2d_{\text{QPU}})$\\
\hline\hline
\end{tblr}
\caption{Qubit resource estimation for the different extended architectures for RSA-2048 as defined in \cref{app:rsa} $^a$Shift automorphisms don't require additional ancilla qubits. \label{table:rsa_qubit_resources}}
\end{table*}

\end{document}